%% file: paper_v19.tex
\documentclass[11pt,a4paper]{article}
\usepackage{epsf,epsfig,psfrag,graphics,color}

\usepackage{amsmath}
\usepackage{amssymb}
\usepackage{appendix}
\usepackage{rotating}
\usepackage{multirow}
\usepackage{geometry}
\usepackage{slashed}
\usepackage{float}
\everymath{\displaystyle}
\usepackage{cite}

\usepackage{graphicx,axodraw}

\usepackage{soul}

\usepackage{color}
\definecolor{nicered}{rgb}{0.7,0.1,0.1}
\definecolor{nicegreen}{rgb}{0.1,0.5,.1}

\usepackage{hyperref}

\begin{document}


\vspace{0.3cm}

\begin{center}
{\bf {\Large Constraining the gauge and scalar sectors of\\
the doublet left-right symmetric model
}}

\vspace{0.5cm}

V\'eronique Bernard$^{a}$, S\'ebastien Descotes-Genon$^{b}$, Luiz Vale Silva$^{c}$

\vspace{0.2cm}

\emph{$^a$ Groupe de Physique Th\'eorique, Institut de Physique 
Nucl\'eaire, UMR 8606,\\
CNRS, Univ. Paris-Sud, Universit\'e Paris-Saclay, 91405 Orsay Cedex, France}

\vspace{0.1cm}

\emph{$^b$ Laboratoire de Physique 
Th\'eorique, UMR 8627,\\
CNRS, Univ. Paris-Sud, Universit\'e Paris-Saclay, 91405 Orsay Cedex, France}

\vspace{0.1cm}

\emph{$^c$ IFIC, Universitat de Val\`{e}ncia - CSIC \\
Parc Cient\'{i}fic, Catedr\'{a}tico Jos\'{e} Beltr\'{a}n 2, E-46980 Paterna, Spain
}

\end{center}

\vspace{2.0cm}

\begin{abstract}
We consider a left-right symmetric extension of the Standard Model where the spontaneous breakdown of the left-right symmetry is triggered by doublets. The 
electroweak $\rho$ parameter is 
protected from large corrections in this Doublet Left-Right Model (DLRM), contrary to the triplet case. This allows in principle for more diverse patterns of symmetry breaking. We consider several constraints on 
the gauge and scalar sectors of DLRM: the unitarity of scattering processes involving gauge bosons with longitudinal polarisations, 
the radiative corrections to the muon $\Delta r$ parameter
and the electroweak precision observables measured at the $Z$ pole and at low energies. Combining these constraints within the frequentist CKMfitter approach, we see that the fit pushes the scale of left-right symmetry breaking up to a few TeV, while favouring an electroweak symmetry breaking triggered not only by the $ SU(2)_L \times SU(2)_R $ bi-doublet, which is the case most commonly considered in the literature, but also by the $ SU(2)_L $ doublet.

\end{abstract}

\vfill
\eject

\section{Introduction}

Left-Right (LR) symmetric models constitute a category of extensions of the SM
that explains the left-handed structure of the SM through the existence of a larger gauge group
$SU(3)_C\otimes SU(2)_L\times SU(2)_R\times U(1)_X$. This group is broken first at a high-energy scale
$\mu_R$ (of the order of the TeV or higher), inducing a difference between left and right sectors, followed by an electroweak symmetry breaking occurring at a lower scale $\mu_W$~\cite{Pati:1974yy,Mohapatra:1974hk,Mohapatra:1974gc,Senjanovic:1975rk,Senjanovic:1978ev}. This extension yields heavy spin-1 $W'$ and $Z'$ bosons, the former predominantly coupling to right-handed fermions, introducing a new ``CKM-like" matrix for right-handed quarks (and similarly for leptons). LR models also lead to new charged and neutral scalar bosons with an interesting pattern of flavour-changing currents~\cite{Chang:1982dp,Zhang:2007da}.
Such a framework has been revived in the recent years for its potential collider implications when parity restoration in the LHC energy reach is considered~\cite{Maiezza:2010ic,Guadagnoli:2010sd}. Interestingly, recent studies of anomalies in rare $b$-decays suggest also the interest of having right-handed currents in order to provide a consistent explanation of all the measurements~\cite{Alguero:2019ptt,Aebischer:2019mlg,
Arbey:2019duh,Ciuchini:2019usw,Alok:2019ufo}.

Stringent constraints come from electroweak precision observables~\cite{Hsieh:2010zr} and from direct searches at LHC~\cite{Aad:2019wvl,Khachatryan:2014dka,Chen:2013fna,Dev:2015kca,Patra:2015bga,Alioli:2017ces}, pushing the scale for LR models to several TeV. Studies in the framework of flavour physics suggest also that the structure for the right-handed ``CKM-like" matrix should be quite different from the left-handed one~\cite{Harari:1983gq,Beall:1981ze,Langacker:1989xa,Barenboim:1996nd,Barenboim:2001vu}. A particularly important indirect constraint comes from kaon-meson mixing, pushing again the mass scale for the new scalar particles up to a few TeV or beyond~\cite{Mohapatra:1977mj,Mohapatra:1983ae,Barenboim:1996wz,Blanke:2011ry,Bertolini:2014sua,Bertolini:2019out}, for which we reassessed short-distance QCD corrections in order to reach the accuracy now requested for such processes~\cite{Bernard:2015boz}.

Various mechanisms can be invoked to trigger the breakdown of the left-right symmetry.
Historically, LR models (LRM) were first considered with doublets in order to break the left-right symmetry spontaneously. Later the focus was set on triplet models (Triplet Left-Right Model, or TLRM), due to their ability to generate both Dirac and Majorana masses for neutrinos and thus introduce a see-saw mechanism~\cite{Mohapatra:1980yp,Deshpande:1990ip}. We would like to reassess the possibility of a left-right symmetry breaking due to doublets (Doublet Left-Right Model, or DLRM). On one hand, it prevents us from providing a see-saw mechanism for neutrinos, for which there is however no experimental evidence yet, but on the other hand, it protects the electroweak $\rho$ parameter from large contributions 
already at the tree level and thus allows in principle a low left-right symmetry breaking scale $\mu_R$ (which might or might not be in contradiction with other phenomenological constraints). 

 In this article, we will discuss three classes of constraints on the DLRM related to the presence of heavy gauge bosons and scalars that affect the dynamics of the light gauge bosons $W$ and $Z$: the unitarity of the processes  involving the scattering of two gauge bosons,
the radiative corrections to the muon $\Delta r$ parameter,
and the electroweak precision observables measured at the $Z$-pole and at low energies.
We will combine these three constraints into a global fit using the frequentist approach of the CKMfitter collaboration~\cite{Charles:2004jd,Charles:2015gya,Charles:2016qtt} in order to constrain 
the parameters of the model. 
We leave the discussion of flavour and the combination of all constraints for future work, due to the large number of additional parameters involved.

In Sec.~\ref{sec:basics}, we discuss the basic features of our model. In Sec.~\ref{sec:bounds}, we consider the constraints coming from the preservation of unitarity for the scattering of gauge bosons. In Sec.~\ref{sec:custodial}, we discuss the breakdown of the custodial symmetry from the $W$ and $Z$ self-energies induced in DLRM. In Sec.~\ref{sec:EWPO}, we discuss the status of electroweak precision observables in these models. In Sec.~\ref{sec:fits}, we perform a fit taking into account all these constraints and discuss the outcome for the parameters of the DLRM, before concluding in Sec.~\ref{sec:concl}. Appendices are devoted to the expressions for EW precision observables, the spectrum, the Feynman rules associated with the DLRM and some aspects of the renormalisation.

\section{The doublet left-right model} \label{sec:basics}

\subsection{Gauge structure and symmetry breaking}

Let us start with the gauge structure of the doublet left-right model (DRLM) and the pattern of its symmetry breaking.
We consider the group  $SU(3)_C\otimes SU(2)_L\times SU(2)_R\times U(1)_{B-L}$ 
with gauge couplings $g_C$, $g_L$, $g_R$ and $g_X$, where $B-L$ is the 
difference between the baryon and the lepton numbers.\footnote{Thus, 
contrary to the SM, baryon and lepton numbers are not accidental symmetries in 
LR models.} At a scale of a few TeV (or higher), a symmetry breaking 
$SU(2)_R\times U(1)_{B-L}\to U(1)_Y$ with $Y$ the hypercharge is triggered by a doublet $\chi_R$ $(1,1,2,1/2)$.   
One has in terms of the component fields of the doublet:
\begin{equation}
\chi_{L,R}=\left(\begin{array}{c}\chi_{L,R}^+\\ \chi_{L,R}^0/\sqrt{2}\end{array}\right)
 \qquad \langle\chi_L\rangle=\left(\begin{array}{c}0\\v_L e^{i\theta_L}/\sqrt{2}\end{array}\right)\qquad
\langle\chi_R\rangle=\left(\begin{array}{c}0\\v_R/\sqrt{2} \end{array}\right)
\end{equation}
with the expected scale $v_R= \mathcal{O} (1$ TeV) and where one has introduced a  doublet $\chi_L$ ($1,2,1,1/2)$ to preserve a left-right symmetric structure
with  $v_L$ at most a few hundred GeV. We will denote the real and imaginary 
parts of these fields as $\chi_L^0=(v_L e^{i\theta_L} + \chi_{L,r}^0 + i 
\chi_{L,i}^0)$ and $\chi_R^0=(v_R + \chi_{R,r}^0 + i \chi_{R,i}^0)$. Note that
performing an obvious extension  of the definition of the electric charge of right-handed fields
in terms of the right weak isospin and the $B-L$ quantum numbers the 
hypercharge acquires a simple meaning in LR models.

At a lower scale, the spontaneous breaking of   
$SU(3)_C\otimes SU(2)_L\times U(1)_Y \to  SU(3)_C\otimes U(1)_Q$ leads to the standard electroweak symmetry breaking. 
It is triggered on one hand by the doublet $\chi_L$ and on the other hand by 
the bidoublet $\phi$ $(1,2,2,0)$ whose presence is mandatory to provide a mass to the fermions, see Sec.~\ref{sec:fermions},

\begin{equation}
\phi=\left(\begin{array}{cc}\phi_1^0/\sqrt{2} & \phi_2^+\\ \phi_1^- & \phi_2^0/\sqrt{2}\end{array}\right)\qquad
\langle\phi\rangle=\left(\begin{array}{cc}\kappa_1/\sqrt{2}  & 0\\ 0 & \kappa_2 e^{i\alpha}/\sqrt{2}\end{array}\right)
\end{equation}
with the conjugate bidoublet $\tilde\phi=\tau_2 \phi^* \tau_2$ transforming similarly. 

It proves useful to  introduce the real ratios
\begin{equation}
 \epsilon=\frac{\kappa_1}{v_R} \qquad r=\frac{\kappa_2}{\kappa_1} \qquad w=\frac{v_L}{\kappa_1} 
\end{equation}
as well as the definitions: 
\begin{equation}
k^2=1+r^2+w^2 \qquad v^2=v_R^2k^2\epsilon^2=v_L^2+\kappa_1^2+\kappa_2^2
\label{eq:paramL}
\end{equation}
Due to the hierarchy of scales involved $\epsilon$ is a small quantity. We will thus perform in the
following an expansion in this parameter. As we will see below, at Leading Order (LO), namely, $\epsilon=0$, the right and the left gauge and scalar fields decouple except for the neutral gauge 
bosons;
mixing starts at Next-to-Leading Order (NLO) $\epsilon=1$ or Next-to-Next-to-Leading Order (NNLO) $\epsilon=2$ depending whether one considers scalars or charged gauge bosons respectively. In the following we will neglect
terms of order $\epsilon^3$ and higher unless specified. We will also consider $r<1$.\footnote{There is some 
subtlety in solving the stability equations discussed in the next section in the limit $r=1$. However as
the hierarchy $m_b \ll m_t$ demands $r$ to be smaller than one, 
see Sec.~\ref{sec:fermions} below, we will not enter into such detail in the following.}

In the case of left-right symmetry breaking triggered by triplets (TLRM), the equivalent of the ratio $ w $ is taken as very small and neglected, on the 
basis of the breaking of the custodial symmetry which already occurs  
at LO. As we will see below, this is not the case anymore when the left-right symmetry breaking is triggered by doublets, and we will thus leave open the possibility that $w$ is of order 1 or larger, letting data constrain its value. Furthermore, for simplicity, we will work under the assumption that
\begin{equation}
 \alpha=\theta_L=0  
\label{eq:cpviol} 
\end{equation}
i.e., no additional sources of CP violation come from the breaking of the gauge symmetries.

\subsection{Spin-0 sector}

Scalar self-interactions are described by the following potential

\begin{eqnarray}
V&=&-\mu_1^2 \langle\phi^\dag.\phi\rangle-\mu_2^2 \langle\tilde\phi.\phi^\dag+\tilde\phi^\dag.\phi\rangle-\mu_3^2 (\chi_L^\dag.\chi_L+\chi_R^\dag.\chi_R)\\\nonumber
&&\quad +
\mu'_1 (\chi_L^\dag.\phi.\chi_R+\chi_R^\dag.\phi^\dag.\chi_L)+\mu'_2(\chi_L^\dag.\tilde\phi.\chi_R+\chi_R^\dag.\tilde\phi^\dag.\chi_L)\\\nonumber
&&\quad +
\lambda_1 \langle\phi^\dag.\phi\rangle^2+\lambda_2 (\langle\tilde\phi.\phi^\dag\rangle^2+\langle\tilde\phi^\dag.\phi\rangle^2)+\lambda_3 \langle\phi^\dag.\tilde\phi\rangle \langle\tilde\phi^\dag.\phi\rangle+\lambda_4 \langle\phi^\dag.\phi\rangle \langle\tilde\phi.\phi^\dag+\phi.\tilde\phi^\dag\rangle\\\nonumber
&&\quad +\rho_1 [(\chi_L^\dag.\chi_L)^2+(\chi_R^\dag.\chi_R)^2]+
\rho_3 [(\chi_L^\dag.\chi_L)(\chi_R^\dag.\chi_R)]\\\nonumber
&&\quad +
\alpha_1 \langle\phi^\dag.\phi\rangle[(\chi_L^\dag.\chi_L)+(\chi_R^\dag.\chi_R)]\\\nonumber
&&\quad +\frac{\alpha_2}{2} e^{i \delta_2}[\langle\tilde\phi.\phi^\dag\rangle(\chi_L^\dag.\chi_L)+\langle\phi.\tilde\phi^\dag\rangle (\chi_R^\dag.\chi_R)] \\\nonumber
&&\quad +\frac{\alpha_2}{2} e^{-i \delta_2}[\langle\tilde\phi^\dag.\phi\rangle(\chi_L^\dag.\chi_L)+\langle\phi^\dag.\tilde\phi\rangle (\chi_R^\dag.\chi_R)]\\\nonumber
&&\quad +
\alpha_3 (\chi_L^\dag.\phi.\phi^\dag.\chi_L+\chi_R^\dag.\phi^\dag.\phi.\chi_R)+
\alpha_4 (\chi_L^\dag.\tilde\phi.\tilde\phi^\dag.\chi_L+\chi_R^\dag.\tilde\phi^\dag.\tilde\phi.\chi_R)
\end{eqnarray}
where $\langle \rangle$ denotes the trace.
For simplicity, we have imposed invariance under the discrete left-right symmetry\footnote{See Refs.~\cite{Branco:1985ng,Ecker:1980at,Maiezza:2010ic} for other discrete symmetries
in the context of LR Models.} in the scalar sector of the theory

\begin{center}
$\chi_L\leftrightarrow \chi_R$, $\phi\leftrightarrow \phi^\dag$ 
\end{center}
The scalar potential for doublet fields has
a slightly different structure compared to the triplet case~\cite{Deshpande:1990ip}, and in particular trilinear terms are allowed. 
In addition to Eq.~(\ref{eq:cpviol}) we will set for simplicity  $\delta_2=0$, i.e. there are no new sources of CP violation from the scalar sector.

One can minimize the potential with respect to the parameters $ v_R, v_L, \kappa_1, \kappa_2 $, giving four conditions for the stability of the vacuum state. These four equations provide relations among the vacuum expectation values $ \kappa_{1,2} $ and $v_{L,R} $, and the underlying parameters of the potential, $ \mu^2_{1,2,3} $, $ \alpha_{1,2,3,4} $, $ \mu'_{1,2} $, $ \rho_{1,3} $ and $ \lambda_{1,2,3,4} $:
\begin{eqnarray}
&&\hspace{-0.65truecm} 0 = \kappa_{2} \left( B + 4 \lambda_{4} \kappa^{2}_{1} \right) + \kappa_{1} \left[ A + \alpha_{4} (v_R^2 + v^{2}_{L}) + 4 \left( \lambda_{3} + 2 \lambda_{2} \right) \kappa^{2}_{2} \right] + \sqrt{2} {\mu'}_{2} v_L v_R + \alpha_{2} \kappa_{2} \left( v_R^2 + v^{2}_{L} \right) , \\
&&\hspace{-0.65truecm} 0 = \kappa_{1} \left( B + 4 \lambda_{4} \kappa^{2}_{2} \right) + \kappa_{2} \left[ A + \alpha_{3} (v_R^2 + v^{2}_{L}) + 4 \left( \lambda_{3} + 2 \lambda_{2} \right) \kappa^{2}_{1} \right] + \sqrt{2} {\mu'}_{1} v_L v_R + \alpha_{2} \kappa_{1} \left( v_R^2 + v^{2}_{L} \right) , \\
&&\hspace{-0.65truecm} 0 = C + 2 \rho v^{2}_{L} + \sqrt{2} \frac{v_L}{v_R} \left( {\mu'}_{1} \kappa_{2} + {\mu'}_{2} \kappa_{1} \right) + 2 \alpha_{2} r  \kappa_{1}^2  , \\
&&\hspace{-0.65truecm} 0 = \frac{v_L}{v_R} \left( C + 2 \rho v_R^2 \right) + \sqrt{2} \left( {\mu'}_{1} \kappa_{2} + {\mu'}_{2} \kappa_{1} \right) + 2 \alpha_{2} r w \kappa_{1}^{2} ,
\end{eqnarray}
where $ \rho \equiv \rho_3 / 2 - \rho_1 $, and
\begin{eqnarray}
A &\equiv& -2 \mu^{2}_{1} + \alpha_{1} (v_R^2 + v^{2}_{L}) + 2 \lambda_{1} (\kappa^{2}_{1} + \kappa^{2}_{2}) , \\
B &\equiv& -4 \mu^{2}_{2} + 2 \lambda_{4} (\kappa^{2}_{1} + \kappa^{2}_{2}) , \\
C &\equiv& -2 \mu^{2}_{3} + 2 \rho_{1} (v_R^2 + v^{2}_{L}) + \alpha_{1} (\kappa^{2}_{1} + \kappa^{2}_{2}) + \alpha_{4} \kappa^{2}_{1} + \alpha_{3} \kappa^{2}_{2} .
\end{eqnarray}
It is useful to further define the combinations of parameters
\begin{equation}
A' \equiv  -2 \mu^{2}_{1} + \alpha_{1} v_R^2\,, \quad
B' \equiv  -4 \mu^{2}_{2}\,, \quad
C' \equiv  -2 \mu^{2}_{3} + 2 \rho_{1} v_R^2 .
\end{equation}
The minimisation conditions at leading order in $ \epsilon \ll 1 $ yield:
\begin{eqnarray}
\mathcal{O}(\epsilon^2) &=& r \frac{B'}{v_R^2} + \left( \frac{A'}{v_R^2} + \alpha_{4} \right) + \sqrt{2} \frac{{\mu'}_{2}}{v_R} w + \alpha_{2} r , \label{eq:equation1simpDoub}\\
\mathcal{O}(\epsilon^2) &=& \frac{B'}{v_R^2} + r \left( \frac{A'}{v_R^2} + \alpha_{3} \right) + \sqrt{2} \frac{{\mu'}_{1}}{v_R} w + \alpha_{2} , \label{eq:equation2simpDoub}\\
\mathcal{O}(\epsilon^2) &=& \frac{C'}{v_R^2} , \label{eq:equation3simpDoub}\\
\mathcal{O}(\epsilon^2) &=& w 2 \rho_{} + \sqrt{2} \left( r \frac{{\mu'}_{1}}{v_R} + \frac{{\mu'}_{2}}{v_R} \right) . \label{eq:equation4simpDoub}
\end{eqnarray}
We have four equalities (up to higher orders in $ \epsilon $) and three parameters $ \{ r, w, v_R \} $ related to the vacuum expectation values of the scalar fields, that we will exploit in order to eliminate the explicit dependence on some parameters of the scalar potential.\footnote{Note that in principle one could solve the stability equations for the parameters $ \{ r, w, v_R \} $ using three out of the four equations, and plugging the solutions into the fourth. In other words a combination of parameters in the scalar potential 
($\mu^2_{1,2} / \mu^2_{3}, \alpha_{1,2,3,4}, \rho_{1,3}, \mu'_{1} / \mu'_{2}$) is $ \mathcal{O} (\epsilon^2)$ and thus vanishes at leading order in $\epsilon$. Whether this resulting combination is stable under radiative corrections, thus requiring
or not a certain amount of tuning, remains to be verified.}

As far as  scalar states are concerned, the theory contains 2 neutral and 2 charged Goldstone bosons 
(linked to the massive $W,Z,W',Z'$ gauge bosons, see App.~\ref{app:scala}),  5 neutral extra scalar bosons (3 CP-even $H_{1,2,3}^0$ and 2 CP-odd $A_{1,2}^0$), and 2 charged  scalar bosons ($H_{1,2}^\pm$),\footnote{Contrary to the TLRM, the DLRM has no doubly charged scalars.}
 as well as a neutral light scalar $h^0$ (corresponding to the light SM-like Higgs boson) of mass $ \mathcal{O} (\epsilon^2)$ given as

\begin{eqnarray}
        &&M_h^2=\frac{v_R^2 \epsilon ^2}{2 k^2} \left(4 \Big( \lambda_1 \left(r^2+1\right)^2+4 r \left(\lambda_4 (1+r^2)+r \lambda^+_{23} \right) 
        +w^2 \left(\alpha_{124}+r^2 (\alpha_1 + \alpha_3)+\alpha_2 r\right)+\rho_1 w^4 \Big) \right .\nonumber\\
        && \qquad \left. - \frac{1}{\rho_1}\left(\alpha_{124}+r^2 (\alpha_1+\alpha_3)+\alpha_2 r+2 \rho_1 w^2\right)^2\right)
\label{eq:massh0}
\end{eqnarray}
with
\begin{equation} 
 \alpha_{124} \equiv \alpha_1+r \alpha_2 + \alpha_4 \, , \quad \lambda^+_{23} 
\equiv 2\lambda_2 + \lambda_3
\label{eq:param}
\end{equation}
In the limit $r,w\to 0^+$,
one gets the simplified formula 
$M_{h}^2=v_R^2\epsilon^2(2\lambda_1-(\alpha_1+\alpha_4)^2/(2\rho_1))$.
The expressions for the masses of the new scalar bosons are  given in App.~\ref{app:scala} at leading order in $ \epsilon $. Note that at that order the CP-even, CP-odd and charged scalars have equal masses $M_i$ for each value of $i$.  One can express some of the parameters of the scalar potential in terms of these $M_i$

\begin{equation}\label{eq:mu2prParameter}
\mu'_2 = \frac{w (-1 \pm X)}{\sqrt{2} v_R (1+r x) (1 + \beta(x) w^2)} (M^2_{1} + M^2_{2})
\end{equation}

\vspace{-0.35truecm}

\begin{eqnarray}
\alpha_{34} &\equiv& \alpha_3 -\alpha_4= - \frac{1}{(r^2 + 1) v_R^2 w} \Bigg( 2 (r^2 - 1) (M^2_{1} + M^2_{2}) w \nonumber\\
&& \qquad + \sqrt{2} \mu'_2 v_R (r^3 x + r^2 + r (2 w^2 - 1) x - 2 w^2 - 1) \Bigg)
\label{eq:alpha34Parameter}
\end{eqnarray}
with
\begin{equation}
X=\sqrt{1- \frac{4 \delta^2}{(1+ \delta^2)^2} \frac{\left(1+r^2\right)\left(1+ \beta(x) w^2\right)}{k^2}}
\label{eq:root}
\end{equation}
and
\begin{equation}
\beta(x)=(1+x^2)/(1+r x)^2  \, , \quad \quad \delta=M_{1}/M_{2} 
 \, , \quad \quad x=\mu_1'/\mu_2'
\label{eq:parax}
\end{equation}

The neutral and charged scalar physical fields decompose as follows up to $ \mathcal{O} (\epsilon^2)$
\begin{eqnarray}
h^0&=&\left(\frac{1}{k}+ \epsilon^2 c_{\phi_1}^{h^0}\right) \phi_{1,r}^0
+\left(\frac{r}{k} + \epsilon^2 c_{\phi_2}^{h^0}\right) \phi_{2,r}^0+\left(\frac{w}
{k}+ \epsilon^2 c_{\chi_{L,r}}^{h^0}\right) \chi_{L,r}^0+\epsilon c_{\chi_{R,r}}^{h^0}\chi_{R,r}^0  
\nonumber\\
H_i^0&=&\left(- \frac{r +t_i w}{u_i}+ \epsilon^2 c_{\phi_1}^{H_i}\right) \phi_{1,r}^0
+\left(\frac{1}{u_i} + \epsilon^2 c_{\phi_2}^{H_i}\right) \phi_{2,r}^0+\left(\frac{t_i}
{u_i}+ \epsilon^2 c_{\chi_{L,r}}^{H_i}\right) \chi_{L,r}^0+\epsilon c_{\chi_{R,r}}^{H_i}\chi_{R,r}^0  
\nonumber\\
A_i^0&=&\left(\frac{r +t_i w}{u_i}+ \epsilon^2 c_{\phi_1}^{A_i}\right) \phi_{1,i}^0
+\left(\frac{1}{u_i} + \epsilon^2 c_{\phi_2}^{A_i}\right) \phi_{2,i}^0+\left(\frac{t_i}
{u_i}+ \epsilon^2 c_{\chi_{L,r}}^{A_i}\right) \chi_{L,i}^0+\epsilon c_{\chi_{R,r}}^{A_i}\chi_{R,i}^0  
\nonumber\\
H_3^0&=& \epsilon ( c_{\phi_1}^{H_3} \phi_{1,r}^0
+  c_{\phi_2}^{H_3} \phi_{2,r}^0+
  c_{\chi_{L,r}}^{H_3} \chi_{L,r}^0) +\bigl(1+\epsilon^2 c_{\chi_{R,r}}^{H_3}\bigr)\chi_{R,r}^0 
\nonumber\\
H_i^\pm&=&\left( \frac{r +t_i w}{u_i}+ \epsilon^2 d_{\phi_1}^{H_i}\right) \phi_{1,r}^0
+\left(\frac{1}{u_i} + \epsilon^2 d_{\phi_2}^{H_i}\right) \phi_{2,r}^0+\left(\frac{t_i}
{u_i}+ \epsilon^2 d_{\chi_{L,r}}^{H_i}\right) \chi_{L,r}^0+\epsilon d_{\chi_{R,r}}^{H_i}\chi_{R,r}^0
\label{eq:dech}
\end{eqnarray}
where the various coefficients of the $\epsilon$ and $\epsilon^2$ terms are combinations
of the parameters of the scalar potential.
The scalar $H_3^0$ is the analogue of the SM Higgs boson in the right sector of the theory at LO in $\epsilon$. The quantities
$t_i$ and $u_i$ arising from the
determination of the mass eigenstates are defined as
\begin{equation}
t_1 \equiv p \, , \quad t_2\equiv q \, , \quad u_i=\sqrt{1 + t_i^2 +(r + w t_i)^2} 
\label{eq:tu}
\end{equation}
where
\begin{equation}
p=-\frac{
   k^2 (1+r x)(1-\delta^2 +\left(1 + \delta^2\right)X)+2 w^2 r(r-x)}{2 w  \left(k^2 (r - x\delta^2  \right)- r^2 (r-x))}
\label{eq:parp}
\end{equation}
The parameter $q$ is related to $p$ via the following relation:
\begin{equation}
q=-\frac{1 + r^2 + p r w}{ r w + p(1+ w^2)}
\end{equation}
and can be obtained from Eq.~\eqref{eq:parp} through the replacement $\delta \to 1/\delta$.
Here and in the following we assume $r<1$ and $\delta >1$.\footnote{If $\delta <1$ or $r>1$ $M_{1}$ and $M_{2}$
have to be swapped and consequently also the expressions for $p$ and $q$.}
In the limit $w \to 0^+$ one has $ \delta={\cal O}(1/\sqrt{w}) $, $p={\cal O}(w^2)$ and $q=-(1+r^2)/(r w)$, with $ \rho_3={\cal O}(1/w) $, see App.~\ref{app:scala},
leading to similar expressions as in Ref.~\cite{Blanke:2011ry} where the TRLM is considered.

As shown in App.~\ref{app:scala},
some coefficients in the above expansion can be expressed in terms of the functions $F_{i=1,2}(r,w,p)$  given by:
\begin{equation}\label{eq:F1}
F_i(r,w,p)=\frac{-1+r^2+ r w t_i}{k u_i}   \\
\end{equation}
These functions also occur in the couplings of the new
Higgs bosons to the quarks, see App.~\ref{app:Hfermions}.
Similarly, one can determine the couplings of the extra Higgs bosons to the gauge bosons, see App.~\ref{app:Hgauge}, which involve another  combination of interest:
\begin{equation}\label{eq:defS}
 S_i =t_i/u_i
\end{equation}

\subsection{Spin-1 sector}\label{sec:spin1}

For the gauge sector (as can be seen for instance in ref.~\cite{Hsieh:2010zr}), one can express
the light $W$ and heavy $W'$ bosons in terms of left and right gauge bosons  up to terms of ${\cal O}(\epsilon^2)$
\begin{equation}
W^\pm_\mu=W^\pm_{L,\mu} +\frac{2c_Ws_R r}{s_W} \epsilon^2 W^\pm_{R,\mu}
\qquad W^{\prime\pm}_\mu=W^\pm_{R,\mu}-\frac{2c_Ws_R r}{s_W} \epsilon^2 W^\pm_{L,\mu} \, ,
\end{equation} 
whereas the physical neutral gauge bosons (massless $A$, light $Z$ and heavy $Z'$) identify as
\begin{eqnarray}
A&=&s_W W^3_L+s_R c_W W^3_R+c_R c_W B \\
Z&=&c_W W^3_L-s_R s_W\left(1-\frac{c_R^2 k z_h}{s_W^2} \epsilon^2\right) W^3_R-c_R s_W\left(1+\frac{s_R^2 k z_h}{ s_W^2}\epsilon^2\right) B \\
Z'&=&-\frac{s_R c_R c_W k z_h}{ s_W}\epsilon^2 W^3_L+c_R\left(1 + s_R^2 k z_h\epsilon^2\right) W^3_R - s_R\left(1-c_R^2 k z_h\epsilon^2\right) B
\end{eqnarray}
with the (sines of the) leading-order mixing angles 
\begin{equation}\label{eq:mixingangles}
s_R=\frac{g_X}{\sqrt{g_X^2+g_R^2}} \qquad s_W=\frac{s_R}{\sqrt{(g_L/g_R)^2+s_R^2}}
\end{equation}
and $c_A=\sqrt{1-s_A^2}$, with an obvious notation for sines and co-sines. We also have
\begin{equation}\label{eq:gRexpression}
\frac{e}{g_L}= s_W \qquad  \frac{g_L}{g_R}=\frac{c_W s_R}{s_W}
\end{equation}

\vspace{3mm}

It is well known that the SM possesses an accidental (global) symmetry called the custodial symmetry. Indeed
before the breaking of the $SU(2)_L \times U(1)_Y$ gauge symmetry, the Higgs potential has a global $SU(2)_L \times SU(2)_R$ symmetry  which  reduces to $SU(2)_V$ when the gauge symmetry  is broken, see for example Ref.~\cite{Willenbrock:2004hu} for a detailed review.  This residual 
custodial symmetry can most easily be seen by  rewriting the Higgs field as a bidoublet under this global symmetry.  Under the assumption that the hypercharge 
gauge coupling vanishes, $g'=0$,  the kinetic part of the Lagrangian is also invariant under the custodial 
symmetry and the gauge bosons $W^\pm$ and $Z$ form a degenerate multiplet. Indeed, in that limit  $s_W=0$ and:
\begin{equation}
\rho \to M_W^2/ M_Z^2 =1
\end{equation}
When $g' \neq 0$ it is easy to see that in the SM the mass matrix for the
gauge bosons can be obtained by replacing $W_\mu^3 \to W_\mu^3 - g'/g B_\mu=
Z_\mu/\cos \theta_W$
so that  the $ \rho$ parameter becomes:
\begin{equation}
\rho \equiv M_W^2/ (M_Z^2 \cos^2 \theta_W) =1
\end{equation} 
Small deviations from $ \rho = 1 $ arise when including radiative corrections. Note that $\sin \theta_W^2$ is renormalisation scheme dependent and various definitions of this parameter exist in the literature with slightly different
numerical values. Correspondingly, there are various definitions of the 
$\rho$ parameter \cite{Tanabashi:2018oca}. In the on-shell renormalisation scheme that will be discussed below, the above equation is promoted to a definition of the renormalised $s_W^2$ at all orders in perturbation theory.

In the DLRM, one can also illustrate this custodial symmetry by writing the two Higgs doublet
fields as two bi-doublets under the $SU(2)_L \times SU(2)_R$ symmetry:
\begin{equation}
\phi_{L,R}=
\begin{pmatrix}
    {\chi_{L,R}^{0 \, *}} & \chi_{L,R}^+\\
         -\chi_{L,R}^- & \chi_{L,R}^0    
  \end{pmatrix}
\end{equation}
with the following transformation properties:
\begin{equation}
\phi_{L,R} \to U_{R,L}\, \phi_{L,R}\, U_{L,R}^\dagger
\end{equation}
with $U_{L,R} \in SU(2)_{L,R}$.

Diagonalising the mass matrix for the gauge bosons, one finds
\begin{eqnarray} 
M_{W,W'}^2&=&\frac{1}{8}\biggl (g_L^2  v^2 + g_R^2  V^2 \mp \sqrt{
   4 g_L^2 g_R^2 \kappa^4 + (g_L^2  v^2 - g_R^2  V^2)^2 -4 g_L^2 g_R^2 (\kappa_1^2 -\kappa_2^2)^2}\biggr)
\\
M_{Z,Z'}^2&=&\frac{1}{8}\biggl( g_L^2 v^2 + g_R^2 V^2 + 
\frac{g_R^2 s_R^2}{c_R^2} \bigl(- 2 \kappa^2 + V^2 + v^2\bigr) \mp \sqrt{4 g_L^2 g_R^2 \kappa^4 + (g_L^2  v^2 -  g_R^2  V^2)^2 +\frac{g_R^2 s_R^2}{c_R^2} \Delta}\biggr) 
\nonumber\\
\Delta&=&2 g_L^2\bigl(\kappa^4 +(v^2-\kappa^2)^2-V^2v^2\bigr)+2 g_R^2\bigl(\kappa^4 +(V^2-\kappa^2)^2-V^2v^2\bigr)+\frac{g_R^2 s_R^2}{c_R^2} \bigl(-2 \kappa^2 +V^2+v^2\bigr)^2 \nonumber
\label{eq:mgaugeb}
\end{eqnarray}
 where the plus (minus) signs
are for the heavy (light) gauge bosons, $v$ is defined in Eq.~\eqref{eq:paramL},
\begin{equation}
V^2= \kappa^2 +v_{R}^2   \quad \quad {\rm and} \quad \quad \quad \kappa^2=\kappa_1^2+\kappa_2^2 .
\end{equation}
In the limit  $g_X \to 0$, we get  $s_W,s_R\to 0$ with a fixed ratio $s_R/s_W =g_L/g_R$. 
Clearly the only difference between the two equations in that limit comes from the last term 
in the square root in the first equation which cancels when  $r\to 1$, leading to degenerate neutral and charge gauge bosons, so that $M_{W'}=M_{Z'}$ and $M_{W}=M_{Z}$. Indeed 
 if $\kappa_1 =\kappa_2$ ($r=1$), the kinetic energy Lagrangian is  invariant
under the custodial symmetry, similarly to the SM.

Expanding the masses, Eq.~(\ref{eq:mgaugeb}) in $\epsilon$,  the first two terms 
read:
\begin{equation}\label{eq:MgaugelightNLO}
M_W=M_W^0\left(1-\epsilon^2 \frac{w_h^2}{2}\right) \quad \quad \quad \quad M_Z=M_Z^0\left(1- \epsilon^2 \frac{z_h^2}{2}\right)
\end{equation}
where
\begin{eqnarray}
M_W^0&=&\frac{1}{2}g_L v \, , \quad M_Z^0=\frac{M_W^0}{c_W}
\nonumber \\ 
w_h&=& \frac{2 r }{k} \, ,\quad  z_h=k \left(c_R^2 -\frac{w^2}{k^2}\right)=w_h +\frac{(1-r)^2 -s_R^2 w^2}{k}
\label{eq:masswzLO}
\end{eqnarray}
The equation for  $z_h$ in terms of  $w_h$ illustrates that $M_W= M_Z$ for $r=1$ and $s_W=s_R=0$.
This leads to the relation:
\begin{equation}
\rho \equiv \frac{M_W^2}{M_Z^2 c_W^2} =1 +\epsilon^2 f(1-r,s_R,w)+ {\cal O}(\epsilon^4)
\end{equation}
with $f$ a function which can be determined from the previous equations,
so that one recovers the SM relation at LO in $\epsilon$, contrary to the TLRM for which the relation
is violated already at LO in $\epsilon$ if $w \neq 0$. Typically in the latter
one has 
$\rho= (1+r^2+2 w^2)/(1+r^2 +4 w^2)+{\cal O}(\epsilon^2)$, see for example \cite{Branco:1999}, so that $w$ has to be much smaller than $1+r^2$ and is usually neglected. Thus the DLRM, which does not trigger a breaking of the custodial symmetry for $v_L\neq 0$ at $\epsilon=0$ contrarily to TRLM, allows an easier fulfilment of electroweak-precision tests for a non-vanishing $v_L$.
 
For the heavy gauge bosons one has 
\begin{equation}\label{eq:MgaugeheavyNLO}
M_{W'}=M_{W'}^0\left(1+\epsilon^2 \frac{1+r^2}{2}\right)\,, \quad \quad \quad \quad M_{Z'}=M_{Z'}^0\left(1+ \epsilon^2 \frac{c_R^4(1+r^2)+s_R^4 w^2}{2}\right)\,,
\end{equation}
with expressions at LO in $\epsilon$ similar to the SM ones up to the replacement $L \to R$ 
\begin{equation}
M_{W'}^0=\frac{1}{2}g_R v_R \, , \quad M_{Z'}^0=\frac{M_{W'}^0}{c_R}
\label{eq:masswpzpLO}
\end{equation}
so that one can define an equivalent relation to the SM case
\begin{equation}
\rho_R \equiv \frac{M_{W'}^2}{M_{Z'}^2 c_R^2} =1 +{\cal O}(\epsilon^2)
\end{equation}


\subsection{Spin-1/2 sector}\label{sec:fermions}

We have focused on the gauge and scalar sectors of the DLRM which are the main focus of our work here. For completeness, we discuss briefly the fermion sector, although further detail would be needed to account for flavour constraints properly.
The Yukawa interactions are given by 
\
\begin{equation}
\mathcal{L}_Y = - \bar{Q}_{L,i} ( y_{ij} \phi + \tilde{y}_{ij} \tilde{\phi} ) Q_{R,j} + h.c.
\end{equation}
in the interaction basis.
After electroweak symmetry breaking,
the diagonalisation of the Yukawa matrices yields the mass matrices  $ M_u $ and 
$ M_d $ and two unitary ``CKM-like" matrices $V_L$ and $V_R$ connecting mass and interaction bases.
One gets the following structure for the Yukawa matrices
\begin{equation}
y=\frac{\sqrt{2}}{(1-r^2)v_R\epsilon}(M_u-r V_L M_d V_R^\dag)\,, \qquad 
\tilde{y}=\frac{\sqrt{2}}{(1-r^2)v_R\epsilon}(V_L M_d V_R^\dag-rM_u )\,.
\qquad 
\end{equation}
As can be seen the limit $r =1$ is not allowed, as it creates additional degeneracies among quark flavours at tree level.
Here the usual choice is made to assign all the redefinition from interaction to mass states to the down-type fermions (of left and right chiralities). The same ``CKM-like" matrices arise in the couplings of the gauge bosons and scalars once fermions are expressed in terms of mass eigenstates.
Note that in the case discussed here and contrary to other realizations of LR models in which discrete symmetries are imposed at the energy scales of spontaneous breaking of the LR gauge group \cite{Senjanovic:2014pva, Senjanovic:2015yea}, the two matrices $ V_L $ and $ V_R $ are not simply related, thus resulting in a large number of new parameters in the flavour sector. 
We will thus leave the important constraints from this sector to a future work, however
we note that the overall good consistency of studies of CP violation with the Kobayashi-Maskawa mechanism embedded into the Standard Model~\cite{Koppenburg:2017mad} is expected to imply important constraints on deviations of $ V_L $ with respect to the SM picture, and on the structure of $ V_R $.

A similar discussion could hold for the lepton part, but in the following we are going to neglect neutrino masses, meaning that no mixing matrix is then needed for the lepton part. Let us however stress that there is no possibility to generate a Majorana mass term for the neutrinos using the doublets, so that the neutrinos are Dirac particles. 

Finally let us just mention that
the couplings of the fermions to the SM-like scalar and the gauge bosons $W$ and $Z$ have the same form as in the SM 
up to corrections of order ${\cal O}(\epsilon^2)$, see Appendix \ref{app:Hfermions}.

\subsection{Parameters}

In the DLRM, one has the following parameters in the Lagrangian:
\begin{itemize}
\item the parameters having an immediate equivalent in the SM, namely the fermion masses (9, corresponding to quarks and charged leptons, since we neglect the neutrino masses) and the CKM-like matrix $ V_L $ (depending on 4 parameters).

\item the analogue of the CKM matrix in the right sector $V_R$
leading to 3 moduli and 6 phases.

\item the gauge couplings $g_C$, $g_R$, $g_L$ and $g_X$. Here, we will allow $g_R$ and  $g_L$ to be independent of one another, i.e., we will not restrict ourselves to the fully left-right symmetric case.

\item the symmetry breaking $SU(2)_R \times U(1)_{B-L} \to U(1)_Y$ scale $v_R$. 

\item the electroweak
breaking scale involving  the three parameters $\epsilon$, $r$, and $w$.

\item the 15 parameters of the scalar potential. At the order at which we work 
and after exploiting the stability conditions, the only ones that are needed for our present study are $\mu'_1$, $\mu'_2$, $\rho_1$, $\lambda_1$, $\lambda_4$,  $\alpha_2$, the combinations $\alpha_{34}$ 
as well as $\alpha_{124}$ and $\lambda^+_{23}$
defined respectively in Eqs.~\eqref{eq:alpha34Parameter} and \eqref{eq:param}, and
\begin{equation}
 \lambda^-_{23} \equiv 2\lambda_2 - \lambda_3
\end{equation}
\end{itemize}

In principle, one could extract constraints on these parameters directly from the data. But it turns out more interesting to re-express some of these parameters in terms of observables. This has been the method used in the Standard Model
where the choice of the input scheme was depending on the observables to
be determined, see for instance \cite{Hollik:1993cg}. One may for example  
trade  $g_L$, $ g_Y $, $\epsilon v_R$ and one of the parameters of the Higgs potential for the $Z$-boson mass $m_Z$,  the electromagnetic constant $\alpha$, the Fermi constant $G_F$ and the light scalar mass $M_{h}$ as done in the SM. 
Instead of using $g_R$ and $v_R$, we may use the co-sine of the mixing angle $c_R$ and the mass of the heavy gauge boson $W'$.
Our final set of parameters will be given in Section~\ref{sec:fits} after having
discussed our strategy to perform the fits.

We aim at constraining some of these parameters from the phenomenology of the weak gauge bosons. Before considering electroweak precision observables, it is interesting to discuss the constraints coming from general requirements, namely, the unitarity of processes involving these gauge bosons.

\section{Constraints from tree-level unitarity}\label{sec:bounds}
  
Assuming a weakly coupled theory, bounds on the parameters of the left-right
models and more specifically on  the masses of additional
scalar bosons
can be obtained from unitarity arguments 
on tree-level scattering amplitudes. Ref.~\cite{Lee:1977eg} investigated such bounds on the mass of 
the  scalar 
boson in the SM from the scatterings of the longitudinally polarized gauge 
bosons $Z$ and $W$. For instance,
expanding the scattering amplitude $T(s, cos \theta )$ in partial waves
\begin{equation}
T(s, cos \theta )=16 \pi \sum_J a_J(s) P_J (cos \theta)
\end{equation}
Ref.~\cite{Lee:1977eg} found at large $s$ and at tree level that in the presence of a scalar $h$,
the coefficient associated with the $J=0$ partial wave amplitude of the $Z Z \to Z Z $ scattering amplitude is given by 
\begin{equation}
a_0 \to -3 \sqrt 2 G_F M_h^2 / (16 \pi) = -3 M_h^2 / (16 \pi v^2) \, ,
\label{eq:a0ZZ}
\end{equation}
where in the second equality one has used the relation between the Fermi constant $G_F$ and the electroweak 
symmetry breaking scale $v$.
Due to unitarity, one must have $|a_0| < 1$ implying a bound
on the SM Higgs mass $M_h^2 < 8 \pi \sqrt 2 /(3 G_F)$.

The effects of multiple scalar bosons have been studied 
some years later in Ref.~\cite{Weldon:1984wt}, and they have been considered extensively in the 
literature for various scenarios of new physics, 
e.g., Refs.~\cite{Senjanovic:1979cta,Olness:1985bg} where radiative
corrections have been considered in some cases. Note that such perturbative bounds have also 
been studied for the TLRM, for instance in Refs.~\cite{Basecq:1985cr,Mondal:2015fja,Maiezza:2016bzp,Chauhan:2018uuy}. The scattering processes for scalar bosons were also discussed, for example in Ref.~\cite{Maiezza:2016bzp} while scattering processes involving both
gauge and scalar bosons were considered  in
Ref.~\cite{Lee:1977eg}.    

In the DLRM of interest here, we will 
focus on the scattering of longitudinally polarized gauge
bosons. We will work in the unitarity gauge and in the limit where $s$ is larger
than the masses of all the particles involved. 
The behaviour of the $T$-matrix at large $E$ (where $E$ denotes the general large energy scale  considered, $s\simeq t\simeq E^2$, where $s$, $t$, $u$ are the usual Mandelstam variables) allows for some checks of the calculation. Indeed the particular structure of some of  the couplings of the DLRM is required to  prevent the presence of $\mathcal{O}(s^2)$ terms in various scattering amplitudes.\footnote{For instance,
the couplings $c^{H_3}$ of $H_3^0$ to $W$ and $Z$, and the coupling $c_{\chi_{R,r}}^{h^0}$ of $h^0$ to $W'$ and $Z'$ must be equal, see Eqs.~(\ref{eq:dech}) and (\ref{eq:coefb}).
Furthermore the following relation must be obeyed
\begin{equation}
\sum _{i=1}^2 S_i^2 = \sum_{i=1}^2 \frac{t_i^2}{1+ t_i^2 +(r + w t_i)^2} =\frac{1 + r^2}{k^2}
\label{eq:rel0}
\end{equation}
where the $S_i$ are defined in Eq.~(\ref{eq:defS}).}

\subsection{Constraints from the scattering of light gauge bosons}

We consider first the scattering of the light gauge bosons and their
modification compared to the SM results. 

\vspace{0.3cm}
$\bullet$ $Z Z \to Z Z$

\vspace{0.1cm}
It is straightforward to generalise the expression obtained in the SM to the DLRM case. One has
in the large $s$ limit
\begin{equation}\label{eq:a0ZZZZ}
a_0 =  -\frac{3 \sqrt 2 G_F }{16 \pi} \left( M_h^2  + \epsilon^2 \frac{(c^{H_3})^2}{k^2}  M_{H_3}^2 +{\cal O}(\epsilon^4) \right)
\end{equation}
We recover the SM expression supplemented by a contribution from the analogue
of the light Higgs boson in the right sector, namely $H_3^0$. The other scalars are further suppressed by $\epsilon^2$ compared to the latter.  Note that the two terms in Eq.~\eqref{eq:a0ZZZZ} are in fact
of the same order in $\epsilon$ since $ M_h^2 = \mathcal{O} (\epsilon^2)$. Using the experimental values of $G_F$ and $M_h$, we see that the first term is very small, $3 M_h^2 \sqrt 2 G_F /(16 \pi) \sim 0.015$, and the condition $|a_0|<1$ essentially gives a constraint on the product $ \epsilon^2 (c^{H_3})^2 M_{H_3}^2/k^2$. 

\vspace{0.3cm}
$\bullet$ $W W \to W W$ 

\vspace{0.1cm}
Additional diagrams involving the  exchange of $Z'$ or $H_3^0$ are  present in the DLRM compared
to the SM ones.  
The computation of this scattering process is a bit more involved since one has to determine the rotation 
matrix to the physical gauge fields up to  $ \mathcal{O} (\epsilon^4)$ in order to 
check  that the $T$-matrix does not grow faster than expected with the energy. One  
gets the following modified expressions  compared to the ones  in Sec.~\ref{sec:spin1}:
\begin{eqnarray}
Z_\mu&=& c_W\left(1 - \frac{c_R ^2 s_R^2}{2 s_W^2} k^2 z_h^2 \epsilon^4\right) W_{L,\mu}^3 + \cdots
\nonumber\\
W_\mu&=&\left(1- \frac{2 c_W^2 s_R^2}{s_W^2} r^2 \epsilon^4\right) W_{L,\mu}^\pm + \cdots
\end{eqnarray} 
where we only show the terms actually needed for our purpose.
 
$E^4$ contributions to the $ T $ matrix come from diagrams with the exchange of gauge bosons and the contact interaction.  They involve new terms compared to the SM proportional to $\epsilon^4 /M_W^4$ and thus 
formally of the same order than the SM ones. However the sum of these contributions cancel and we are left in the $u$ channel with:
\begin{equation}
T_{\rm vector}=-\frac{g_L^2 u}{4 M_W^2}(1 - 3 \epsilon^2 w_h^2)
\end{equation}
The contributions from the Higgs sector in the $s$ and $t$ channels are given by:
\begin{equation}
T_{\rm scalar}=-\frac{g_L^2}{4 M_W^2} \biggl[\biggl(1- \epsilon^2(3 w_h^2 -\frac{2}{k} c^{h^0})\biggr)\biggl( \frac{s^2}{s -M_h^2} + \frac{t^2}{t -M_h^2}\biggr)
	-\epsilon^2 \frac{(c^{H_3})^2}{k^2}  \biggl( \frac{s^2}{s -M_{H_3}^2} + \frac{t^2}{t -M_{H_3}^2}\biggr)\biggr]
\end{equation}
It is easy to check that the $E^2$ growth cancels when summing up these two
types of 
contributions using the relation Eq.~(\ref{eq:cH3h0}) between $c^{h^0}$ and $ c^{H_3}$, 
 so that one finally gets in the large $s$ limit
\begin{equation}
a_0=-\frac{2 \sqrt 2 G_F }{16 \pi} \bigl( M_h^2  + \epsilon^2 \frac{(c^{H_3})^2}{k^2}  M_{H_3}^2 +{\cal O}(\epsilon^4) \bigr)
\end{equation}
which leads to a weaker bound than $Z Z$ scattering. 

\subsection{Constraints from the scattering of heavy gauge bosons}

Let us consider now the scatterings of heavy gauge bosons with
longitudinal polarisations.

\vspace{0.3cm}
$\bullet$ $Z' Z' \to Z' Z'$

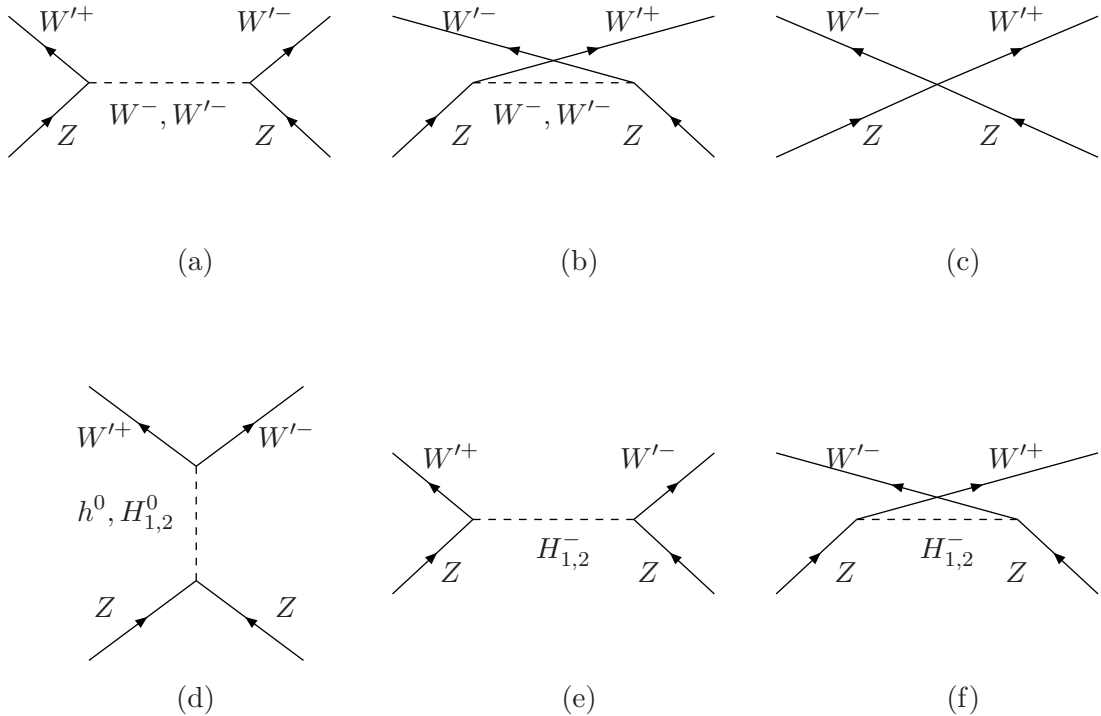
\begin{figure}[t!]
\begin{picture}(140,50)(0,0)
\ArrowLine(130,-30)(100,-2)
\ArrowLine(40,-2)(10,23)
\Text(70,-14)[]{$ W^{-}, W'^{-}$}
\ArrowLine(100,-2)(130,23)
\Text(32,22)[]{$W'^+$}
\DashLine(40,-2)(100,-2){3}
\ArrowLine(10,-30)(40,-2)
\Text(106,22)[]{$W'^-$}
\Text(32,-22)[]{$Z$}
\Text(106,-22)[]{$Z$}
\Text(80,-70)[]{(a)}
\end{picture}
\begin{picture}(140,50)(0,0)
\ArrowLine(130,-30)(100,-2)
\ArrowLine(100,-2)(10,23)
\Text(70,-14)[]{$ W^{-}, W'^{-}$}
\ArrowLine(40,-2)(130,23)
\Text(39,22)[]{$W'^-$}
\DashLine(40,-2)(100,-2){3}
\ArrowLine(10,-30)(40,-2)
\Text(100,22)[]{$W'^+$}
\Text(36,-22)[]{$Z$}
\Text(100,-22)[]{$Z$}
\Text(80,-70)[]{(b)}
\end{picture}
\begin{picture}(140,50)(0,0)
\Text(100,22)[]{$W'^+$}
\Text(90,-22)[]{$Z$}
\Text(46,-22)[]{$Z$}
\Text(39,22)[]{$W'^-$}
\ArrowLine(70,-2.65)(10,23)
\ArrowLine(130,-30)(70,-2.65)
\ArrowLine(10,-30)(70,-2.65)
\ArrowLine(70,-2.65)(130,23)
\Text(80,-70)[]{(c)}
\end{picture}

\vspace{4.cm}

\begin{picture}(140,50)(0,0)
\ArrowLine(40,-55)(80,-25)
\ArrowLine(80,18)(40,48)
\Text(54,0)[]{$h^0, H_{1,2}^0$}
\ArrowLine(80,18)(120,48)
\Text(46,30)[]{$W'^+$}
\Text(114,30)[]{$W'^-$}
\DashLine(80,18)(80,-25){3}
\ArrowLine(120,-55)(80,-25)
\Text(46,-35)[]{$Z$}
\Text(114,-35)[]{$Z$}
\Text(80,-70)[]{(d)}
\end{picture}
\begin{picture}(140,50)(0,0)
\ArrowLine(130,-30)(100,-2)
\ArrowLine(40,-2)(10,23)
\Text(74,-14)[]{$  H_{1,2}^-$}
\ArrowLine(100,-2)(130,23)
\Text(32,22)[]{$W'^+$}
\DashLine(40,-2)(100,-2){3}
\ArrowLine(10,-30)(40,-2)
\Text(106,22)[]{$W'^-$}
\Text(32,-22)[]{$Z$}
\Text(106,-22)[]{$Z$}
\Text(80,-70)[]{(e)}
\end{picture}
\begin{picture}(140,50)(0,0)
\ArrowLine(130,-30)(100,-2)
\ArrowLine(100,-2)(10,23)
\Text(74,-14)[]{$ H_{1,2}^-$}
\ArrowLine(40,-2)(130,23)
\Text(39,22)[]{$W'^-$}
\DashLine(40,-2)(100,-2){3}
\ArrowLine(10,-30)(40,-2)
\Text(100,22)[]{$W'^+$}
\Text(36,-22)[]{$Z$}
\Text(100,-22)[]{$Z$}
\Text(80,-70)[]{(f)}
\end{picture}

\vspace{3.cm}
\caption{{\it Diagrams contributing to $Z Z \to W' W'$. The other scatterings discussed
in the text can be obtained from these diagrams changing the labelling of the lines
accordingly. No contact term is present for $Z' Z' \to W W$ and the first line is
absent when only $Z$ or  $Z'$ gauge bosons scatter.}}
\label{fig:gbscatt}
\end{figure}

\vspace{0.1cm}
This is the analogue of the $Z Z$ scattering in the SM, so that this process is
expected to constrain the mass of $H_3^0$.
Like in the case of the SM, no exchanges of gauge bosons are possible and only the three
neutral scalar exchange diagrams (d, e, f) shown in Fig.~\ref{fig:gbscatt} contribute. The $E^2$ terms
cancel among the diagrams
due to the relation between the Mandelstam variables $s +  t +u =4 M_{Z'}^2 $. 
The  partial wave amplitude in the large $s$ limit up to 
order ${\cal O}(\epsilon^2)$  reads
\begin{equation}
a_0^{Z'Z'}=-\frac{3}{16  \pi v_R^2 } \biggl( M_{H_3^0}^2 +\frac{\epsilon^2} { k^2 }
\biggl( \bigl(c_R^2 k z_h + s_R^2 w^2 + k c^{h^0}_{\chi_{R,r}}  \bigr)^2 M_{h^0}^2  +\sum_{i=1}^2 ( (-c_R^2 +s_R^2) w S_i +c^{H_i}_{\chi_{R,r}})^2  M_{H_i^0}^2 \biggr) \biggr)
\end{equation}
As noted previously one gets at LO in $\epsilon$ a similar relation to the SM case,  Eq.~(\ref{eq:a0ZZ}), but in the right sector, with  $H_3^0$ being the equivalent of the Standard Model Higgs boson.  One can also rewrite
 $M_{H_3^0}$ at LO in $\epsilon$ in terms of the $\rho_1$ parameter of the scalar potential,
 Eq.~(\ref{eq:M3rho1}), leading to the following range:
\begin{equation}
0<\rho_1 <  \frac{8 \pi}{3} \, ,
\end{equation}
where the lower bound comes from its relation to $M_{H_3}$.

\vspace{0.9cm}

\vspace{0.3cm}
$\bullet$ $W' W' \to W' W'$

\vspace{0.1cm}

This scattering is the right-sector analogue
of $W W \to W W$ scattering. There are two $Z$-, two $Z'$- and two $\gamma$-exchange diagrams together with  a contact diagram.
The triple and quadruple gauge couplings involving the $W'$ gauge boson
are proportional to:
\begin{equation}
i  \frac{ e^2} {c_W^2 s_R^2}
\end{equation}
All these diagrams grow like $E^4$ at high energy but their sum yields
\begin{equation}
T_{\rm vector}=\frac{g_R^2}{4 M_{W'}^2}u
\end{equation}
Furthermore, at lowest order in $\epsilon$, only $H_3^0$ contributes. Its $E^2$ growth 
cancels that from the gauge bosons so that one finally gets
\begin{equation}
T_{\rm scalar}=-\frac{M_{H_3^0}^2}{ v_R^2}\biggl(\frac{s}{s - M_{H_3^0}^2}+
\frac{t}{t - M_{H_3^0}^2} \biggr)
\end{equation}
As expected, this scattering essentially gives a bound on the mass of $H_3^0$
which is somewhat weaker than in the previous case.

\subsection{Constraints from scatterings involving both light and heavy gauge bosons}

Finally we discuss the cases coming from the scattering of heavy and light gauge bosons, both with longitudinal polarisations.
 
\vspace{0.3cm}
$\bullet$ $W' W' \to W W$

\vspace{0.1cm}
This scattering gets contributions from $Z$, $Z'$,  $\gamma$ and scalars in the
$s$ and $t$ channels, but there is no contact term. Summing all the gauge
bosons diagrams using the couplings given in App.~\ref{app:feynmanrules} yields:
\begin{equation}
T_{{\rm vector}} =\frac{s \left(1+r^2\right)}{k^2 v_R^2} +2 t \frac{ \left(1+r^4 +(1+r^2)w^2\right)}{k^4 v_R^2}
\label{eq:tmat}
 \end{equation}
On the other hand the scalar exchange leads to:
\begin{eqnarray}
T_{{\rm scalar}} &=& - \frac{1}{k^2 v_R^2} \biggl((1+r^2 + k c^{h^0}_{\chi_{R,r}}) \frac{ s^2}{s-M_{h^0}^2} +\frac{4 r^2}{ k^2 }  \frac{t^2}{t-M_{h}^2} 
\biggr.
\nonumber\\
&&\qquad\qquad\biggl.
+ \sum F_i^2 k^2 t^2
\biggl(\frac{1}{t-M_{H_i^0}^2} + 
\frac{1}{t-M_{A_i^0}^2} \biggl) +c^{H_3} \frac{ s^2}{s-M_{H_3^0}^2}\biggr) 
 \end{eqnarray}
Using Eq.~(\ref{eq:coefb}) and the fact that
\begin{equation}
\sum_i F_i^2 k^2 =1 +r^2\left(1- \frac{4}{k^2}\right)
\label{eq:rel}
\end{equation}
it is easy to verify that the leading terms in 
$s$ and $t$ cancel exactly the ones 
in Eq.~(\ref{eq:tmat}), so that one gets in the large $s$ limit
\begin{equation}
a_0^{W' W}=-\frac{1}{16 \pi k^2  v_R^2} \biggl(\frac{k^2(1+ r^2+ k c^{h^0}_{\chi_{R,r}}) +4 r^2}{k^2} M_{h^0}^2  + \sum^2_{i=1} F_i^2 k^2 (M_{H_i^0}^2+ M_{A_i^0}^2) + c^{H_3} M_{H_3^0}^2\biggr)
\end{equation} 
When $w=0$, only one of the coupling $F_i$ does not vanish and thus the sum is limited to a single value of $i$. Furthermore at LO in $\epsilon$,
the  mass $M_{h}$ of the light Higgs boson can be neglected 
while  the masses of the CP-even and CP-odd
scalars are equal. This leads to the following bound 
\begin{equation}
\frac{ M_{H^0}^2 }{8  \pi v_R^2 } \frac{(1-r^2)^2}{(1+r^2)^2}+\frac{1}{16 \pi}\frac{(1+r^2)\alpha_1+2 r \alpha_2+r^2 \alpha_3 +\alpha_4}{1+r^2} < 1 
\end{equation}
which translates into a bound on a specific combination of four
parameters of the Higgs potential, see Eq.~(\ref{eq:param}):
\begin{equation}
\alpha_{124} - r \alpha_2 +\frac{1}{1+r^2} \alpha_{34} < 16  \pi  
\end{equation}
 using the relation between $M_{H_i^0}^2$ and $\alpha_{34}$ at LO in $\epsilon$.

\vspace{0.3cm}
$\bullet$ $Z' Z \to Z' Z$

\vspace{0.1cm}
At lowest order there are three contributions to the amplitude

\hspace{-0.85truecm}\begin{eqnarray}
\hspace{-0.85truecm}T&=&-\frac{1}{  v_R^2 k^2 } \biggl[ z_h^2 \biggl( \frac{s^2}{s- M_{h^0}^2}+\frac{t^2}{t- M_{h^0}^2}\biggr) 
 +\sum_{i=1}^2  w^2 S_i^2\biggl(\frac{s^2}{s- M_{H_i^0}^2}
+\frac{t^2}{t- M_{H_i^0}^2}\biggr)\biggr.
\nonumber\\
&&\qquad\qquad\biggl.+
\bigl(c_R^2 k z_h +  s_R^2 w^2+ k c^{h^0}_{\chi_{R,r}}\bigl) \frac{u^2}{u- M_{h^0}^2} +c^{H_3} \frac{u^2}{u- M_{H_3^0}^2}\biggr] \nonumber
\end{eqnarray}
It is easy to check that the $E^4$ terms cancel each other
using  Eq.~(\ref{eq:rel0}), the relations between the coefficients $c$, and the fact that
\begin{equation}
-z_h^2-w^2 (1 - \frac{ w^2}{k^2})+(c_R^2 k z_h +  s_R^2 w^2)=0 \, .
\label{eq:rel1}
\end{equation}
Thus one  finally gets in the large $s$ limit
\begin{eqnarray}
a_0^{Z' Z}&=&-\frac{1}{16 \pi  v_R^2 k^2 } \biggl(\bigr(2 z_h^2+ (c_R^2 k z_h +  s_R^2 w^2+ k c^{h^0}_{\chi_{R,r}})\bigr) M_{h^0}^2 \biggl. 
\nonumber \\
&&\qquad\qquad\qquad \biggl. +\sum_{i=1}^2   2 w^2 S_i^2 M_{H_i^0}^2 +c^{H_3} M_{H_3^0}^2 \biggr )
\label{eq:aZpZ}
\end{eqnarray}
Imposing the tree unitarity bound $|a_0^{Z' Z}| < 1$
constrains the combination of parameters present in the equation above.

\vspace{0.3cm}
$\bullet$ $Z' Z' \to W W$

\vspace{0.1cm}
The diagrams contributing to this scattering process are shown in Fig.~\ref{fig:gbscatt}. 
Contrary to $ Z W \to Z W $ no contact diagrams are present. At lowest order in
$\epsilon$, the $W'$ exchange diagram does not contribute and one gets:
\begin{equation}
T_{{\rm vector}} =e^2 \frac{c_R^2 s_R^2 c_W^2}{s_W^4}  k^2 z_h^2 \frac{M_{Z'}^2 s}
{M_W^4} \epsilon^4 
\end{equation}  
Using the expressions for the gauge boson masses, this equation 
reads
\begin{equation}
T_{{\rm vector}} = k^2 z_h^2 \frac{s}{v_R^2 k^2}
\end{equation}
The exchanges of the two neutral Higgs bosons $h^0$ and $H_3^0$ in the $s$  channel and
charged scalars $H_i^+$ in the $t$ and $u$  channels give
\begin{eqnarray}
T_{{\rm scalar}}& =& -\frac{1}{ v_R^2 k^2 }\biggl( 
(c_R^2 k z_h + s_R^2 w^2 + k c^{h^0}_{\chi_{R,r}}) \frac{s^2}{s -M_{h^0}^2} +c^{H_3} \frac{s^2}{s -M_{H_3^0}^2} \biggr.
\nonumber \\
&&\biggl.+ w^2 \sum_i S_i^2 \biggl(\frac{t^2}{t -M_{H_i^\pm}^2}
 + \frac{u^2}{u -M_{H_i^\pm}^2}\biggr) \biggr)
\end{eqnarray}
The $E^2$ growth of this amplitude cancels the vector exchange due to 
the relations, Eqs.~(\ref{eq:rel0}) and (\ref{eq:rel1}).
One finally gets for $a_0^{Z' W}$ in the large $s$ limit
\begin{equation}
a_0^{Z'W}=- \frac{1}{16  \pi v_R^2 k^2 }
\biggl((c_R^2 k z_h + s_R^2 w^2 +k c^{h^0}_{\chi_{R,r}}) M_{h^0}^2+ 2 w^2 \sum_iS_i^2 M_{H_i^\pm}^2 +c^{H_3} M_{H_3^0}^2\biggr)
\end{equation}
At LO the charged and neutral scalar masses are equal  so that neglecting
the mass of the light Higgs boson one obtains the same constraint as
in Eq.~(\ref{eq:aZpZ}) at that same order.

\vspace{0.3cm}
$\bullet$ $Z Z \to W' W'$

\vspace{0.1cm}
This case involves a contact diagram  and the contribution of the $W'$ exchange. The latter
 is similar to the SM one and thus involves the ratio $M_Z^2 s /M_{W'}^4$. Consequently the sum of 
these two contributions is 
 $\mathcal{O}(\epsilon^2)$. The $W$ exchange yields at leading order in $\epsilon$
\begin{equation}
T_{{\rm vector}} = 4 e^2 r^2  \frac{s_R^2}{s_W^4} \frac{M_{W'}^2 s}{M_Z^2 M_W^2} \epsilon^4
=4 r^2  \frac{s}{v_R^2 k^4}+{\cal O}(\epsilon^2)
\label{eq:scattZW'}
\end{equation}
where the expressions of the masses at leading order in $\epsilon$, Eqs.~(\ref{eq:masswzLO}) and (\ref{eq:masswpzpLO}) have been used in the
second line of Eq.~(\ref{eq:scattZW'}).
The exchange of the neutral light Higgs boson $h^0$ in the $s$  channel and
charged scalars in the $t$ and $u$  channels gives
\begin{eqnarray}
T_{{\rm scalar}}& =& -\frac{1}{  v_R^2 k^2 }\biggl( 
 (1+r^2 + k c^{h^0}_{\chi_{R,r}}) \frac{s^2}{s -M_{h^0}^2}+c^{H_3} \frac{s^2}{s -M_{H_3^0}^2} \biggr.
\nonumber \\
&&\biggl.+  \sum_i F_i^2 k^2 \biggl(\frac{t^2}{t -M_{H_i^0}^2}  + \frac{u^2}{u -M_{H_i^0}^2}\biggr) \biggr)
\end{eqnarray}
In order to determine the $Z W'$ coupling of the charged scalar boson in
Table \ref{tab:feynGBWZ} one needs to compute the coefficient of
the term proportional
to $\epsilon \, \chi_R$ in the decomposition of $H_1^\pm$ given in Eq.~(\ref{eq:dech}).
It contributes to the coupling with a multiplicative factor $\sin^2 \theta_W$
while the LO terms in the decomposition  contributes with $\cos^2 \theta_W$.

Using Eq.~(\ref{eq:rel})
the $E^2$ terms cancel in the sum of the vector and scalar contributions
and one gets for $a_0^{Z W'}$ in the large $s$ limit:
\begin{equation}
a_0^{Z W'}= -\frac{1}{16  \pi k^2 v_R^2 }
\biggl( (1+r^2 + k c^{h^0}_{\chi_{R,r}}) M_{h^0}^2+ 2 \sum_i F_i^2 k^2 M_{H_i^0}^2+c^{H_3} M_{H_3^0}^2 \biggr)
\end{equation}
Neglecting the mass of the light Higgs boson, one gets the same expression  and
consequently the same unitarity bounds as
for  $W'W' \to W W$ scattering.

\begin{figure}[t!]
\hspace{-1.45cm}
\begin{picture}(140,50)(0,0)
\Photon(80,60)(80,19.5){3}{5}
\PhotonArc(80,0.225)(17.5,90,270){3}{7}
\PhotonArc(80,0.225)(17.5,270,450){3}{7}
\Photon(80,-17.5)(80,-55){3}{5}
\Text(80,-80)[]{(a)}
\Text(65,40)[]{$Z$}
\Text(42,0)[]{$W,W'$}
\Text(110,0)[]{$W'$}
\Text(65,-40)[]{$Z$}
\end{picture}
\hspace{-1.15cm}
\begin{picture}(140,50)(0,0)
\Photon(80,60)(80,-55){3}{12}
\PhotonArc(60,1.5)(17.15,0,360){3}{15}
\Text(80,-80)[]{(b)}
\Text(65,40)[]{$Z$}
\Text(30,0)[]{$W'$}
\Text(65,-40)[]{$Z$}
\end{picture}
\hspace{-1.15cm}
\begin{picture}(140,50)(0,0)
\Photon(80,60)(80,19.5){3}{5}
\PhotonArc(80,0.225)(17.5,90,270){3}{7}
\PhotonArc(80,0.225)(17.5,270,450){3}{7}
\Photon(80,-17.5)(80,-55){3}{5}
\Text(80,-80)[]{(c)}
\Text(65,40)[]{$W$}
\Text(50,0)[]{$W$}
\Text(110,0)[]{$Z'$}
\Text(65,-40)[]{$W$}
\end{picture}
\hspace{-1.15cm}
\begin{picture}(140,50)(0,0)
\Photon(80,60)(80,19.5){3}{5}
\PhotonArc(80,0.225)(17.5,90,270){3}{7}
\PhotonArc(80,0.225)(17.5,270,450){3}{7}
\Photon(80,-17.5)(80,-55){3}{5}
\Text(80,-80)[]{(d)}
\Text(65,40)[]{$W$}
\Text(50,0)[]{$W'$}
\Text(110,0)[]{$Z$}
\Text(65,-40)[]{$W$}
\end{picture}
\vspace{3.cm}
\caption{\small\it Self-energy diagrams for the $W$ and $Z$ bosons in Left-Right models:
only gauge bosons are considered in the loops.}
\label{fig:diagselfg}
\end{figure}
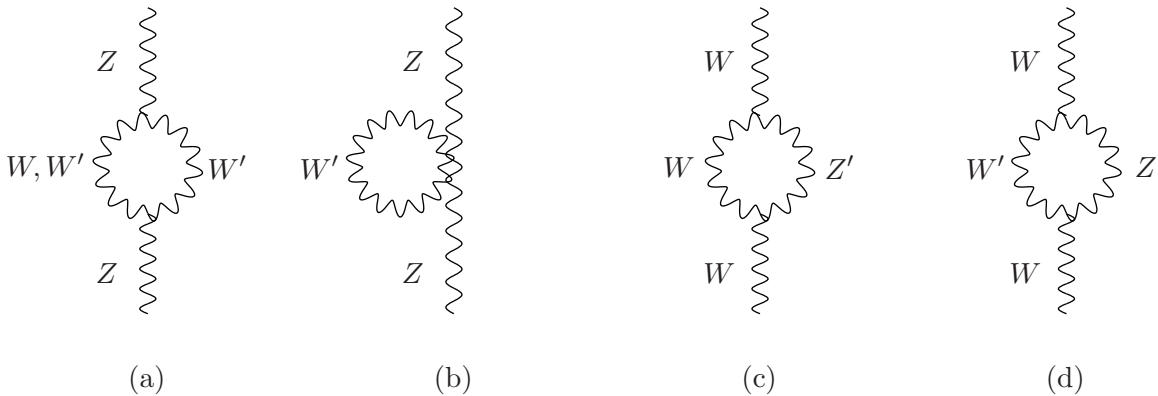

\subsection{Summary}

Summarizing our results one has  altogether four 
bounds from the unitarity conditions on the masses of the scalar bosons at LO in the $\epsilon$ expansion:\footnote{We do not distinguish between the LO masses and their expansion up to ${\cal O}(\epsilon^2)$ here.}

\begin{eqnarray}
	\frac{(c^{H_3})^2}{k^4} \frac{M_{H_3^0}^2} {v_R^2}&<& 16 \frac{\pi}{3} \nonumber\\
	\frac{M_{H_3^0}^2} {v_R^2}&<& 16 \frac{\pi}{3} \nonumber\\
	2 \sum_i F_i^2 \frac{M_{H_i^\pm}^2}{v_R^2} +\frac{c^{H_3}}{k^2} \frac{M_{H_3^0}^2}{v_R^2} &<&16 \pi \nonumber\\
	2 \frac{w^2}{k^2} \sum_i S_i^2 \frac{M_{H_i^\pm}^2}{v_R^2} +\frac{c^{H_3}}{k^2} \frac{M_{H_3^0}^2}{v_R^2} &<&16 \pi
	\label{eq:resumeUnitarity}
\end{eqnarray}

\noindent
Note that we have divided all the masses by 
the characteristic scale $ v_R $ of LR symmetry breaking. Indeed this scale is unknown.
A way of probing indirectly such a scale is to use the ElectroWeak Precision Observables (EWPO), as will be 
discussed in the following. Assuming that these observables will allow for a precise
determination of $v_R$ one can, from the second equation above,
extract an upper bound on the mass of $H_3^0$ at LO in $\epsilon$ in exactly 
the same way as the knowledge of $G_F$ in the SM allowed to put bounds on the mass of the SM Higgs boson. 
The 
 masses of the other scalar bosons, namely $ M_{H^0_i} \sim M_{A^0_i} \sim M_{H^\pm_i} $ (equal at LO in $ \epsilon $) ($i=1,2$) involve
 some extra free parameters  
 among 
 which there are combinations of parameters from
 the scalar potential,
so that in practice the EWPO alone might not be sufficient to extract them.
We will come back to the role played by unitarity in probing the scalar sector of the model after introducing the EW precision observables that will be explored in a global analysis.

\section{\texorpdfstring{$\Delta r$}{Dr} and the mass of the \texorpdfstring{$W$}{W} at one loop} \label{sec:custodial}

Another important constraint in any electroweak model comes from the
mass of the $W$ and its connection with the (muon) Fermi decay constant, encoded in $\Delta r$.  In the SM significant progress has been made in the 
computation
of these quantities as well as of the electroweak precision observables which will be discussed in Section~\ref{sec:EWPO}, leading to reduced theoretical uncertainties.  The state of the art for the mass of the $W$ is a full  two-loop electroweak evaluation 
with higher order QCD corrections and resummation of
reducible tems, see Ref.~\cite{Erler:2019hds} for a status report on precision theoretical 
calculations within the SM 
(one of the earliest computations can be found in Ref.~\cite{Sirlin:1980nh}). In  
theories beyond the Standard Model, some one-loop determinations 
of the $W$ mass
have also been performed, see for example Ref.~\cite{Czakon:2002wm} for the TLRM or more recently in Ref.~\cite{Stal:2015zca} for the Next-to-Minimal Supersymmetric extension of the Standard Model. 

There are also computations
of the related quantity $\Delta \rho$ which is defined as the difference of  the $Z$ and $W$ self-energies at zero momentum transfer, each being weighted by 
the inverse of the square of the mass of the respective gauge boson,
 for example in the framework of the two Higgs doublet model \cite{LopezVal:2012zb,Hessenberger:2016atw}.
 In the SM the loop corrections to  $\Delta \rho$  
 are finite, but this feature 
  is not necessarily true in models beyond the SM. Here we will perform a one-loop calculation
of the muon $\Delta r$ parameter in the DLRM 
focusing mainly on the contributions involving ratios of the heavy gauge bosons and the new scalars  to the light particles   
since one may expect contributions proportional to large logarithms of the form 
$ \log \epsilon^2$.
In order to perform such a computation
we first need to discuss the renormalisation of the model.

\subsection{Renormalisation}\label{sec:ren}

In the SM a number of popular renormalisation schemes are used to compute radiative corrections to observables. One of the 
mostly used when dealing with ElectroWeak Precision Observables (EWPO) is the on-mass shell 
scheme,\footnote{This is in particular the scheme used in the Fortran package Zfitter \cite{Arbuzov:2005ma}, one of the earliest 
open source software projects for the computation
of fermion pair production and radiative corrections at high energy $e^+ e^-$ colliders, which we  will use to perform our fits to electroweak precision observables.}
in which a set for parameter renormalisation is given in terms of the electric charge and the masses of the various particles, the gauge bosons, the Higgs and the quarks. In this scheme the tree-level formula $s_W^2=1-M_Z^2/M_W^2$ is promoted to a definition of the renormalised
$s_W$ to all orders in perturbation theory.

We will compute  the (renormalised) $W$ self-energy in the on-mass shell scheme. Following e.g. Ref.~\cite{Hollik:1993cg}, we associate
multiplicative renormalisation constants to each free parameter and each multiplet of fields in the Lagrangian:
\begin{eqnarray}
W^{L,a}_\mu \to (Z_2^W)^{1/2} W^{L,a}_\mu  &&  \quad \quad \quad W^{R,a}_\mu \to (Z_2^{W'})^{1/2}  W^{R,a}_\mu 
\nonumber \\
B_\mu \to   (Z_2^B)^{1/2} B_\mu &&  \quad \quad \quad g_X \to (Z_1^{B})(Z_2^B)^{-3/2} g_X
\nonumber \\
g_L \to (Z_1^{W})(Z_2^W)^{-3/2} g_L   &&  \quad \quad \quad g_R \to (Z_1^{W'})(Z_2^{W'})^{-3/2} g_R
\nonumber \\
v_L \to (Z^{H_L})^{1/2} v_L( 1 - \delta v_L) &&  \quad \quad \quad v_R \to (Z^{H_R})^{1/2} v_R( 1 - \delta v_R) 
\nonumber \\
\kappa_1 \to (Z^\phi)^{1/2} \kappa_1( 1 - \delta \kappa_1) &&  \quad \quad \quad
\kappa_2 \to (Z^\phi)^{1/2} \kappa_2( 1 - \delta \kappa_2)
\end{eqnarray}
Introducing
the renormalised constants in the Lagrangian and choosing the eigenmass state basis, we can define new renormalised 
quantities up to ${\cal O} (\epsilon^2)$, namely
\begin{equation}
  \begin{pmatrix}
    \delta Z_i^\gamma\\
    \delta Z_i^Z \\
    \delta Z_i^{Z'} 
  \end{pmatrix}=
  \begin{pmatrix}
    s_W^2& c_W^2 s_R^2 &  c_W^2 c_R^2\\
      c_W^2&  s_R^2( s_W^2 -2 c_R^2 k z_h \epsilon^2) &c_R^2( s_W^2 +2 s_R^2 k z_h \epsilon^2)\\
0 & c_R^2( 1 +2 s_R^2 k z_h \epsilon^2)& s_R^2( 1 -2 c_R^2 k z_h \epsilon^2)
  \end{pmatrix}
 \begin{pmatrix}
     \delta Z_i^W\\
   \delta Z_i^{W'} \\
\delta Z_i^{B}
  \end{pmatrix}
\end{equation}
with the standard definition $\delta Z_i^X =Z_i^X -1$. The SM result is recovered at LO in $\epsilon$  with $c_R=1$ for which the right handed fields decouple from the left handed ones, see for example Ref.~\cite{Hollik:1993cg}.

One can express the self-energies of the gauge bosons in terms of these renormalised quantities, leading 
with obvious notations:
\begin{eqnarray}
\hat \Sigma^{\gamma \gamma}(q^2)&=& \Sigma^{\gamma \gamma}(q^2) +\delta Z_2^\gamma q^2 \nonumber \\ 
\hat \Sigma^{Z Z}(q^2)&=& \Sigma^{ZZ}(q^2)- \delta M_Z^2 +\delta Z_2^Z( q^2-M_Z^2) 
\nonumber \\
\hat \Sigma^{W W }(q^2)&=& \Sigma^{WW}(q^2)- \delta M_W^2 +\delta Z_2^W( q^2-M_W^2) \nonumber \\ 
\hat \Sigma^{Z' Z'}(q^2)&=& \Sigma^{Z'Z'}(q^2)- \delta M_{Z'}^2 +\delta Z_2^{Z'}( q^2-M_{Z'}^2) 
\nonumber \\
\hat \Sigma^{W' W' }(q^2)&=& \Sigma^{W' W'}(q^2)- \delta M_{W'}^2 +\delta Z_2^{W'}( q^2-M_{W'}^2) 
\end{eqnarray}
\begin{eqnarray}
\hat \Sigma^{Z Z'}(q^2)&=& \Sigma^{Z Z'}(q^2)- \delta M_{Z Z'}^2+\delta Z_2^{Z Z'}\left( q^2-\frac{1}{2}(M_{Z}^2+ M_{Z'}^2)\right)  
\nonumber \\
\hat \Sigma^{W W'}(q^2)&=& \Sigma^{W W'}(q^2)- \delta M_{W W'}^2+\delta Z_2^{W W'}\left( q^2-\frac{1}{2}(M_{W}^2+M_{W'}^2)\right)  
\nonumber \\
\hat \Sigma^{\gamma Z}(q^2)&=& \Sigma^{\gamma Z}(q^2)- \delta M_{\gamma Z}^2+\delta Z_2^{ \gamma Z} \left(q^2-\frac{1}{2} M_{Z}^2\right)  
\nonumber \\
\hat \Sigma^{\gamma Z'}(q^2)&=& \Sigma^{\gamma Z'}(q^2)- \delta M_{\gamma Z'}^2
+\delta Z_2^{\gamma Z'}\left( q^2-\frac{1}{2}M_{Z'}^2\right) 
\label{eq:renself}
\end{eqnarray}
where the quantities with a hat define the renormalised self-energies.
Furthermore we can use the following on-shell renormalisation conditions:
\begin{eqnarray}
&&{\rm Re} \hat \Sigma^{W W}(M_W^2)={\rm Re} \hat \Sigma^{Z Z}(M_Z^2)= 
{\rm Re} \hat \Sigma^{W' W'}(M_{W'}^2)= {\rm Re} \hat \Sigma^{Z' Z'}(M_{Z'}^2)=0
\nonumber \\
&&\hat \Sigma^{Z Z'}(0)=\hat \Sigma^{W W'}(0)=0
\label{eq:onsb}
\end{eqnarray}
as well as the QED-like conditions
\begin{eqnarray}
\hat \gamma^{\gamma e e}_\mu (q^2=0,\slashed{q}_1=\slashed{q}_2=m_e)&=& i e \gamma_\mu
\nonumber \\
\hat \Sigma^{\gamma Z}(0)&=&\hat \Sigma^{\gamma Z'}(0)=0 \nonumber \\
\frac{ d \hat \Sigma^{\gamma \gamma}}{d q^2}(0)&=&0 
\label{eq:onsg2}
\end{eqnarray}
where one has imposed the absence of mixing at $q^2=0$.

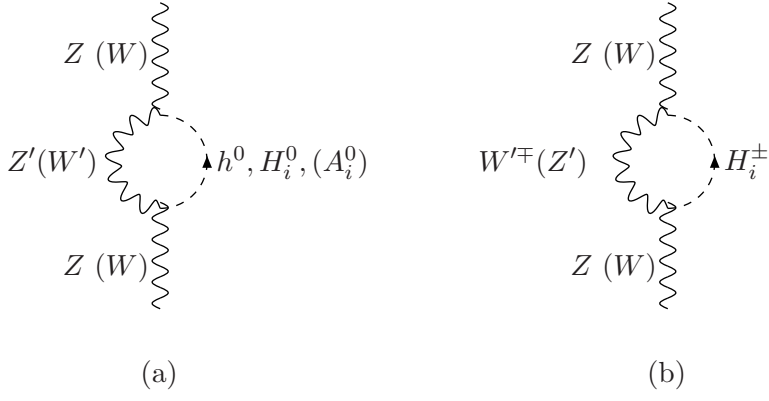
\begin{figure}[t!]
\hspace{1.cm}
\begin{picture}(140,50)(0,0)
\Photon(80,60)(80,20){3}{5}
\PhotonArc(80,0.225)(17.5,90,270){3}{7}
\DashArrowArc(80,0.225)(17.5,270,450){3}
\Photon(80,-15)(80,-55){3}{5}
\Text(80,-80)[]{(a)}
\Text(60,40)[]{$Z\,\, (W)$}
\Text(40,0)[]{$Z'(W')$}
\Text(130,0)[]{$h^0,H_i^0,(A_i^0)$}
\Text(60,-40)[]{$Z \,\, (W)$}
\end{picture}
\hspace{1.5cm}
\begin{picture}(140,50)(0,0)
\Photon(80,60)(80,20){3}{5}
\PhotonArc(80,0.225)(17.5,90,270){3}{7}
\DashArrowArc(80,0.225)(17.5,270,450){3}
\Photon(80,-15)(80,-55){3}{5}
\Text(80,-80)[]{(b)}
\Text(60,40)[]{$Z \,\,(W)$}
\Text(30,0)[]{$W'^\mp (Z') $}
\Text(110,0)[]{$H_i^\pm$}
\Text(60,-40)[]{$Z\,\, (W)$}
\end{picture}
\vspace{3.cm}
\caption{\small\it Same as in Fig.~\ref{fig:diagselfg}
but with one heavy gauge boson and one scalar boson in the loops.}
\label{fig:diagselfgH}
\end{figure}

A relation between $\delta Z_1^\gamma$ and $ \delta Z_2^\gamma $ can be obtained from the renormalisation of
the charge discussed in  Appendix \ref{sec:append}.  Finally the various $\delta Z_i$ are given up to NLO in $\epsilon$ by:
\begin{eqnarray}
\delta Z_2^\gamma&=&-\Pi^\gamma(0)\equiv  \frac{d \Sigma^{\gamma Z}}{d q^2}(0)
\nonumber \\
\delta Z_1^\gamma&=&-\Pi^\gamma(0)+\frac{s_W}{c_W} \frac{\Sigma^{\gamma Z}(0)}{M_Z^2}\bigl(1+ \frac{s_R^2 k z_h}{ s_W^2}\epsilon^2\bigr) 
+\frac{s_R}{c_R c_W}\frac{\Sigma^{\gamma Z'}(0)}{M_Z'^2} \bigl(1+ \frac{s_W^2  z_h}{k s_R^2}(c_R^2 k z_h +s_R^2 w^2)\epsilon^2\bigr)
\nonumber\\
\delta Z_2^W&=& \delta Z_2^{LO} +\epsilon^2 \delta Z_2^{NLO}
\nonumber\\
\delta Z_1^W&=&\delta Z_1^{LO} +\epsilon^2 \left(\delta Z_2^{NLO} - z_h c_W 
\left(w^2\frac{s_R}{k c_R} +\frac{c_R}{ s_R} z_h\right) \frac{\Sigma^{\gamma Z'}(0)}{M_Z'^2}\right) 
\label{eq:ZfacW}
\end{eqnarray}
with
\begin{eqnarray}
\delta Z_2^{NLO}&=&\frac{c_W} {k^2 s_W^2} \left[
(4 r^2 - k^2 z_h^2)\left(\frac{1}{c_W}\delta Z_{Z W \gamma}
+c_W\biggl(
\frac{\delta M_{W}^2}{M_{W}^2} - \frac{\delta M_{W'}^2}{M_{W'}^2}\biggr)
 - c_W \delta Z_{Z' W' \gamma}\right) \right.
\nonumber\\
&&\left.\quad
-2 c_R^2 c_W k^3 z_h\biggl( \frac{\delta M_{Z'}^2}{M_{Z'}^2}-\frac{\delta M_{W'}^2}{M_{W'}^2}\biggr) \right.
\\
&&\left.\quad
+s_R \left(\frac{4 k^2  r  c_W^2}{s_W} \frac{\Sigma^{W W'}(0)}{M_{W}^2}
-2 k^3 z_h \frac{c_R c_W}{s_W} \delta Z_{Z Z' \gamma} 
 -\frac{2}{c_R} k z_h (2 w^2 +3 k z_h) \frac{\Sigma^{\gamma Z'}(0)}{M_Z'^2}  \right) \right] \nonumber
\label{eq:Z2NLO}	
\end{eqnarray}
and one has defined
\begin{eqnarray}
\delta Z_{Z W \gamma}&=&2 \frac{c_W}{s_W} \frac{\Sigma^{\gamma Z}(0)}{M_Z^2}+ \frac{c_W^2}{s_W^2}\biggl(\frac{\delta M_Z^2}{M_Z^2}-\frac{\delta M_W^2}{M_W^2}\biggr)
\nonumber\\
\delta Z_{Z' W' \gamma}&=&2 \frac{c_R}{s_R c_W}\frac{\Sigma^{\gamma Z'}(0)}{M_Z'^2}+ \frac{c_R^2 }{s_R^2}\biggl(
\frac{\delta M_{Z'}^2}{M_{Z'}^2}-\frac{\delta M_{W'}^2}{M_{W'}^2}\biggr)
\nonumber\\
\delta Z_{Z Z' \gamma}&=& \frac{\Sigma^{\gamma Z'}(0)}{M_{Z'}^2} + \frac{c_W}{ s_W} \frac{\Sigma^{Z Z'}(0)}{M_{Z'}^2}
\end{eqnarray}
The LO terms are  the SM like expressions in the limit $s_R=0$:
\begin{eqnarray}
\delta Z_2^{LO}&=&-\Pi^\gamma(0) +\delta Z_{Z W \gamma}
\nonumber\\
\delta Z_1^{LO}&=&-\Pi^\gamma(0) +\frac{3 -2 s_W^2}{s_W c_W} \frac{\Sigma^{\gamma Z}(0)}{M_Z^2}
+\frac{s_R}{c_R c_W} \frac{\Sigma^{\gamma Z'}(0)}{M_{Z'}^2}
+ \frac{c_W^2}{s_W^2}\biggl(\frac{\delta M_Z^2}{M_Z^2}-\frac{\delta M_W^2}{M_W^2}\biggr) 
\label{eq:renorm}
\end{eqnarray}
The renormalisation conditions, Eqs.~(\ref{eq:onsb}) and (\ref{eq:onsg2}),
have been used to derive these expressions. For completeness  the wave-function renormalisation relevant for 
the heavy gauge bosons is given in  App.~\ref{sec:append} as well as the
renormalisation of $s_W$.
Note that in all these expressions  the self-energies and the counterterms are
divided by the squared mass of a gauge boson in such a way that the ratio is a quantity of order ${\cal O}(\epsilon^0)$ at LO. In principle $\Sigma^{\gamma Z'}(0)$ and  $\Sigma^{Z Z'}(0)$ are also quantities of  order one, however at LO the former 
is equal to  $ -c_W/s_W  \Sigma^{Z Z'}(0)$ so that the combination of 
these two quantities appearing  in $\delta Z_{Z Z' \gamma}$ turns out to be of order ${\cal O}(\epsilon^2)$. 

Few last remarks are in order.
On the right-hand side
of the expressions above, the self-energies contain contributions from the heavy particles. Moreover,
the heavy particles will only start to contribute at ${\cal O}(\epsilon^2)$ 
in the SM-like contributions, so that one recovers the SM result at leading order in $\epsilon$. Finally one can perform a similar on-shell subtraction for the scalar self-energies and
replace eight of the scalar parameters by the eight scalar masses,
the remaining
parameters being taken as $\overline{\rm MS}$ running  parameters and thus renormalisation  scale dependent. However, the renormalisation of the Higgs sector is not really needed for our purposes here.


\begin{figure}[t!]
\hspace{1.cm}
\begin{picture}(140,50)(0,0)
\Photon(80.225,60)(80.225,19.8){3}{5}
\DashArrowArc(80,0.225)(17.5,90,270){3}
\DashArrowArc(80.,0.225)(17.5,270,450){3}
\Photon(80.225,-16)(80.225,-55){3}{5}
\Text(80,-80)[]{(a)}
\Text(60,40)[]{$Z$}
\Text(30,0)[]{$A_i^0,\, (H_i^\pm)$}
\Text(145,0)[]{$h^0,\,\,H_i^0,\,H_3^0,\, (H_i^\mp)$}
\Text(60,-40)[]{$Z $}
\end{picture}
\hspace{1.5cm}
\begin{picture}(140,50)(0,0)
\Photon(80.225,60)(80.225,19.8){3}{5}
\DashArrowArc(80,0.225)(17.5,90,270){3}
\DashArrowArc(80,0.225)(17.5,270,450){3}
\Photon(80.225,-16)(80.225,-55){3}{5}
\Text(80,-80)[]{(b)}
\Text(60,40)[]{$W^\pm$}
\Text(40,0)[]{$H_i^\mp $}
\Text(140,0)[]{$h^0, \,\,H_i^0,\,H_3^0\,A_i^0$}
\Text(60,-40)[]{$W^\pm$}
\end{picture}
\vspace{3.cm}
\caption{\small\it Self-energy diagrams for the $W$ and $Z$ boson in Left-Right models
involving  scalars only.}
\label{fig:diagselfH}
\end{figure}
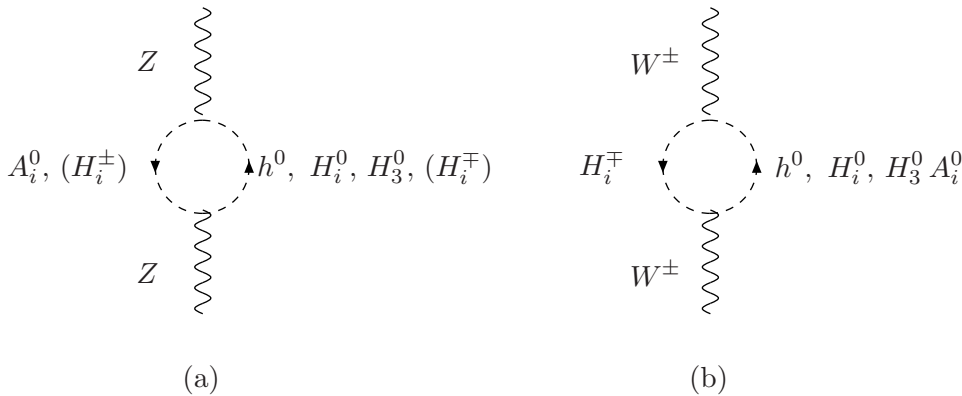

\subsection{Computation of \texorpdfstring{$\Delta r$}{Dr}}\label{sec:deltar}

Let us now turn to  the  muon decay amplitude and the determination of $\Delta r$. The amplitude can
be decomposed as:
\begin{equation}
{\cal M}= {\cal M}_{LL} + {\cal M}_{LR}+ {\cal M}_{RL}+{\cal M}_{RR}
\end{equation}
where each part is proportional to
\begin{equation}
{\cal M}_{AB} \propto {{\cal M}^0}_{AB} = (\bar u_{\nu_\mu} \gamma_\mu P_A 
u_\mu) \, (\bar u_{e} \gamma_\mu P_B v_{\nu_e})
\end{equation}
It involves in principle the $W$ and $W'$ self-energies.
It turns out that at the order $\epsilon^2$ of interest, only the LL part
survives, so that only  the propagation of the $W$ boson will contribute. Therefore, like in the SM case, the matrix element of the loop
diagrams can be written as a proportionality factor multiplying the Born matrix
element
\begin{equation}
\Delta {\cal M}=\Delta r \times {{\cal M}}^{\rm Born}
\end{equation}
leading to a similar expression to the SM case for the (muon) Fermi constant  $G_F$:
\begin{equation}
\frac{G_F}{\sqrt 2} = \frac{e^2}{8 s_W^2 M_W^2}(1+ \Delta r)
\label{eq:fermict}
\end{equation}
with
\begin{equation}
\Delta r = \frac{\hat \Sigma_{WW}(0)}{M_W^2} + {\rm vertex \,\, corrections \,\, +
box \,\,contribution}
\label{eq:fermiSetOfContrs}
\end{equation}

There are two kinds of potentially  large contributions to
$\Delta r$.  The  first one comes from terms
involving ratios of a heavy mass and a light mass, and in particular logarithms of these ratios.
The second one stems 
from the terms inversely
proportional to $c_R $ and/or $s_R$.
Indeed
if one of these two becomes very small, these contributions (a priori of order one) would be enhanced.
For processes
with only external light particles, we can focus on the contributions where at least one particle present in the loop is heavy. Indeed,
the  ${\cal O}(\epsilon^2)$ corrections from the light particles can be safely neglected as  they only involve mass ratios of order one leading to small 
logarithms and they feature no factors inversely proportional to $c_R $ and/or $s_R $.

Let us start with the self-energies.
There are three types of  diagrams to compute involving: (i) only intermediate gauge bosons, (ii) one  gauge boson and one scalar,  (iii) only scalars. These diagrams are shown in  Fig.~\ref{fig:diagselfg} for the  self-energies of the light gauge bosons, with at least one heavy particle in the loop. Similar diagrams can
easily be drawn for the other self-energies, either with one light and one heavy gauge bosons or with two heavy gauge bosons. 
Note that we do not show tadpole diagrams here:
as in the SM case, we performed the renormalisation of the scalar vacuum
expectation values so that one can omit all tadpole diagrams in the renormalised
amplitudes and Green functions \cite{tayl}.
The self-energy contributions to $\Delta r$ in Eq.~\eqref{eq:fermiSetOfContrs} can thus easily be obtained from Eqs.~\eqref{eq:renself}, \eqref{eq:ZfacW}, \eqref{eq:Z2NLO}. It can be decomposed as:
\begin{equation}
\Delta r|_{SE}=\Delta r^{LO}|_{LL}+\epsilon^2 ( \Delta r^{NLO}|_{LL} +\Delta r |_{LH}+\Delta r |_{HH} +\cdots)
\label{eq:deltarexp2}
\end{equation}
where the lower indices denote whether the particles on the external lines
are light (L) or heavy (H) and the ellipsis denotes neglected contributions to be discussed below.
The structure of these various contributions can easily be inferred from the SM 
calculation of the self-energies, see for example \cite{Hollik:1993cg} where their expressions  are explicitly given. It involves sums of terms which are products 
of
couplings of the internal particles to the external ones, which are summarized for the case of the 
DLRM 
in Tables~\ref{tab:feyn3GB} with the various coefficients appearing 
in the case of the gauge bosons to two scalars defined in App.~\ref{app:Hgauge},
 and known loop functions,
modulo some extra factors. Only the following three loop functions appear
\begin{equation}
	\int \frac{d^4 k}{(2 \pi)^4} \frac{1}{k^2 -m^2}  \, ; \quad \quad   \int \frac{d^4 k}{(2 \pi)^4} \frac{1; k_\mu k_\nu}{(k^2 -m^2)
	((k-q)^2-m'^2)}
\end{equation}
where $m$ and $m'$ are the masses of the internal particles and $q^2$ is the four-momentum
squared of the external particle which is either on or off-shell. The 
loop function with  $k_\mu k_\nu$ in the numerator can be expanded into two Lorentz covariants times scalar coefficients after integration, and only the coefficient  
proportional to $g_{\mu\nu}$  is needed here. In the case of
$\Delta r|_{HH}$ one straightforwardly replaces light masses and couplings with their heavy counterparts. 
$\Delta r|_{LL}$ contains contributions which are LO and NLO (i.e., ${\cal O}(\epsilon^2)$) with respect to $\epsilon$. It turns out that only SM-like contributions arise at LO, namely  contributions from the light particles with exactly the same couplings
as in the SM, so that
\begin{equation}
\Delta r^{LO}|_{LL} \equiv \Delta r|_{``SM"}
\end{equation}

We can then consider 
the vertex corrections and the box contribution. The final result for their sum reads:
\begin{equation}
\Delta_{vb}= \Delta_{vb}|_{``SM"}-\frac{3 g_L^2 s_R^2 s_W^2}{2 c_W^2} \frac{ M_W^2}{M_{W'}^2} \log \frac{M_{Z'}^2}{M_W^2}  
\end{equation}  
where $\Delta_{vb}|_{``SM"}$ are SM like contributions and one has again only considered the contributions which involve ratios
of light to heavy particles.

Up to now we have focused on the contributions from the  gauge bosons and scalars. 
Other contributions might be numerically relevant, in particular from the top quark which are very important in the SM as its contributions are proportional to its mass squared
\begin{equation}
	\Delta r_{SM} \sim -\rho_{SM}^{top}/ \tan^2 \theta_W \, ,\quad \quad
	\rho_{SM}^{top}= \frac{3 G_F}{8 \sqrt 2 \pi^2 } m_t^2
	\label{eq:topContribution}
\end{equation}
This explains in particular why this quantity has been used to constrain the mass of the top quark within the SM before more direct measurements, e.g., \cite{Ellis:1990zu}.
However it has been found
that in theories where $\rho \ne 1$  the dependence of $\rho^{top}$
on the top quark mass
 can be very different from the SM (and 
 much weaker): 
for example in models with an extra $U(1)'$ symmetry it is logarithmic \cite{Czakon:1999ha}. In TLRM $\rho_{SM}^{top}$ is multiplied by a factor $M_W^2/(M_{W'}^2-M_W^2)$ leading to a decreasing contribution 
from the top as the mass of the $W'$ increases~\cite{Czakon:1999ue}. 
Indeed, in presence of new physics in theories with $\rho \neq 1$ the entire structure of loop corrections is modified and the  Appelquist-Carazzone decoupling was found not to hold. This casts
 some doubts about the validity 
 of the usual implementation of new physics corrections, which amounts to
 combining the loop corrections to the SM with modifications from new physics at tree level in this case \cite{Jegerlehner:1991ed,Czakon:1999ha,Chen:2003fm,Chen:2005jx}. But this issue has been
 later
 discussed
in detail in Ref.~\cite{Chankowski:2006jk}
in the framework of a  NP model with an extra $U(1)'$ gauge symmetry:
introducing a renormalisation scheme with manifest decoupling,
$\rho^{top}$  takes its SM form up to terms vanishing as $M_{Z'} \to \infty$. It has thus
been pointed out in that reference that a renormalisation scheme can indeed be chosen in such
a way that new physics effects can be treated as corrections to the well established SM results.
The main difference between such a renormalisation scheme and the one in \cite{Czakon:1999ha} for example
lies in the way the
couplings related to the extended sectors are treated. In the latter they are expressed in terms of 
some low energy observables leading to  uncertainties becoming larger with the mass of the additional gauge boson  while in the former they are taken as $\overline{\rm MS}$ running parameters, see \cite{Chankowski:2006jk}
for more details. The fact that the dependence of $\rho$ with $m_t$ differs with the renormalisation scheme
is of course due to a different absorption of the $m_t$ dependence in some of the renormalized couplings. 
Actually,
in our case, the deviation from $ 1 $ only appears at order $\epsilon^2$ contrarily to the TLRM and  
our renormalisation scheme does fulfil the Appelquist-Carazzone decoupling so that 
the loops involving the top quark  will only give a small ${\cal O}(\epsilon^2)$ correction to the (SM-like) quadratic result. We can thus safely neglect the new physics contributions related to the top quark.

Adding up all contributions described above, our  
final expression for $\Delta r$ can be schematically written as
\begin{equation}
\Delta r = \Delta r|_{``SM"} + \epsilon^2 \Delta r|_{NLO}
\label{eq:deltarexp}
\end{equation}
where $\Delta r|_{``SM"}$ is identical to the SM contribution, so that one
recovers the SM expectations in the limit $\epsilon=0$. The expression for
$\Delta r|_{NLO}$ is quite lengthy and can be provided to the interested reader
as a \textit{Mathematica} notebook upon request. 
Up to now we have only considered coefficients of the fields up to 
${\cal O}(\epsilon^2)$. In principle one would have to compute 
them  up to $ \epsilon^4 $ since terms of order $\epsilon^4$ in the self-energies could 
in principle contribute to $\Delta r$ at $\epsilon^2$, but we consider that this task goes beyond the scope of this 
article.
Even though $\Delta r$ itself has no dependence on this scale $\mu$ in principle, there is however still  one in practice since
we did not perform a complete calculation. We will come back to this dependence when discussing our results.

\section{Electroweak Precision Observables} \label{sec:EWPO}

\subsection{Computation in left-right models}

Apart from the mass of the $W$ and $\Delta r$ other important constraints on the SM and its extensions are provided by the
precision electroweak measurements at the Z resonance which were performed
at LEP and SLC~\cite{ALEPH:2005ab}, as well as the weak charges. Note that the field of electroweak precision test
is very active and will remain so in the future with the advent of new high 
energy lepton colliders which would make it possible to increase the precision
of the electroweak fits by an order of magnitude or more, and hence allow to probe  the effect of higher operators in the Standard Model effective
field theory at an unprecedented level, see for instance \cite{deBlas:2019wgy}.

Here we will consider the following observables ${\cal O}$:
\begin{itemize}
\item the mass of the $Z$-boson, its total width as well as the hadronic cross section;
\item various ratios of cross sections
\begin{equation}
R(l)=\frac{\Gamma_Z(had)}{\Gamma_Z(l \bar l)} \,\, \, {\rm for}\,\, \, l =e \,\, , \mu \,
\, ,\tau
\quad R(q)=\frac{\Gamma_Z(q \bar q)}{\Gamma_Z(had)} \,\, \, {\rm for} \,\,\, q =c \,\, , b  \, ;
\end{equation} 
\item the unpolarized forward backward asymmetries $A_{FB}(l)$, $ l = e, \mu, \tau $, $A_{FB}(q)$, $ q = c, b $, and the
final state couplings $\mathcal{A}_{l}$, $ l = e, \mu, \tau $, $\mathcal{A}_{q}$, $ q = c, b $;
\item the weak charges $Q_W$ measured from atomic parity experiments for Caesium and Thallium, as well as for the proton.
\end{itemize}

In order to determine these observables one needs to know the vector and axial vector couplings of the $Z$ gauge boson to fermions in the DLRM. They read at tree level
\begin{eqnarray}
g_V(f)&=&\frac{e}{2 s_W c_W}\biggl((T_{3 L}^f-2 Q^f s_W^2)+ k z_h  
\epsilon^2 (T^f_{3 R} c_R^2 - (X_L^f + X_R^f) s_R^2)\biggr)
\nonumber\\
g_A(f)&=&\frac{e}{2 s_W c_W}\biggl(T_{3 L}^f - k z_h 
\epsilon^2 (T^f_{3 R} c_R^2 + (X_L^f - X_R^f) s_R^2)\biggr)
\end{eqnarray}
where $T_{3 (L,R)}$ are respectively the left/right weak isospin and $ Q^f_{L,R} = T^f_{3(L,R)} + X_{L,R}^f$ the charges of the left and right handed fermions, with $ Q^f_{L} = Q^f_{R} = Q^f $.
Detailed expressions of the observables in terms of these couplings are found in App.~\ref{sec:expsEWPO}.
In the limit where $\epsilon=0$ one recovers the SM expressions, and in
the case where $r=w=0$ these expressions agree with Ref.~\cite{Hsieh:2010zr}.
However,
$ Q_W (p) $ and $ Q_W (n) $ lead to an atomic parity violation for $ ^A {\rm Cs}_Z $ different from the one
found in \cite{Hsieh:2010zr}, which can be traced back to an improper value of $ 2 A - Z $ used in that reference~\cite{JHY}.

As can be seen, most of the observables depend only on the 
two combinations of DLRM parameters $s_R^2 \epsilon^2$ and $ k z_h \epsilon^2$. Only the $W$ width and the weak charges depend on $r^2 \epsilon^2 $ and $w^2 \epsilon^2$.

\subsection{Parametrization of the observables}

Let us consider all the observables we have discussed previously, namely 
\begin{equation}
{\cal O}^{{\rm DLRM}}=\{G_F, M_W, \Gamma_W, {\rm EWPO \, \, \, at \, \, \,  the \, \, \, Z \, \, \, pole}, \text{weak charges} \},
\label{eq:data}
\end{equation}
to which one has added the total width of the $W$ gauge boson, $\Gamma_W$.
These will be used as data in a global fit in the next section. 
They have the general form: 
\begin{equation}\label{eq:EWPOseparation}
{\cal O}^{{\rm DLRM}} ={\cal O}^{[\rm ``SM"]}+ \epsilon^2 {\cal O}^{[\rm ``DLRM"]}
\end{equation}
where ${\cal O}^{[\rm ``SM"]}$ are the LO contributions in the series expansion in $\epsilon$, and ${\cal O}^{[\rm ``DLRM"]}$ are the corrections at NLO. As we have seen the former contributions
are all SM-like. One can take advantage of this fact to use the developed tools in the SM to compute  
these quantities
and to incorporate the radiative corrections, which are known to be very important in order to reproduce for example
the mass of the $W$ gauge boson in the SM.  
For the NLO contributions ${\cal O}^{[\rm ``DLRM"]}$  we will assume that their typical size is such that one can 
keep only their tree-level contributions, as loop corrections would be counted
 as higher-order contributions compared to the order up to which we are working, cf. the discussion 
 after Eq.~\eqref{eq:topContribution} in Section~\ref{sec:deltar} concerning the validity of 
this  procedure to implement new physics corrections.

For the SM-like contributions, we will use the Zfitter package\cite{Bardin:1999yd,Arbuzov:2005ma}.\footnote{We have used the version Zfitter 6.42. The flag ``IALEM" of Zfitter is set to 2 to use $\Delta \alpha^{(5)}_{had}(M_Z^0)$ as input. In Zfitter the value of $G_F^0$ is fixed to its physical value, we thus modified the programs so as to let this parameter free, see also \cite{Wells:2014pga}.  Otherwise, we use the same flags as in the subroutine DIZET.} The input of this package is the set of parameters $\mathcal{S^{[\rm ``SM"]}}$
\begin{equation}
	\mathcal{S^{[\rm ``SM"]}} =\{G_F^0,\, M_Z^0, \,M_h^0, \,m_t^0,\,\alpha_s(M_Z^0),\,\Delta\alpha_{{\rm had}}^{(5)}(M_Z^0)\}
\end{equation}
where the superscript 
$0$ denotes 
the combination of parameters of the theory corresponding to the LO expressions 
($\epsilon=0$) of the observables under
consideration. Contrarily to the fit done in the SM, they differ from their physical values by order $\epsilon^2$ corrections. This allows us to determine $M_W^0$ as: 
\begin{equation}
(M_W^0)^2=  (M_Z^0)^2 \biggl(\frac{1}{2}+\sqrt{\frac{1}{4}-\frac{\alpha (M_Z^0) \pi}{\sqrt 2 G_F^0  (M_Z^0)^2}(1 +\Delta r|_{LO})}\biggr)
\end{equation}
as well as $s_W \equiv \sqrt{1 - (M^0_W / M^0_Z)^2} $ and the EWPO at leading order in $\epsilon$.

Calling Zfitter in the course
of the global fit is far  too much time consuming so that in practice it is 
advantageous to parametrize  the observables ${\cal O}^{[\rm ``SM"]}$.
Varying the input parameters $\mathcal{S}^{[\rm ``SM"]}$  by $2.5\%$ around their experimental central values, i.e.,

\begin{eqnarray}
&122.7 \, {\rm GeV}  < M_h^0 < 128.7 \, {\rm GeV}\nonumber \\
&169 \, {\rm GeV}<m_t^0<177.4 \, {\rm GeV}\nonumber\\
&89 {\rm GeV} <M_Z^0< 93.4 \, {\rm GeV}\nonumber\\
&0.116<\alpha_s(M_Z^0)<0.121 \nonumber\\
&0.0269< \Delta \alpha_{{\rm had}}^{(5)}(M_Z^0)<0.0282\nonumber\\
&1.139\times 10^{-5} \, {\rm GeV}^{-2}<G_F^0<1.194\times 10^{-5} \, {\rm GeV}^{-2}
\end{eqnarray}
 and allowing the observable to vary by at most $4\%$,\footnote{We have also considered a variation of the input parameters $\mathcal{S}^{[\rm ``SM"]}$ by $ 10\% $ and of the observable by at most $ 15\% $, with similar results for our analyses of SM and DLRM.} we obtain a rather accurate parametrization of ${\cal O}^{[\rm ``SM"]}$ as:
\begin{eqnarray}
\mathcal{{\cal O}^{[\rm ``SM"]}} &=& c_{0} + c_{1}  L_{H} + c_{2} \Delta_{t} + c_{3}  \Delta_{\alpha_{s}} + c_{4}  \Delta^{2}_{\alpha_{s}} + c_{5}  \Delta_{\alpha_{s}} \Delta_{t} + c_{6}  \Delta_{\alpha} + c_{7}  \Delta_{Z} 
\nonumber \\
&+& c_{8}  \Delta_{H} \Delta_{t} +c_9   L_{H}^2 +c_{10} \Delta_{t}^2 + 
c_{11} \Delta_{\alpha}^2+c_{12} \Delta_{Z}^2 +c_{13} \Delta_{Z}\Delta_{\alpha_{s}}+c_{14}\Delta_{Z} \Delta_{t}+ c_{15} \Delta_{G}
\nonumber \\
&+& c_{16} \Delta_{G}^2+ c_{17} \Delta_{G}  \Delta_{Z} +c_{18} \Delta_{Z}^3+
c_{19} \Delta_{G}^3+c_{20} \Delta_{G}  \Delta_{Z}^2 +c_{21}\Delta_{\alpha}\Delta_{Z}+c_{22} \Delta_{G}^2  \Delta_{Z} 
\nonumber \\
&+&c_{23} \Delta_{G}^4+c_{24} \Delta_{G}^3  \Delta_{Z}+c_{25} \Delta_{G}^2  \Delta_{Z}^2 +c_{26} \Delta_{G} \Delta_{\alpha} + c_{27} \Delta_{G} \Delta_{\alpha}^2 +c_{28} \Delta_{G}^2 \Delta_{\alpha}
\nonumber \\
&+&c_{29} \Delta_{Z}^4+c_{30} \Delta_{Z} \Delta_{\alpha}^2+c_{31} \Delta_{Z}^2 \Delta_{\alpha}+c_{32} \Delta_{Z}^2 \Delta_{\alpha}^2+c_{33} \Delta_{G} \Delta_{Z}^3 +c_{34} \Delta_{G}^4 \Delta_{Z}
\label{eq:paramZfitter}
\end{eqnarray}
\noindent
where
\begin{eqnarray} 
\Delta_{\alpha} &=& \frac{\Delta \alpha(M_Z^0)}{0.059} - 1 \, ,
\quad \quad  \Delta_{\alpha_{s}} = \frac{\alpha_{s} (M_{Z}^0)}{0.1184} - 1 \, , 
\nonumber \\ 
 L_{H} &=& \log \frac{M_{h}^0}{125.7 \, \operatorname{GeV}} \, , \quad \quad 
\Delta_{H} = \frac{M_{h}^0}{125.7 \, \operatorname{GeV}} - 1 \, ,
\nonumber \\
\Delta_{t} &=& \left( \frac{m_{t}^0}{173.2 \, \operatorname{GeV}} \right)^{2} - 1
\, ,\quad \quad \Delta_{Z} = \frac{M_{Z}^0}{91.1876 \, \operatorname{GeV}} - 1  \, ,
\nonumber \\
\Delta_{G} &=& \frac{G_F^0}{1.16637 \times 10^{-5} \, \operatorname{GeV^{-2}}}  - 1  
\end{eqnarray}
Note that we have traded here the parameter $ \Delta \alpha^{(5)}_{had} (M_Z) $ for $ \Delta \alpha (M_{Z}) $:
\begin{equation}
\Delta \alpha (s) = \Delta \alpha^{(5)}_{\rm had} (s) + \Delta \alpha_{e \mu \tau} (s) + \Delta \alpha_{top} (s) \, ,
\end{equation}
where \cite{Steinhauser:1998rq, Kuhn:1998ze}
\begin{equation}
\Delta \alpha_{e \mu \tau} (M_{Z}) = 0.031497686 \, , \quad \Delta \alpha_{top} (M_{Z}) = - 0.000072 \, ,
\end{equation}
with negligible uncertainties.

The coefficients $c_i$  as well as the maximal difference (in percent) between the Zfitter value and our parametrization are collected in  Tables \ref{tab:coef1}-\ref{tab:coef3}. They have been determined using a grid of 15 points in each 
direction of the parameter space ${\cal S^{[\rm ``SM"]}}$. Note that the maximal deviations are of the order
or smaller than one percent except for the forward backward asymmetries $A_{FB}(c)$ and $A_{FB}(l)$ which are of the order of $10\%$. We have tested the stability of the results with the number of points. It turns out that the result for $A_{FB}(b)$ is rather unstable.  Thus  
for the three asymmetries we will use their definitions in terms of ${\cal A}_e$ and  ${\cal A}_f$ \cite{Tanabashi:2018oca} to parametrize them, namely, $ A_{FB} (f) = 3 \mathcal{A}_e \mathcal{A}_f/4 $.

\section{Global Fits} \label{sec:fits}

We now have all the ingredients to perform a global fit to the parameters of the DLRM using the information on the EWPO discussed above with further constraints from unitarity and perturbativity.

\subsection{Method} \label{sec:method}

We want to perform the statistical combination of the various observables and constraints.
We will follow the CKMfitter statistical approach used in flavour physics to combine constraints in a frequentist framework~\cite{Charles:2004jd,Charles:2011va,Charles:2016qtt}, building a $\chi^2$ function from the likelihoods of the various observables.
The theoretical uncertainties are treated following the Rfit scheme corresponding to a modification of the likelihood including a bias parameter left free to vary within the quoted range for the theoretical uncertainty~\cite{Charles:2016qtt}. For a parameter of interest, the $\chi^2$ is considered at different values of this parameter and minimised with respect to the other parameters of the fit. The result is interpreted as a $p$-value associated to each possible value of the parameter, which can be used to determine confidence intervals for the parameter within a particular model. The compatibility of the various measurements with the model considered can also be assessed through the computation of the corresponding pull.

We will consider both the SM and the DLRM, which will allow a direct comparison between the two models.  
In the SM case, the fit parameters are the ones from $S ^{[\rm ``SM"]}$. In the DLRM, one adds to these parameters
 $c_R$, $r$, $w$, the 5 Higgs potential parameters $\alpha_{124}$, $\alpha_2$,
 $\lambda_{23}^{\pm}$, $\lambda_4$ and the ratio $x=\mu_1'/\mu_2'$ as well as 
 the three dimensionless quantities $M_{H_2}/v_R$, $\delta$ defined in 
 Eq.~\eqref{eq:parax} and $\delta_3=M_{H_3}/M_{W'}$ (though, of course, 
 different choices will not change the physical results).

In the parameters of the DLRM defined above we have discarded $\epsilon$. We can 
rewrite Eq.~(\ref{eq:fermict}) 
to exhibit a structure similar to Eq.~(\ref{eq:EWPOseparation})
\begin{equation}
\frac{G_F}{\sqrt 2} = \frac{{M_W^0}^2}{M_W^2} \frac{G^0_F}{\sqrt 2} +\epsilon^2 \frac{e^2}{8 s_W^2 M_W^2} \Delta r|_{NLO}
\label{eq:fermictlr}
\end{equation}
with
\begin{equation}
\frac{G^0_F}{\sqrt 2}=\frac{e^2}{8 s_W^2 {M_W^0}^2}(1+ \Delta r|_{LO})
\label{eq:fermictlo}
\end{equation}
Since  $G_F$ is determined with such a high precision we will fix it to its central value and use Eq.~(\ref{eq:fermictlr}) to determine
the parameter $\epsilon$. One has to solve an equation of the type
$a +  \epsilon^2( c+  \log \epsilon^2) =0$ where the logarithm comes from
the contribution of one heavy and one light particle in the loops.  Its solution can be expressed
as $\epsilon^2= -a / W(-a e^{c})$ where $W(x)$ is the Lambert function.\footnote{
The Lambert function $W(x)$ is multivalued except at zero. The
real-valued W is defined only for $x \geq -1/e$ and is double-valued on the 
interval $-1/e<x<0$. The additional constraint $W \geq-1$ defines the principal
 branch $W_0(x)$ which is single valued, while the lower branch $W_{-1}(x)$ 
decreases from $W_{-1}(-1/e)=-1$ to $W_{-1}(0)=-\infty$. We will concentrate
here on the principal branch  for which the solutions to $\epsilon$ are the largest. These are indeed the most interesting ones since they should
lead to the largest deviation to the SM results.} We thus obtain $\epsilon$ in terms of $G_F$ (and all the other parameters defined above) using our computation of $\Delta r$.

Our fit will thus include the following constraints:
\begin{itemize}
\item
We can straightforwardly use the above discussion concerning the EWPO expressed in the DLRM 
 ${\cal O}^{[``{\rm DLRM}"]}$, Eq.~\eqref{eq:data};
 \item We include the bounds required in order to satisfy perturbative unitarity at tree level discussed in Sec.~\ref{sec:bounds};
 \item We also impose perturbativity constraints in the sense that $ \epsilon^2 $ corrections to any of the observables discussed are limited to be at most half of the LO ($ \epsilon^0 $) terms in the same quantity.
  \end{itemize}

\begin{figure}[t!]
\begin{minipage}[b]{0.5\linewidth}
\epsfig{figure=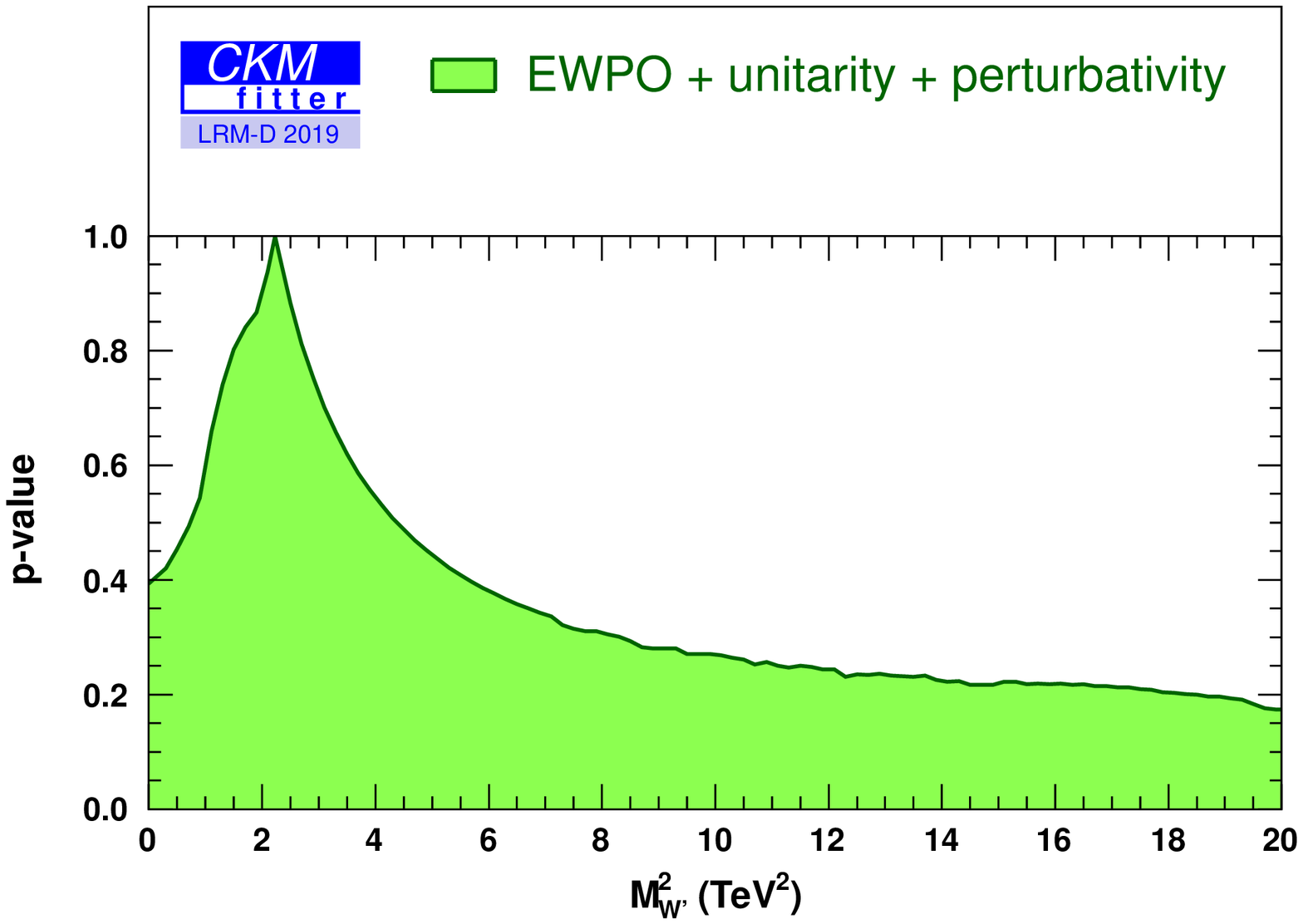, height=6cm}
\end{minipage}
\begin{minipage}[b]{0.5\linewidth}
\epsfig{figure=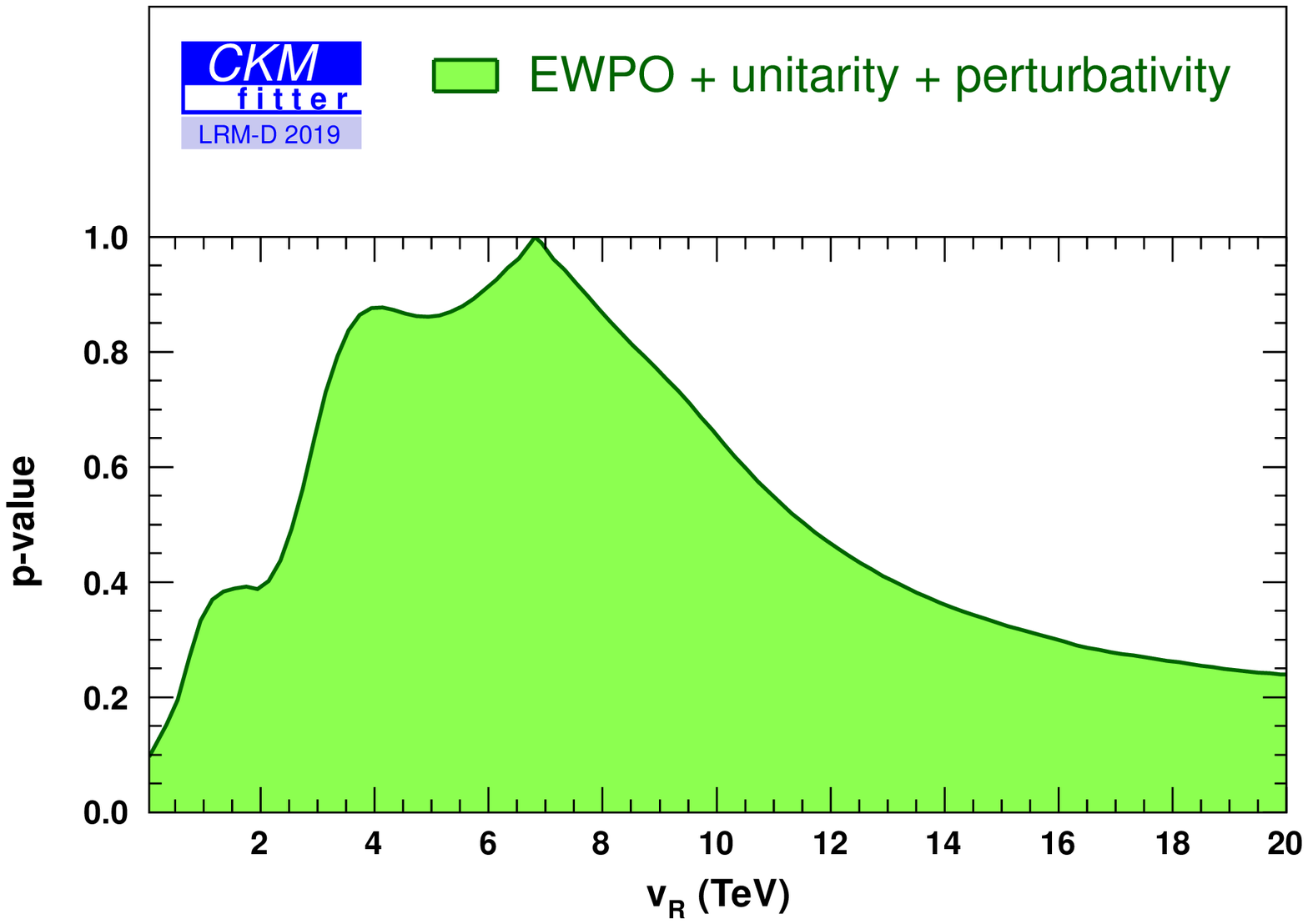, height=6cm}
\end{minipage}
\begin{minipage}[b]{0.5\linewidth}
\epsfig{figure=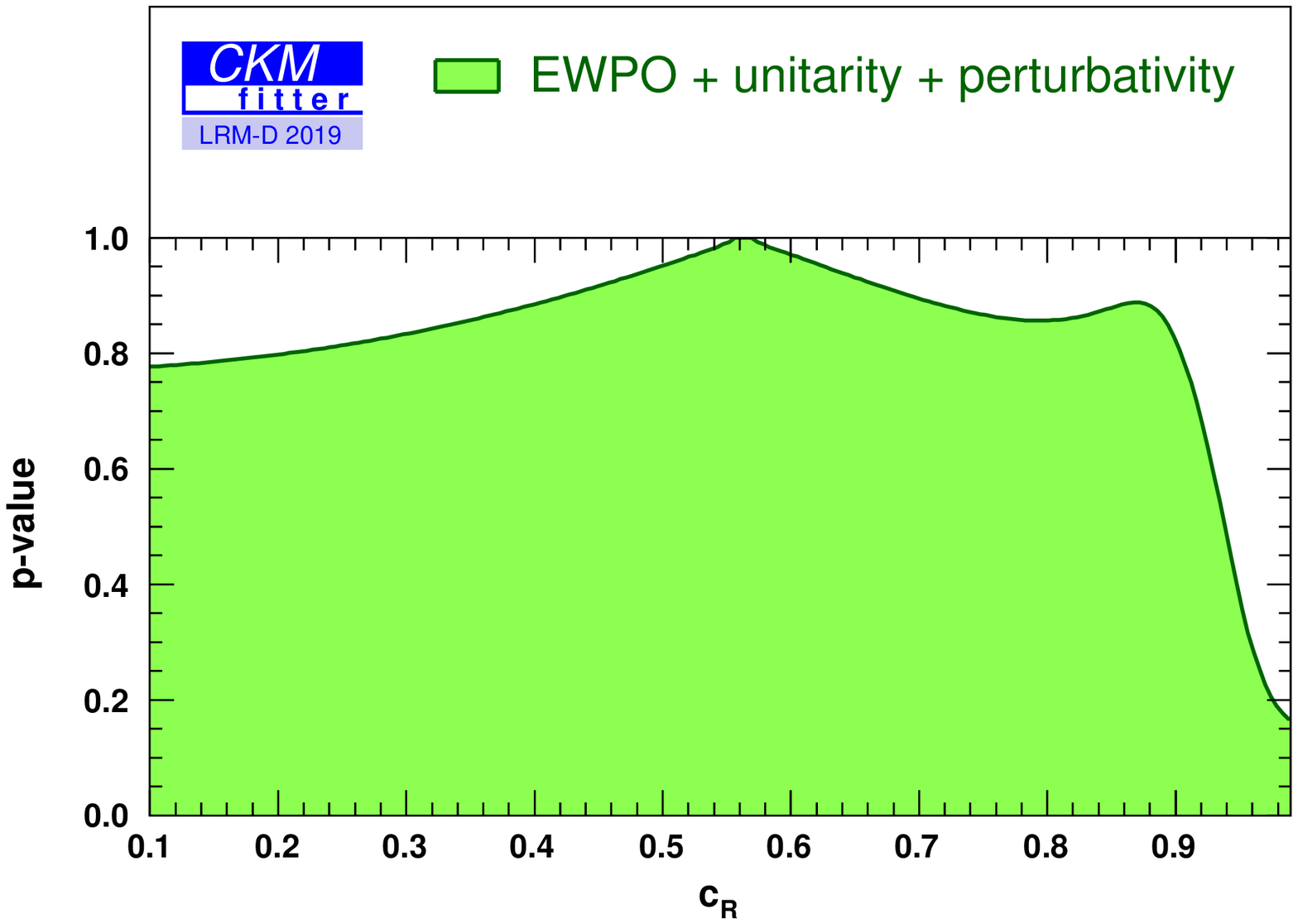, height=6cm}
\end{minipage}
\begin{minipage}[b]{0.5\linewidth}
\hspace{1mm}\epsfig{figure=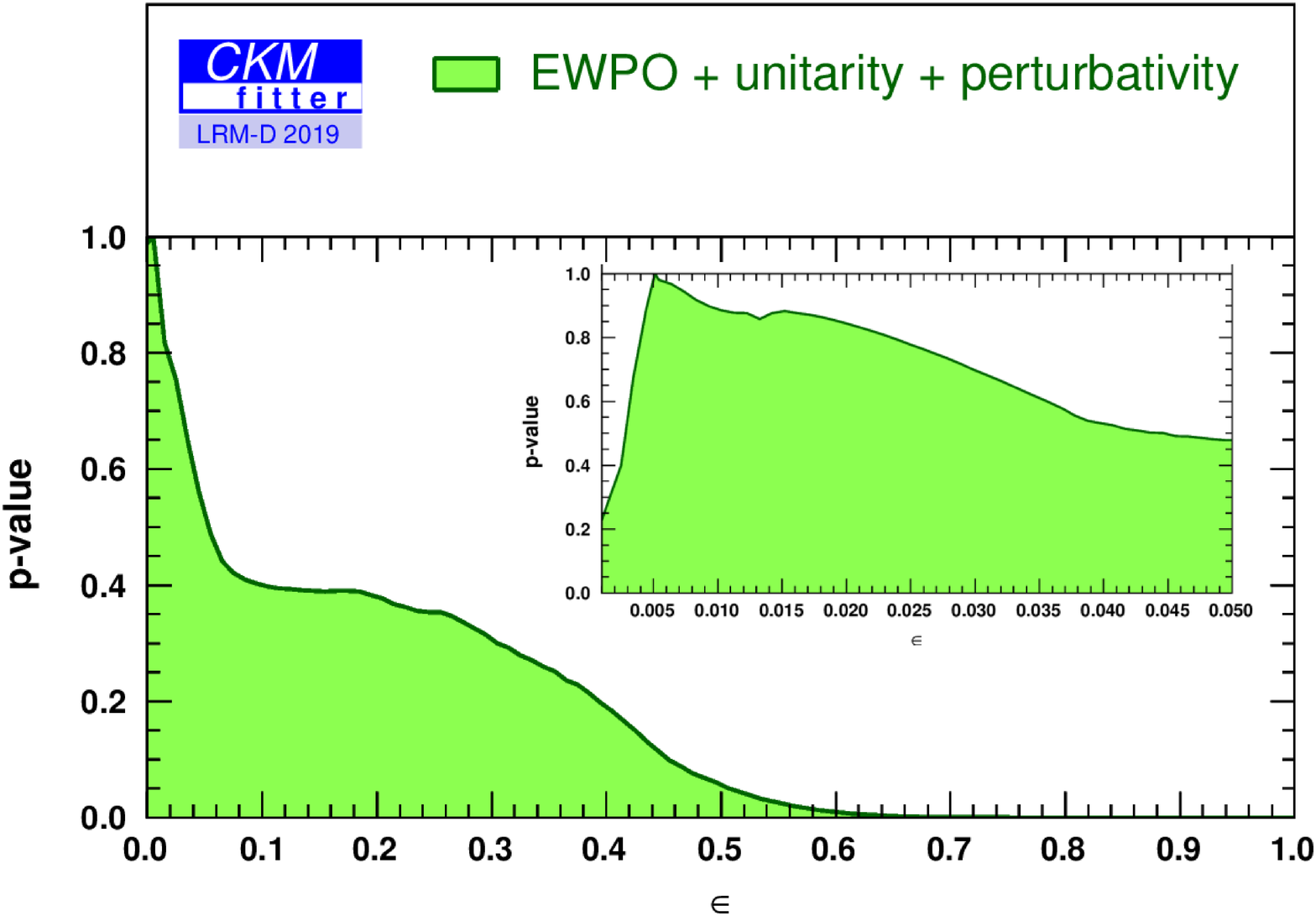,height=5.8cm}
\end{minipage}
\begin{minipage}[b]{0.5\linewidth}
\epsfig{figure=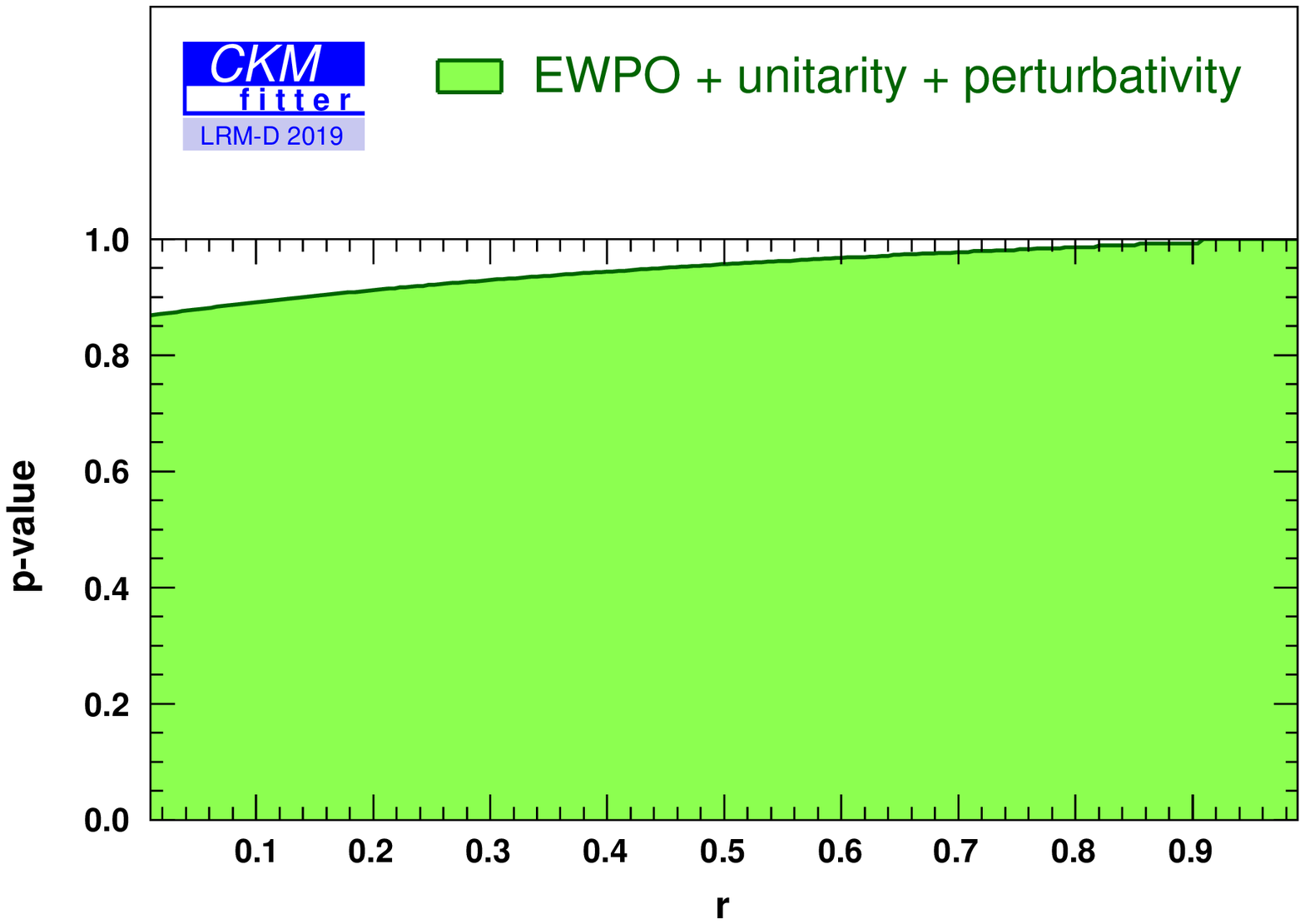, height=6cm}
\end{minipage}
\begin{minipage}[b]{0.5\linewidth}
\epsfig{figure=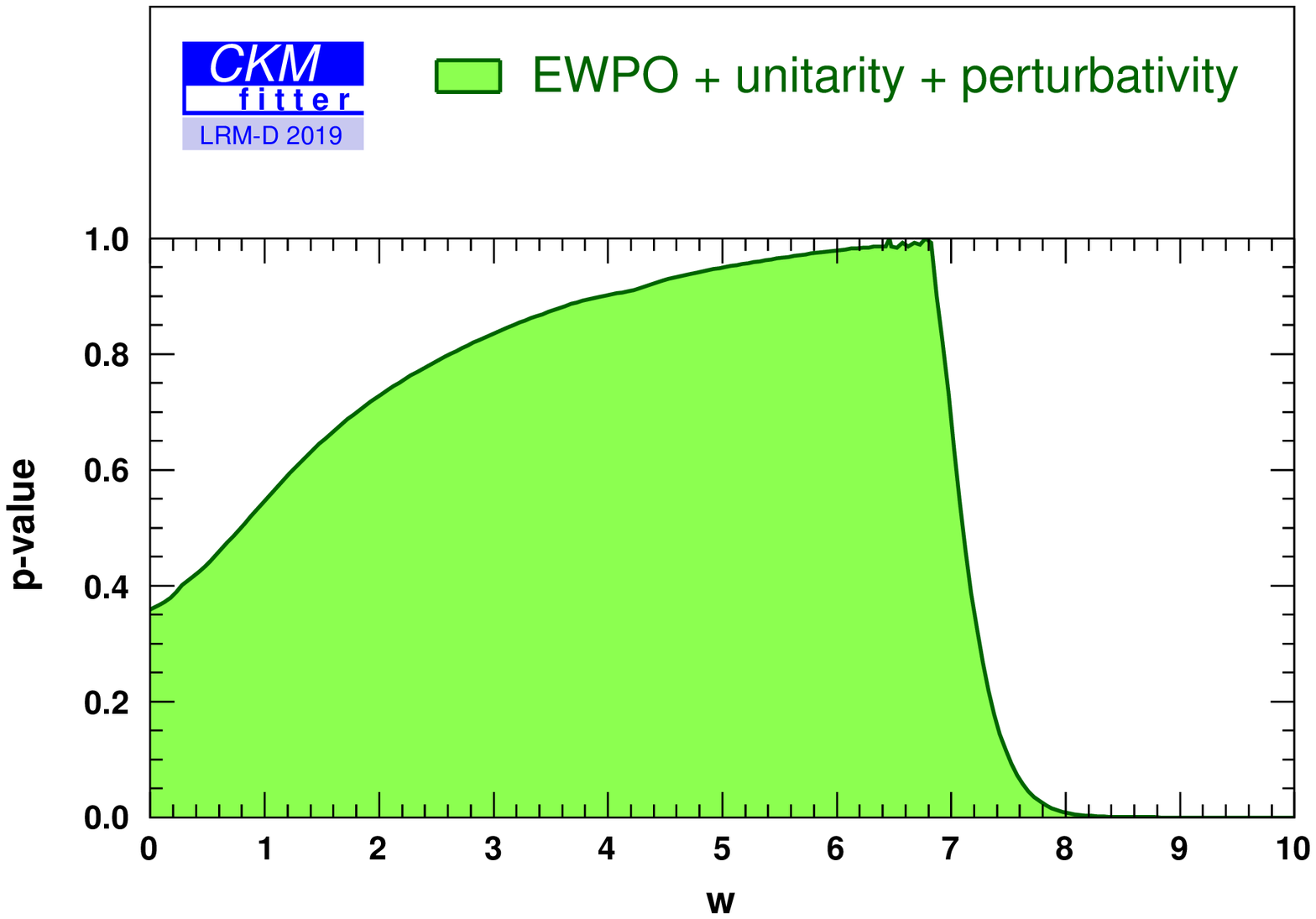, height=6cm}
\end{minipage}
\caption{The $p$-values for some parameters of the DLRM: the value of the parameter at the best-fit point corresponds to a $p$-value of 1, while different Confidence Level (CL) regions are read from different values of $p$-value (a $ 1 \sigma $ CL region corresponds to a $p$-value of $ \sim 0.33 $). The figure for the $p$-value of $ \epsilon $ has an insert of a zoomed region.}
\label{fig:param}
\end{figure}

\begin{figure}[t!]
{
\begin{minipage}[b]{0.5\linewidth}
\epsfig{figure=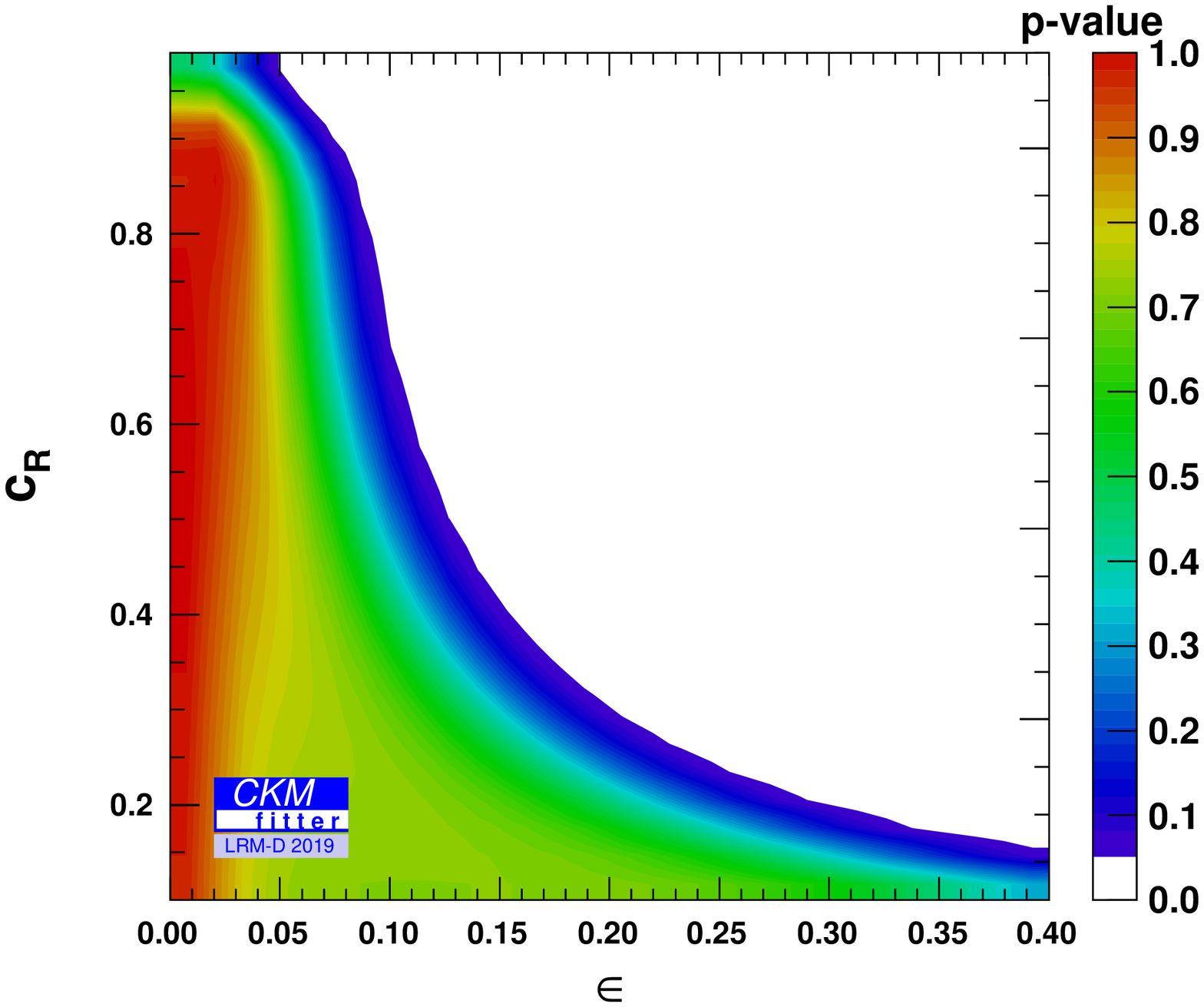, height=6cm}
\end{minipage}
\begin{minipage}[b]{0.5\linewidth}
\epsfig{figure=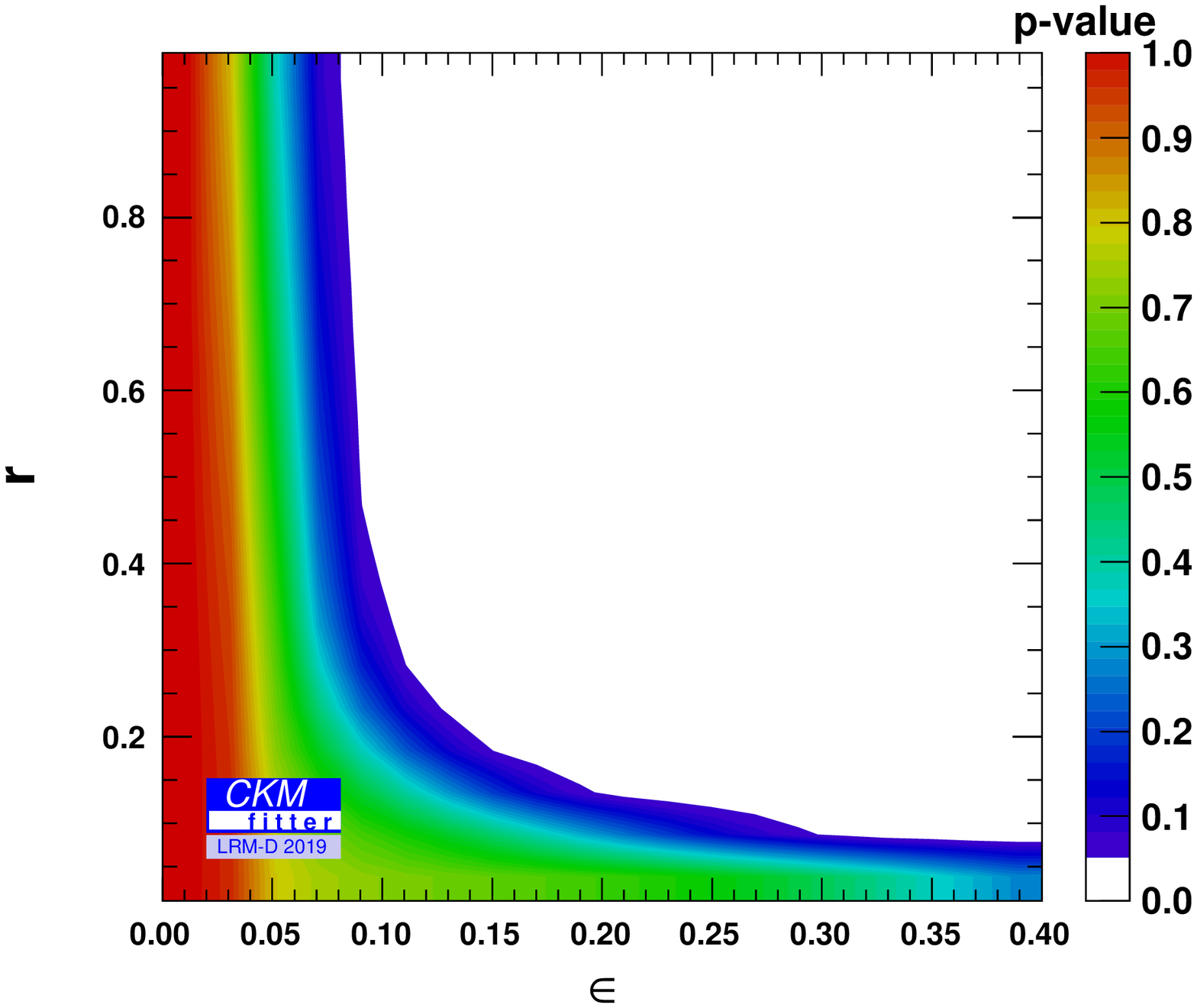, height=6cm}
\end{minipage}
\begin{minipage}[b]{0.5\linewidth}
\epsfig{figure=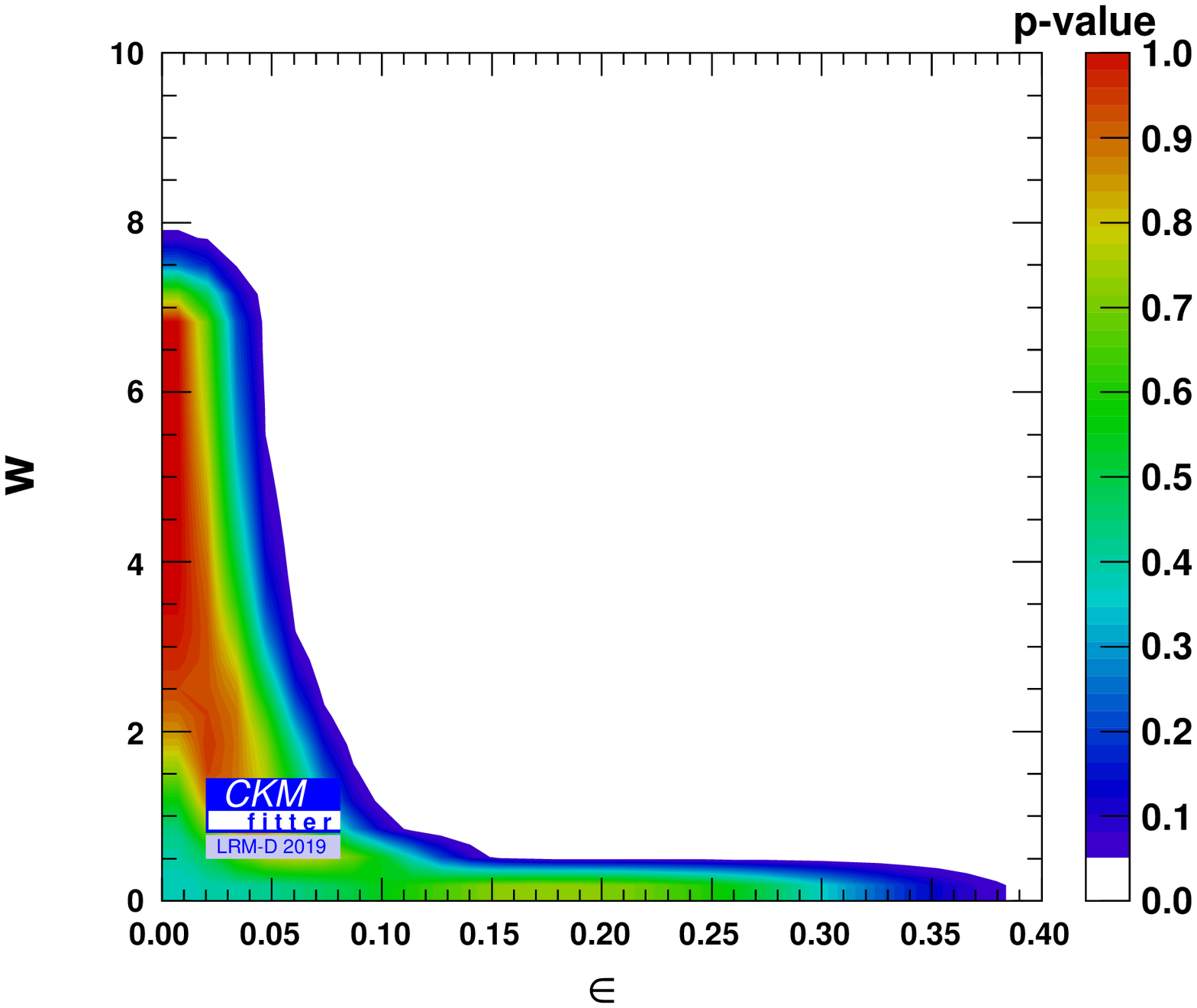, height=6cm}
\end{minipage}
\begin{minipage}[b]{0.5\linewidth}
\epsfig{figure=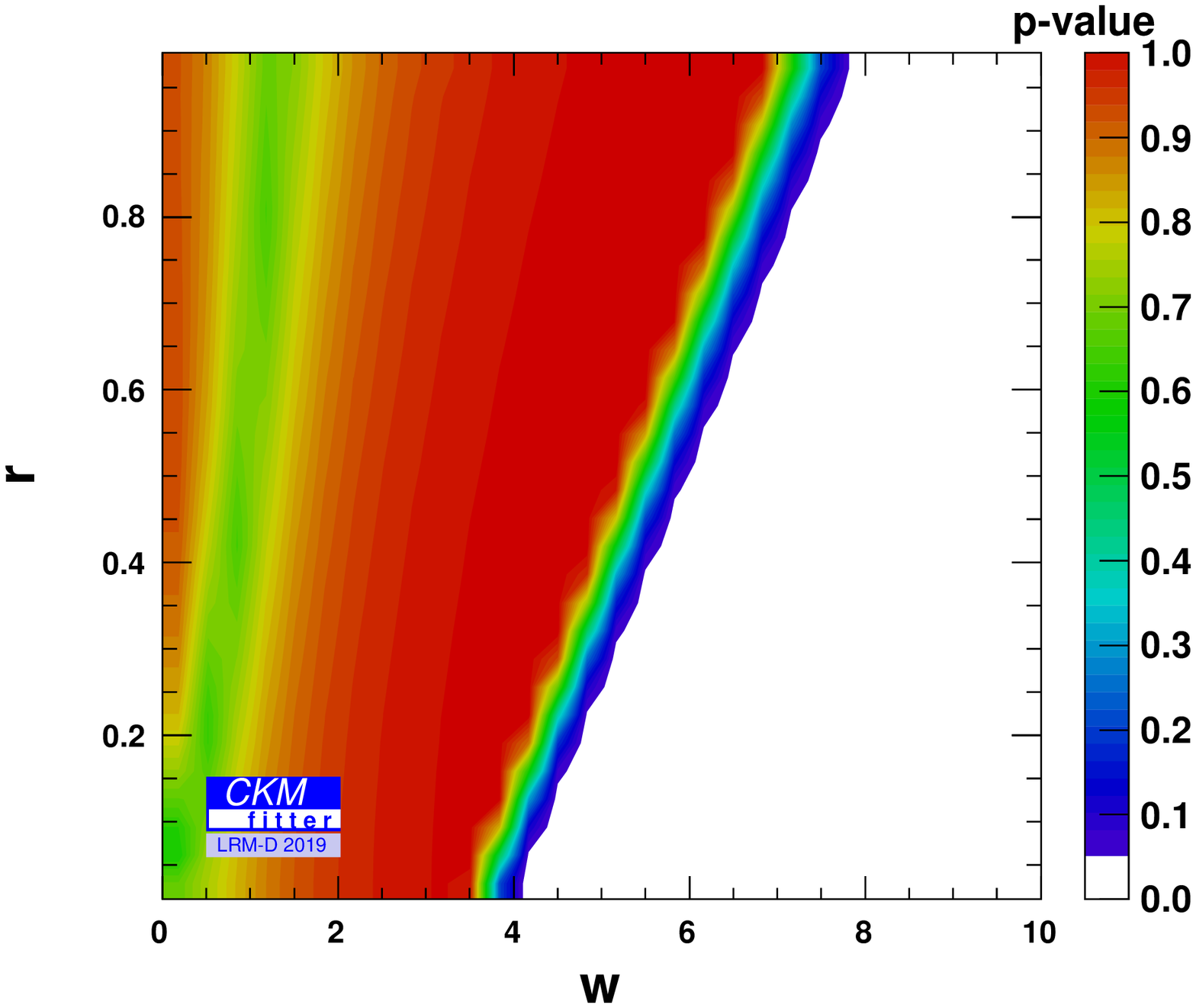,height=6cm}
\end{minipage}
\begin{minipage}[b]{0.5\linewidth}
\epsfig{figure=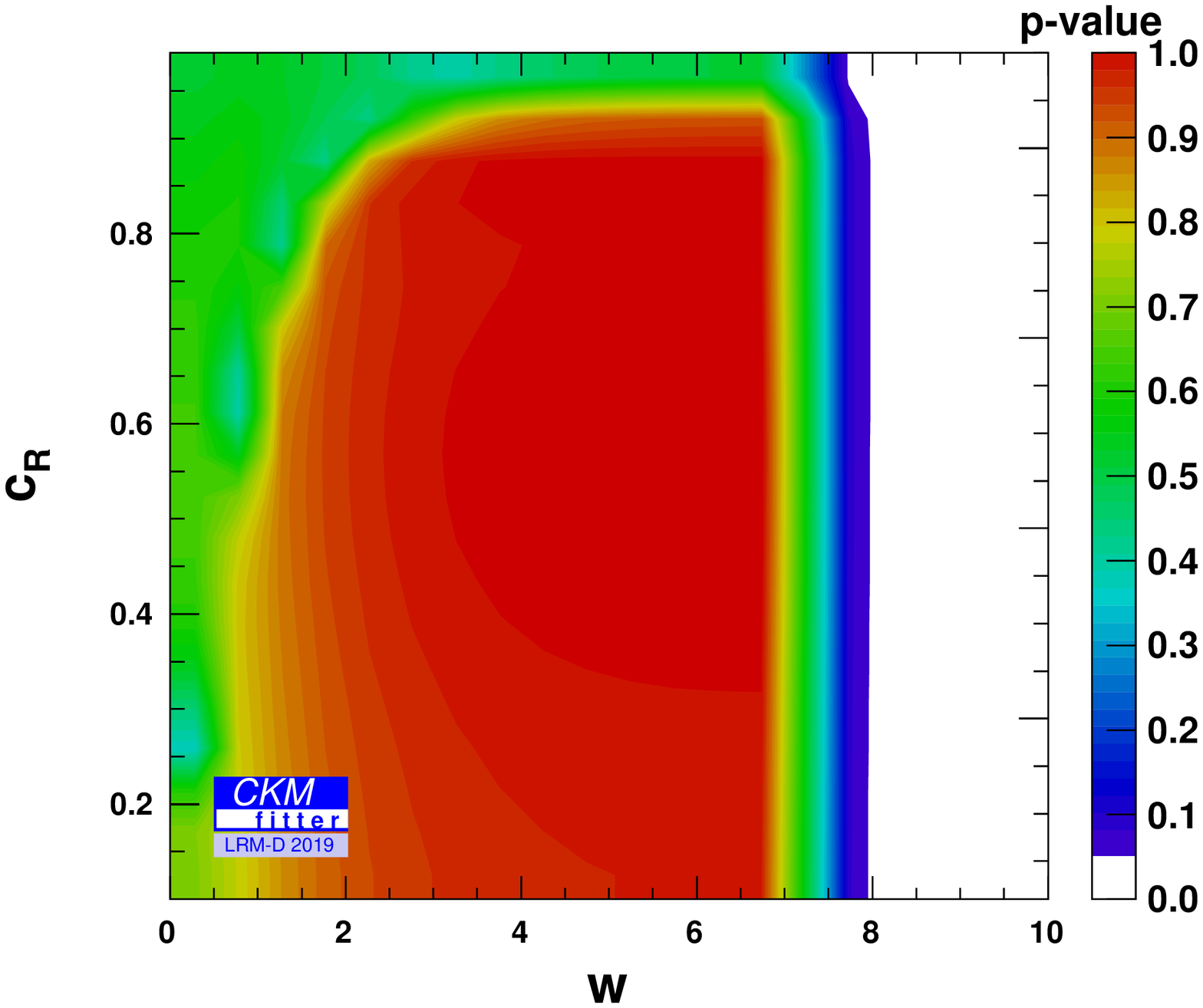, height=6cm}
\end{minipage}
\begin{minipage}[b]{0.5\linewidth}
\epsfig{figure=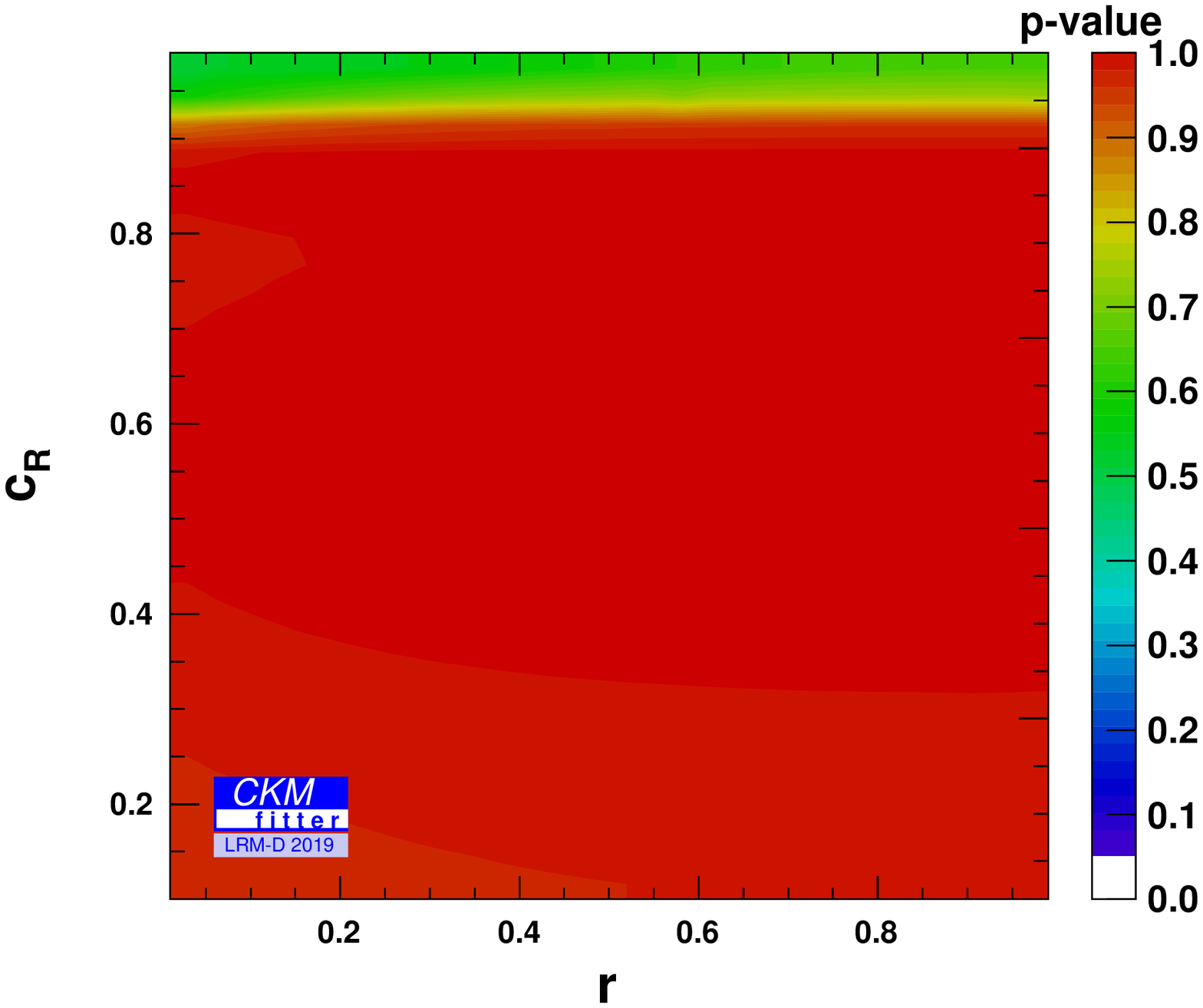, height=6cm}
\end{minipage}
}
\caption{Correlations between the parameters $\epsilon, r, w$ and $c_R$ of the DLRM.}
\label{fig:param2D}
\end{figure}

\subsection{Results}\label{subsec:results}

We start by discussing the results of our fit assuming the SM, given in Table \ref{tab:fitsm}.
The input for $ M_W $ includes an estimation of $ \pm 4 $~MeV for the theoretical error of missing higher-order perturbative calculations.
Note that we have taken for $\alpha_s(M_Z^2)$ the value of
$0.1184 \pm 0.0012$ \cite{Tanabashi:2018oca}, to which we will come back later.

The minimum value of the $ \chi^2 $ is $\chi^2_{min}|_{SM}=22.4$
with a (naive) number of degrees of freedom equal to 23,\footnote{As discussed in Ref.~\cite{Charles:2004jd,Lenz:2010gu}, the precise number of degrees of freedom can be difficult to assess in the presence of theoretical uncertainties and constraints depending only on some of the parameters of the fit.} thus resulting in a $ p $-value of $ \sim 0.5 $.
The compatibility for a given observable within the model considered can be assessed using the one-dimensional pull{~\footnote{Note that in the context of EWPO a different definition is usually found in the literature.}}
defined as
\begin{equation}
pull =\sqrt{\chi^2_{min}-\chi^2_{min,!o}}
\label{eq:pull}
\end{equation}
where $!o$ means that the $\chi^2$ is built and minimised without the experimental 
information on the observable under consideration: $ \chi^2_{!o} $ thus leads to an indirect prediction for this observable. As it is well known, there are a few tensions among the EWPO in the SM, notably $A_{FB}(b)$ and $A_e^{SLD}$, which exhibit an important correlation among their pulls 
\cite{Tanabashi:2018oca}. 
Note that our results 
differ slightly for some observables, as for example the mass of the $W$, from the ones of the 
global fit by the Gfitter collaboration~\cite{Haller:2018nnx}. However we did not use exactly the same inputs and 
the calculations of the observables are not done with the same level of sophistication here.

\vspace{1.5mm}

Let us now turn to the DLRM. The results of the global fit are obtained as follows:

\vspace{0.5mm}

i)  In 
Eq.~\eqref{eq:mu2prParameter} for $ \mu'_2 $ the negative sign has been chosen,
but the positive one gives essentially the same results;

ii) No bound on $ M_{W'} $ is considered (we will come back to this point at the end of the section);

iii) The parameters of the Higgs Lagrangian are restricted within the range $ [-20, 20] $, in order to avoid non-perturbative regimes related to strong couplings, i.e., we impose $ \alpha_2 / (4 \pi) \lesssim 1 $, etc.
Similarly, we require that $ g_X^2 / (4 \pi) < 1 $ and $ g_R^2 / (4 \pi) < 1 $, thus implying $ 0.1 < c_R < 0.99 $. Together with the conditions based on Eq.~\eqref{eq:dech} discussed at the end of Section~\ref{sec:method}, these requirements are collectively called ``perturbativity" in our analysis;

iv) We  exclude the case $ r = 1 $ in our analysis, which is not allowed by the hierarchy of the masses of the fermions, by imposing $ 0 \leq r \leq 0.99 $. Following the discussion of Sec.~\ref{sec:fermions}, such a range of values for $ r $ does not guarantee that the hierarchy of masses is respected, but as we will see $ r $ plays a minor 
role in the fit, so that narrower 
ranges could be chosen with no impact on the analysis;

v) As discussed in Section~\ref{sec:custodial}, there is a residual dependence on the 
renormalisation scale
 in the expression of $\Delta r$.
In principle, we should assign a theoretical uncertainty typically of order one so as to take into account the missing contributions in our computation. However, given the large number of free parameters, for instance the ones of the scalar potential, the fit is insensitive to the presence of this extra uncertainty and the results remain unchanged;

vi) We  take $ \mu = M_Z $, which is the natural scale of the problem. This choice is in agreement with the one for $\alpha_s $, which we remind the reader is  an $ \overline{\rm MS} $ scheme running parameter.
In the DLRM, where $ M_Z^0 \neq M_Z $ in general, we implement the running of $ \alpha_s $ between these two scales at the leading log;

vii) Given that DLRM corrections to $ \Delta \alpha^{(5)}_{had}$, $ m_{top}^{pole} $ have not been computed, 
we add a $ 10\% $ uncertainty to their inputs so that they are allowed to receive additional contributions in the DRLM. This could naturally be improved by a computation of these quantities in the DRLM, however at the prize of adding new parameters in our global fit, which lies beyond the scope of this article.  We follow the same procedure for
$M_h^0$ since we do not consider the $\epsilon^2$ corrections to this quantity. The latter 
involves parameters of the scalar potential which (as we will see 
below) are very badly constrained in our analysis.

\vspace{1.5mm}

\begin{table}
\renewcommand{\arraystretch}{1.2}
\begin{center}
\begin{tabular}{|cc|c|c||c|}
\hline
Observable  & input & full fit ($ 1~\sigma $) & prediction ($ 1~\sigma $) & pull \\
\hline
\hline
  $ 10^5 \times G_F $ [GeV$ ^{-2} $]  & $ 1.1663787(6) $ & $ 1.1663787(6) $ & $ 1.1668 \pm 0.0011 $ & 0.4 \\
  $ M_Z $ [GeV] &  $91.1876 \pm 0.0021$ & $ 91.1875 \pm 0.0021 $ &$ 91.186 ^{+0.018}_{-0.016} $ & 0.1 \\
  $ M_h $ [GeV]  & $ 125.1 \pm 0.2 $ & $ 125.10 \pm 0.20 $ & $ 108^{+45}_{-29} $ & 0.4 \\
  $ m_{top}^{pole} $ [GeV] &  $ 172.47 \pm 0.46 \pm 0.50 $& $ 173.00^{+0.46}_{-1.36} $ & $ 174.5 ^{+4.8}_{-3.4} $ & 0.4 \\
  $ \Delta \alpha (M_Z^2) $ & -- & $ 0.05865^{+0.00040}_{-0.00049} $ & $ 0.05865 ^{+0.00040}_{-0.00049} $ & - \\ 
  $ \alpha_s(M_Z^2) $  & $ 0.1184 \pm 0 \pm 0.0012 $& $ 0.11876^{+0.00085}_{-0.00157} $ & $ 0.1188 \pm 0.0026 $ & 0.0 \\
\hline
\hline
  $ \Gamma_Z $ [GeV]  & $2.4952 \pm 0.0023$& $ 2.49566^{+0.00081}_{-0.00115} $ & $ 2.4954 ^{+0.0011}_{-0.0010} $ & 0.3 \\
  $ \sigma_{had} $ [nb]  & $41.541 \pm 0.037$& $ 41.4763^{+0.0099}_{-0.0068} $ & $ 41.4720 ^{+0.0111}_{-0.0026} $ & 1.7 \\
  $ R_b $  & $0.21629 \pm 0.00066$ & $ 0.215835^{+0.000047}_{-0.000017} $ & $ 0.215835 ^{+0.000046}_{-0.000017} $ & 0.4 \\
  $ R_c $  & $0.1721 \pm 0.0030$ & $ 0.172242^{+0.000025}_{-0.000044} $ & $ 0.172242 ^{+0.000025}_{-0.000044} $ & 0.1 \\
  $ R_e $  & $20.804 \pm 0.050$ & $ 20.7439^{+0.0086}_{-0.0123} $ & $ 20.7384 ^{+0.0133}_{-0.0074} $ & 0.8 \\
  $ R_\mu $  & $20.785 \pm 0.033$ & $ 20.7441^{+0.0086}_{-0.0123} $ & $ 20.7335 ^{+0.0172}_{-0.0025} $ & 1.2 \\
  $ R_\tau $   & $20.764 \pm 0.045$ & $ 20.7910^{+0.0086}_{-0.0123} $ & $ 20.7966 ^{+0.0033}_{-0.0170} $ & 0.9 \\
  $ A_{FB} (b) $  & $0.0992 \pm 0.0016$ & $ 0.10348^{+0.00079}_{-0.00077} $ & $ 0.1048 \pm 0.0010 $ & 2.8 \\
  $ A_{FB} (c) $  & $0.0707 \pm 0.0035$ & $ 0.07397 \pm 0.00060 $ & $ 0.07399 \pm 0.00060 $ & 0.6 \\
  $ A_{FB} (e) $  & $0.0145 \pm 0.0025$ & $ 0.01634 \pm 0.00024 $ & $ 0.0163 \pm 0.0002 $ & 0.4 \\
  $ A_{FB} (\mu) $  & $0.0169 \pm 0.0013$ & $ 0.01634^{+0.00025}_{-0.00024} $ & $ 0.0163 \pm 0.0002 $ & 0.3 \\
  $ A_{FB} (\tau) $  & $0.0188 \pm 0.0017$ & $ 0.01634^{+0.00025}_{-0.00024} $ & $ 0.0163 \pm 0.0002 $ & 1.4 \\
  $ \mathcal{A}_b $  & $0.923 \pm 0.020$ & $ 0.934717^{+0.000105}_{-0.000092} $ & $ 0.93472 ^{+0.00010}_{-0.00009} $ & 0.4 \\
  $ \mathcal{A}_c $  & $0.670 \pm 0.027$ & $ 0.66813^{+0.00049}_{-0.00048} $ & $ 0.66812 \pm 0.00048 $ & 0.2 \\
  $ \mathcal{A}_e^{SLD} $ & $0.1516 \pm 0.0021$ & $ 0.1476 \pm 0.0011 $ & $ 0.148 \pm 0.001 $ & 2.3 \\
  $ \mathcal{A}_e (P_\tau) $ & $0.1498 \pm 0.0049$& $ 0.1476 \pm 0.0011 $ & $ 0.148 \pm 0.001 $ & 0.5 \\
  $ \mathcal{A}_\mu^{SLD} $  & $0.142 \pm 0.015$ & $ 0.1476 \pm 0.0011 $ & $ 0.148 \pm 0.001 $ & 0.4 \\
  $ \mathcal{A}_\tau^{SLD} $  & $0.136 \pm 0.015$ & $ 0.1476 \pm 0.0011 $ & $ 0.148 \pm 0.001 $ & 0.8 \\
  $ \mathcal{A}_\tau (P_\tau) $  & $0.1439 \pm 0.0043$& $ 0.1476 \pm 0.0011 $ & $ 0.148 \pm 0.001 $ & 0.9 \\
\hline
  $ M_W $ [GeV]  & $ 80.379 \pm 0.014 \pm 0.004 $ & $ 80.3696^{+0.0079}_{-0.0095} $ & $ 80.3669 ^{+0.0094}_{-0.0134} $ & 0.5 \\
  $ \Gamma_W $ [GeV]  & $2.085 \pm 0.042$ & $ 2.09144^{+0.00087}_{-0.00123} $ & $ 2.09144 ^{+0.00087}_{-0.00122} $ & 0.2 \\
\hline
  $ Q_W (p) $  & $0.0719 \pm 0.0045$ & $ 0.07350 \pm 0.00058 $ & $ 0.07352 \pm 0.00058 $ & 0.4 \\
  $ Q_W (Cs) $ & $-72.62 \pm 0.43$ & $ -72.928^{+0.042}_{-0.036} $ & $ -72.930 ^{+0.041}_{-0.036} $ & 0.7 \\
  $ Q_W (Tl) $  & $-116.4 \pm 3.6$ & $ -115.423^{+0.063}_{-0.053} $ & $ -115.423 ^{+0.062}_{-0.053} $ & 0.3 \\
\hline
\end{tabular}
\end{center}
\caption{Results of the global fit in the SM. The first row gives the parameters of the fit, $ \{ G_F, M_Z, M_H, m_{top}^{pole}, \Delta \alpha, \alpha_s(M_Z^2) \} $. The third column, ``full fit", gives the result from the fit, the fourth one, ``prediction",
is the value of the observable predicted in the SM without knowledge of its experimental value, while in 
the last column the pull is defined as in Eq.~\eqref{eq:pull}. The inputs are taken from:
\cite{Tanabashi:2018oca} for $ G_F $ and $ \alpha_s $, \cite{Haller:2018nnx} for  $ m_{top}^{pole} $, \cite{Aad:2015zhl} for $M_h$,   \cite{ALEPH:2005ab} for 
$M_Z$ and the EWPO in the second row,  \cite{Aaltonen:2013iut,Aaboud:2017svj,Awramik:2003rn,Haller:2018nnx} for $ M_W $,  \cite{ALEPH:2010aa} for $\Gamma_W $, \cite{Androic:2018kni} for 
$ Q_W (p) $,  \cite{Wood:1997zq,Guena:2004sq,Tanabashi:2018oca}  for  $ Q_W (Cs) $ and 
\cite{Edwards:1995zz,Vetter:1995vf} for  $ Q_W (Tl) $. When two uncertainties are present, the first is statistical while the second is theoretical, treated in the Rfit scheme of \cite{Charles:2004jd,Charles:2011va,Charles:2016qtt}.}
\label{tab:fitsm}
\end{table}

\begin{table}
\renewcommand{\arraystretch}{1.1}
\begin{center}
\begin{tabular}{|c|c|c|c||c|}
\hline
Observable  & input & full fit ($ 1~\sigma $) & prediction ($ 1~\sigma $) & pull \\
\hline
\hline
   $ 10^5 \times G_F^{(0)} $ [GeV$ ^{-2} $]  & -- & $ 1.166 \pm 0.001 $ & -- & - \\
  $ M_Z^{(0)} $ [GeV] & -- & $ 91.21 \pm 0.03 $ & -- & - \\
  $ M_h^{(0)} $ [GeV] &  $125.1 \pm 12 $ & $ 125 \pm 12 $ & $ 25^{+38}_{-9} $ & - \\
  $ M_W^{(0)} $ [GeV] & -- & $ 80.37 \pm 0.02 $ & -- & - \\
  $ m_{top}^{pole,(0)} $ [GeV]  & $ 172.47 \pm 17 $ & $ 172^{+7}_{-4} $ & $ 172 ^{+8}_{-4} $ & - \\
  $ \Delta \alpha (M_Z^2) $ & $ 0.05898 \pm 0.0032 $ & $ 0.059^{+0.002}_{-0.003} $ & $ 0.058 ^{+0.004}_{-0.001} $ & - \\
  $ \alpha_s(M_Z^2) $ &  $ 0.1184 \pm 0 \pm 0.0012 $ & $ 0.1196^{+0.0001}_{-0.0018} $ & $ 0.130 \pm 0.006 $ & 1.8 \\
  $ c_R $ & $ [0.1, 0.99] $ & $ 0.56^{+0.39}_{-0.46} $ & -- & - \\
  $ \epsilon$  & $ \geq 0 $ & $ 0.01^{+0.32}_{-0.01} $ & -- & - \\
  $ r $ & $ [0, 0.99] $ & no bound & -- & - \\
  $ w $ & $ \geq 0 $ & $ 6.9^{+0.4}_{-6.8} $ & -- & - \\
  $ \alpha_{124}, \alpha_2, \lambda_{1}, \lambda^\pm_{23}, \lambda_{4} $ & $ [-20, 20] $ & no bound & -- & - \\
  $ M_{H_1}, M_{H_2} $ & $ \geq 0 $ & no bound & -- & - \\
  $ M_{H_3} $ [TeV] & $ \geq 0 $ & $ 20^{+50}_{-20} $ & -- & - \\
  $ x_{} = \mu'_1 / \mu'_2 $ & -- & no bound & -- & - \\
\hline
\hline
  $ 10^5 \times G_F $ [GeV$ ^{-2} $]  & $ 1.1663787 $ fixed & $ 1.1663787 $ fixed & -- & - \\
  $ M_Z $ [GeV]  & $91.1876 \pm 0.0021$ & $ 91.187^{+0.006}_{-0.005} $ & $ 91.17 \pm 0.06 $ & 0.2 \\
\hline
  $ \Gamma_Z $ [GeV] &  $2.4952 \pm 0.0023$ & $ 2.496 \pm 0.003 $ & $ 2.491 ^{+0.016}_{-0.009} $ & 0.3 \\
  $ \sigma_{had} $ [nb]  & $41.541 \pm 0.037$ & $ 41.51 \pm 0.03 $ & $ 41.42 \pm 0.07 $ & 1.4 \\
  $ R_b $ &  $0.21629 \pm 0.00066$ & $ 0.2158^{+0.0001}_{-0.0002} $ & $ 0.2158 ^{+0.0001}_{-0.0004} $ & 0.5 \\
  $ R_c $ & $0.1721 \pm 0.0030$ & $ 0.17232^{+0.00006}_{-0.00007} $ & $ 0.17232^{+0.00006}_{-0.00007} $ & 0.1 \\
  $ R_e $  & $20.804 \pm 0.050$ & $ 20.73 \pm 0.01 $ & $ 20.73 \pm 0.01 $ & 1.1 \\
  $ R_\mu $  & $20.785 \pm 0.033$ & $ 20.73 \pm 0.01 $ & $ 20.73 ^{+0.01}_{-0.02} $ & 1.6 \\
  $ R_\tau $ &  $20.764 \pm 0.045$ & $ 20.78 \pm 0.01 $ & $ 20.78 \pm 0.01 $ & 0.6 \\
  $ A_{FB} (b) $  & $0.0992 \pm 0.0016$ & $ 0.1033 \pm 0.0009 $ & $ 0.105 \pm 0.001 $ & 2.8 \\
  $ A_{FB} (c) $  & $0.0707 \pm 0.0035$ & $ 0.0738^{+0.0007}_{-0.0008} $ & $ 0.0739 ^{+0.0007}_{-0.0008} $ & 0.6 \\
  $ A_{FB} (e) $ & $0.0145 \pm 0.0025$ & $ 0.0165^{+0.0004}_{-0.0005} $ & $ 0.0165 ^{+0.0004}_{-0.0005} $ & 0.3 \\
  $ A_{FB} (\mu) $ &  $0.0169 \pm 0.0013$ & $ 0.0165 \pm 0.0005 $ & $ 0.0162 ^{+0.0007}_{-0.0003} $ & 0.4 \\
  $ A_{FB} (\tau) $ & $0.0188 \pm 0.0017$ & $ 0.0165 \pm 0.0005 $ & $ 0.0162 ^{+0.0005}_{-0.0003} $ & 1.5 \\
  $ \mathcal{A}_b $  & $0.923 \pm 0.020$ & $ 0.9342 \pm 0.0004 $ & $ 0.9342 ^{+0.0004}_{-0.0003} $ & 0.4 \\
  $ \mathcal{A}_c $ &  $0.670 \pm 0.027$ & $ 0.6674^{+0.0008}_{-0.0007} $ & $ 0.6674^{+0.0008}_{-0.0007} $ & 0.2 \\
  $ \mathcal{A}_e^{SLD} $  & $0.1516 \pm 0.0021$ & $ 0.148 \pm 0.001 $ & $ 0.145 \pm 0.002 $ & 2.5 \\
  $ \mathcal{A}_e (P_\tau) $  & $0.1498 \pm 0.0049$ & $ 0.148 \pm 0.001 $ & $ 0.147 \pm 0.001 $ & 0.5 \\
  $ \mathcal{A}_\mu^{SLD} $  & $0.142 \pm 0.015$ & $ 0.148 \pm 0.001 $ & $ 0.148 \pm 0.001 $ & 0.4 \\
  $ \mathcal{A}_\tau^{SLD} $  & $0.136 \pm 0.015$ & $ 0.148 \pm 0.001 $ & $ 0.148 \pm 0.001 $ & 0.8 \\
  $ \mathcal{A}_\tau (P_\tau) $  & $0.1439 \pm 0.0043$ & $ 0.148 \pm 0.001 $ & $ 0.148 \pm 0.001 $ & 0.9 \\
\hline
  $ M_W $ [GeV]  & $ 80.379 \pm 0.014 \pm 0.004 $ & $ 80.37^{+0.02}_{-0.01} $ & $ 80.2 \pm 0.2 $ & 0.8 \\
  $ \Gamma_W $ [GeV]  & $2.085 \pm 0.042$ & $ 2.091^{+0.002}_{-0.003} $ & $ 2.091 ^{+0.002}_{-0.003} $ & 0.1 \\
\hline
  $ Q_W (p) $  & $0.0719 \pm 0.0045$ & $ 0.074^{+0.002}_{-0.006} $ & $ 0.074 ^{+0.003}_{-0.011} $ & 0.5 \\
  $ Q_W (Cs) $  & $-72.62 \pm 0.43$ & $ -72.8^{+0.6}_{-0.2} $ & $ -72.9 ^{+1.4}_{-0.1} $ & 0.5 \\
  $ Q_W (Tl) $ & $-116.4 \pm 3.6$ & $ -115.2^{+0.9}_{-0.2} $ & $ -115.0^{+0.6}_{-0.5} $ & 0.4 \\
\hline
  $ M_{W'}^2 $ [TeV$^2$] & -- & $ 2^{+6}_{-2} $ & -- & - \\
  $ M_{Z'}^2 $ [TeV$^2$] & -- & $ 7^{+730}_{-7} $ & -- & - \\
\hline
\end{tabular}
\end{center}
\vspace{-6mm}
\caption{Same as in the Table~\ref{tab:fitsm} but in the DLRM except for  $ m_{top}^{pole,(0)} $ and $ M_h^{(0)} $ for which we have increased the error bars of the input value, see the discussion of the results in Section~\ref{sec:fits}.}
\label{tab:fitlrm}
\end{table}

Comparing the results from the DLRM shown in Table \ref{tab:fitlrm} with Table \ref{tab:fitsm}
one observes the same tensions as in the SM, so that there is no
concrete improvement from the DLRM as far as the EWPO are concerned. The minimum of the $ \chi^2 $ is $\chi^2_{min}|_{DLRM}=20.2$, with a (naive) number of degrees of freedom of 20,\footnote{This number of degrees of freedom is the same as 
in the SM case minus the parameters that we constrained in the DLRM, namely $ \{ \epsilon, w, c_R \}$. Obviously, the DRLM contains many more parameters, but as they 
 are not constrained by the fit, we do not include them in our counting for the degrees of freedom relevant for a statistical interpretation of the value of $\chi^2_{min}$.} leading to a $ p $-value of $ \sim 0.4 $.
One can consider the SM as a limiting case of the DLRM, so that both can be seen as two nested hypotheses~\cite{Lenz:2010gu,Charles:2016qtt}. 
 It then follows that
the quantity $\chi^2_{min}|_{SM}-\chi^2_{min}|_{DLRM}$ is distributed as following a 
$\chi^2$ law with a (naive) number of degrees of freedom of 3. It can be interpreted as a $0.5\sigma $ deviation, not large enough for preferring the DLRM hypothesis over the SM one.

As seen from Table~\ref{tab:fitlrm}, among the parameters 
specific to DLRM the set of observables in Eq.~\eqref{eq:data} constrains $ \epsilon $ and $ w $
while  $ c_R $ and $ r $ remain essentially unconstrained. Values of
$\epsilon \lesssim 0.3 $ and large values of the parameter $w \sim \mathcal{O} (1) $ are favoured. Consequently the doublet $ \chi_L$ plays an important role in triggering the spontaneous EW symmetry breaking, whereas it is essentially absent in the TLRM 
due to the breaking of the custodial symmetry and  the bi-doublet $ \phi $ alone triggers this breaking. The role played by $ w $ in the DLRM can
further be seen from Table~\ref{tab:MW'}. The results just discussed are also illustrated in Fig.~\ref{fig:param}, from which one reads $ 1\sigma $ confidence level regions by taking for each parameter the intervals
corresponding to $p$-values higher than
$ \sim 0.33 $. Also shown in this figure is the LR breaking scale $ v_R $, which is constrained by the EWPO to be $ v_R = 6.8^{+8.6}_{-0.8} $~TeV at $ 1\sigma $. We note that in the analysis of the EWPO in the framework of the  TLRM with $w=0$~\cite{Blanke:2011ry}, small values of  the ratio of the
EW and LR symmetry breaking scales (equivalent to the quantity $ \sqrt{2 k}  \epsilon  $ discussed  here)  have been favoured with a preference for $ c_R < 0.7$.

\begin{table}
        \renewcommand{\arraystretch}{1.1}
        \centering
        \begin{tabular}{|c|c|}
        \hline
        $ w $ & $ \chi^2_{min} $ \\
        \hline
        free & $ 20.2 $ \\
        \hline
        $ 1 $ & $ 20.5 $ \\
        \hline
        $ 0.01 $ & $ 22.2 $ \\
        \hline
        \end{tabular}
        \hspace{2cm}
        \begin{tabular}{|c|c|}
        \hline
        $ M_{W'}^2 $ [TeV$^2$] & $ \chi^2_{min} $ \\ 
        \hline
        $ > 0 $ & $ 20.2 $ \\ 
        \hline
        $ > 20 $ & $ 21.7 $ \\ 
        \hline
        $ > 50 $ & $ 22.0 $ \\ 
        \hline
        \end{tabular}
        \caption{Impact on the $\chi^2$ from (left) $ w $, and (right) $ M_{W'}$.}
\label{tab:MW'}
\end{table}

\begin{figure}[t!]
\begin{minipage}[b]{0.5\linewidth}
\epsfig{figure=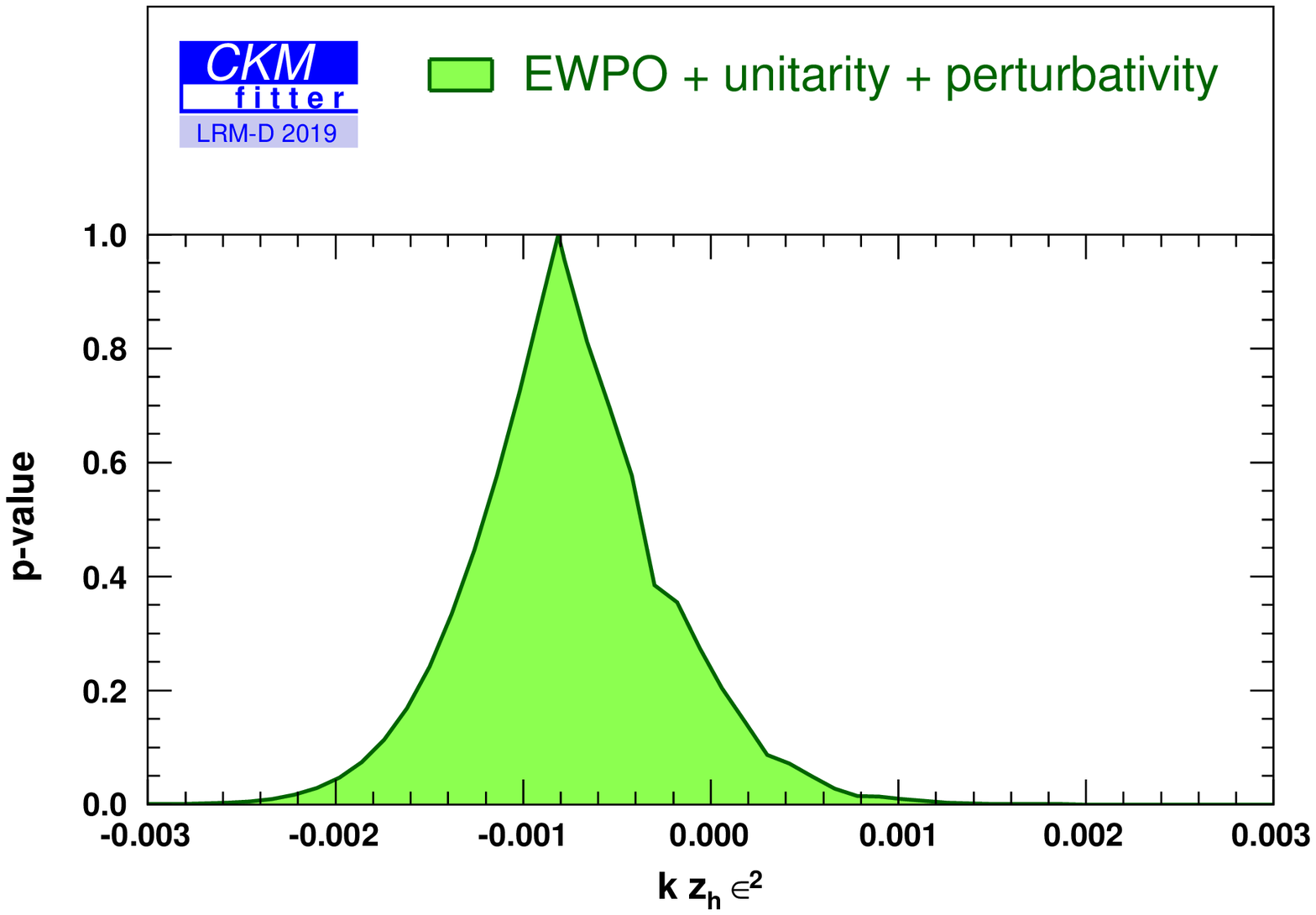, height=6cm}
\end{minipage}
\begin{minipage}[b]{0.5\linewidth}
\epsfig{figure=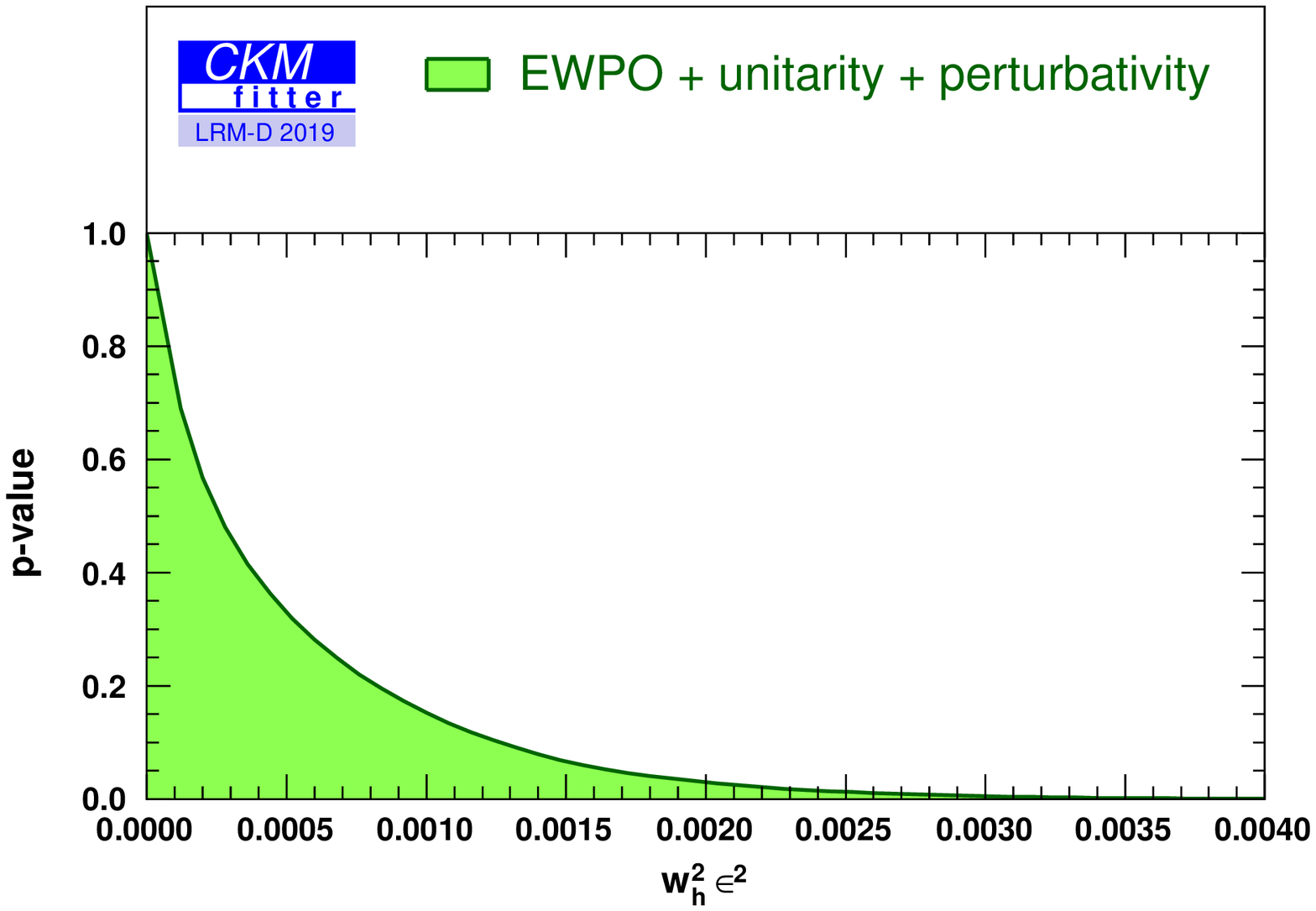, height=6cm}
\end{minipage}
\begin{minipage}[b]{0.5\linewidth}
\epsfig{figure=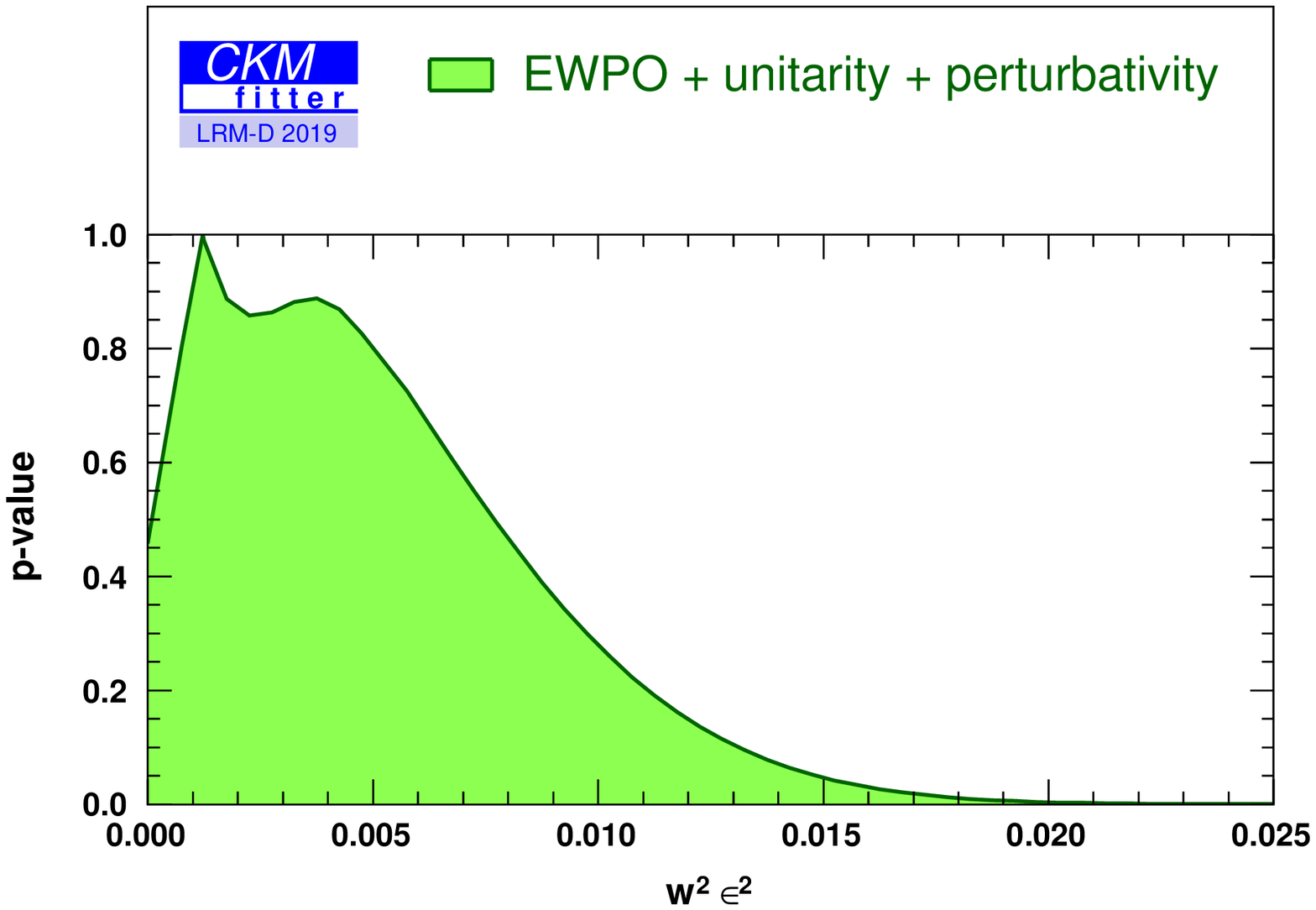, height=6cm}
\end{minipage}
\begin{minipage}[b]{0.5\linewidth}
\epsfig{figure=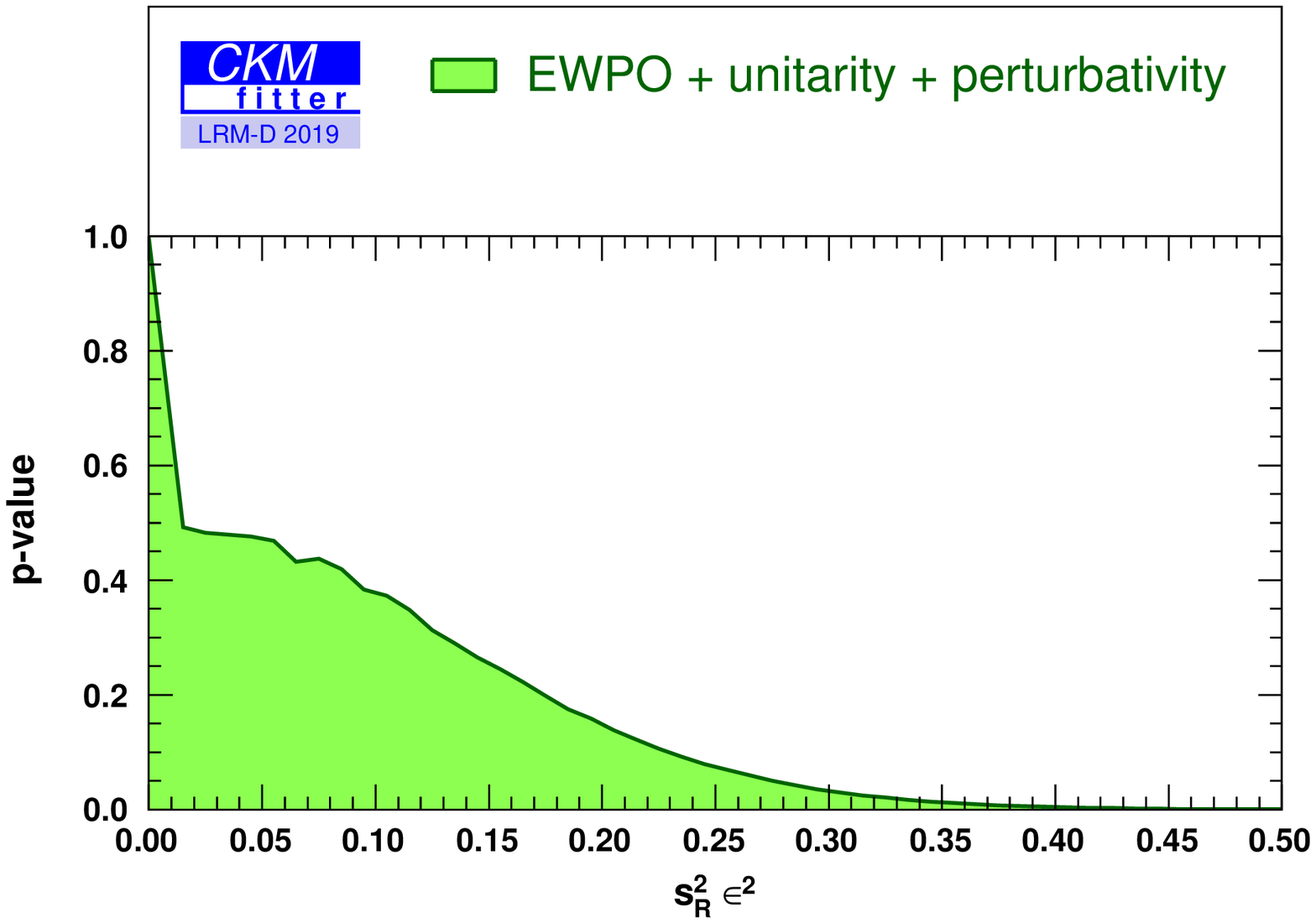,height=6cm}
\end{minipage}
\caption{The $p$-values for some parameters of the DLRM.}
\label{fig:paramNew}
\end{figure}

\begin{figure}[t!]
{
\begin{minipage}[b]{0.5\linewidth}
\epsfig{figure=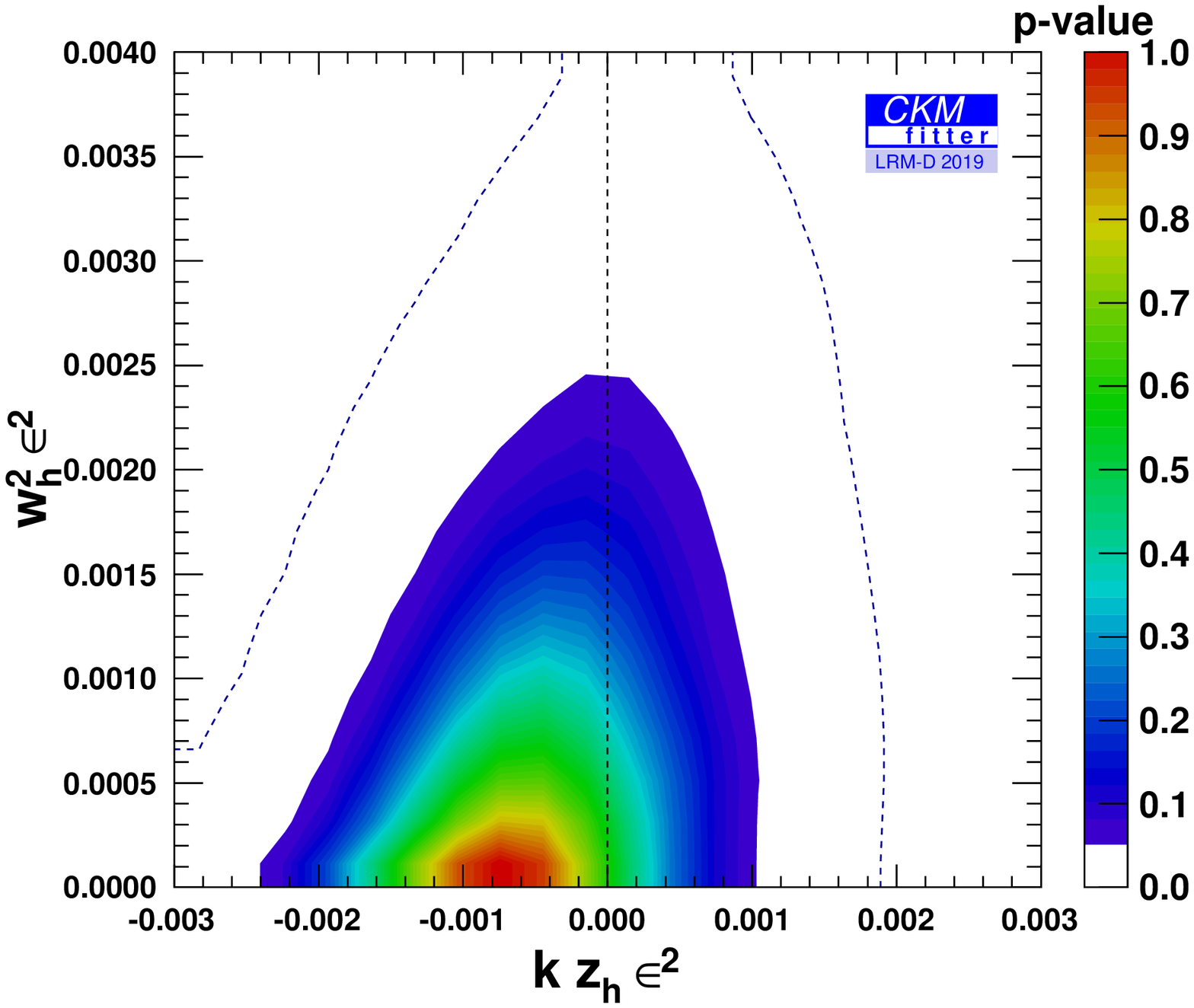, height=6cm}
\end{minipage}
\begin{minipage}[b]{0.5\linewidth}
\epsfig{figure=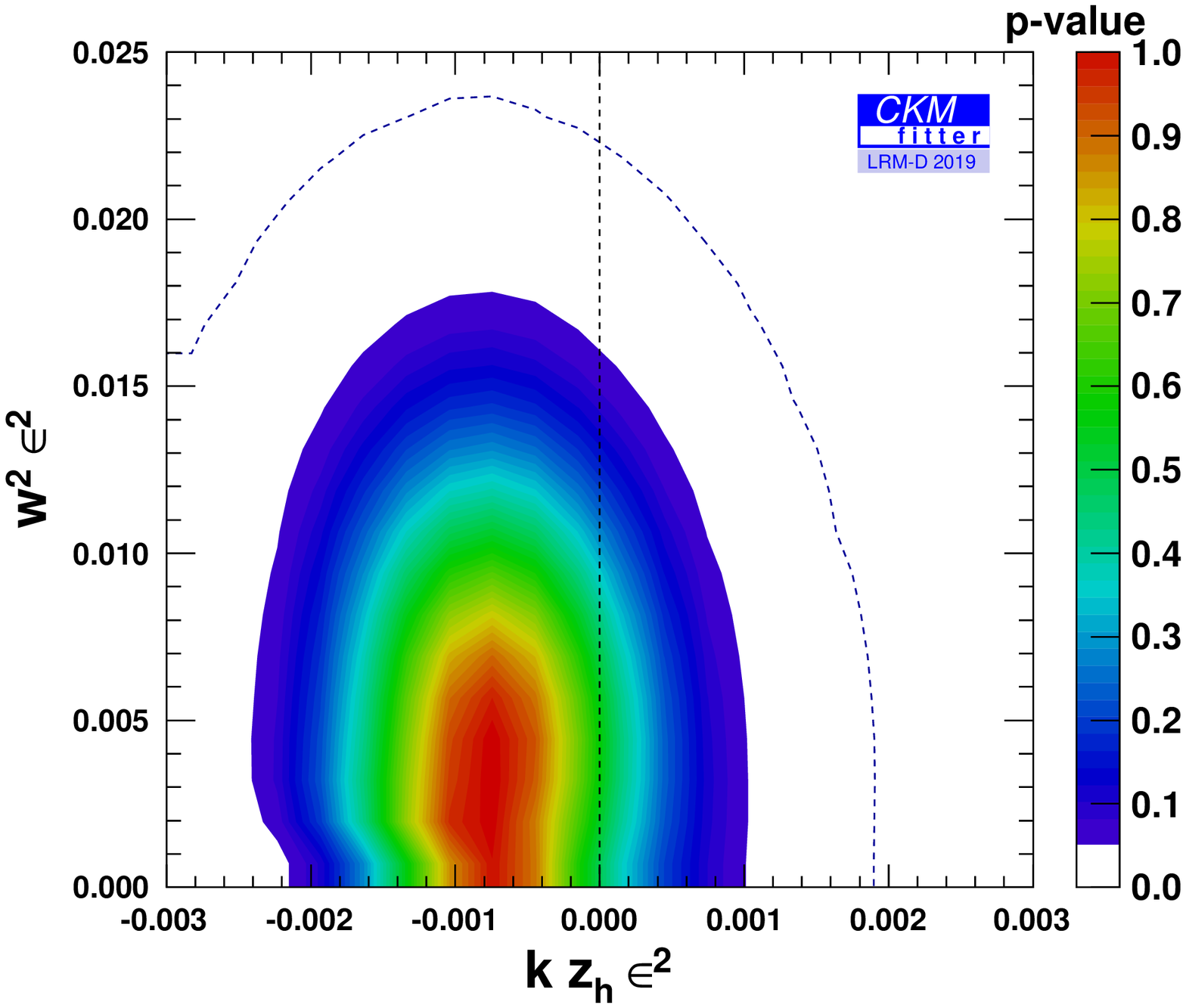, height=6cm}
\end{minipage}
\begin{minipage}[b]{0.5\linewidth}
\epsfig{figure=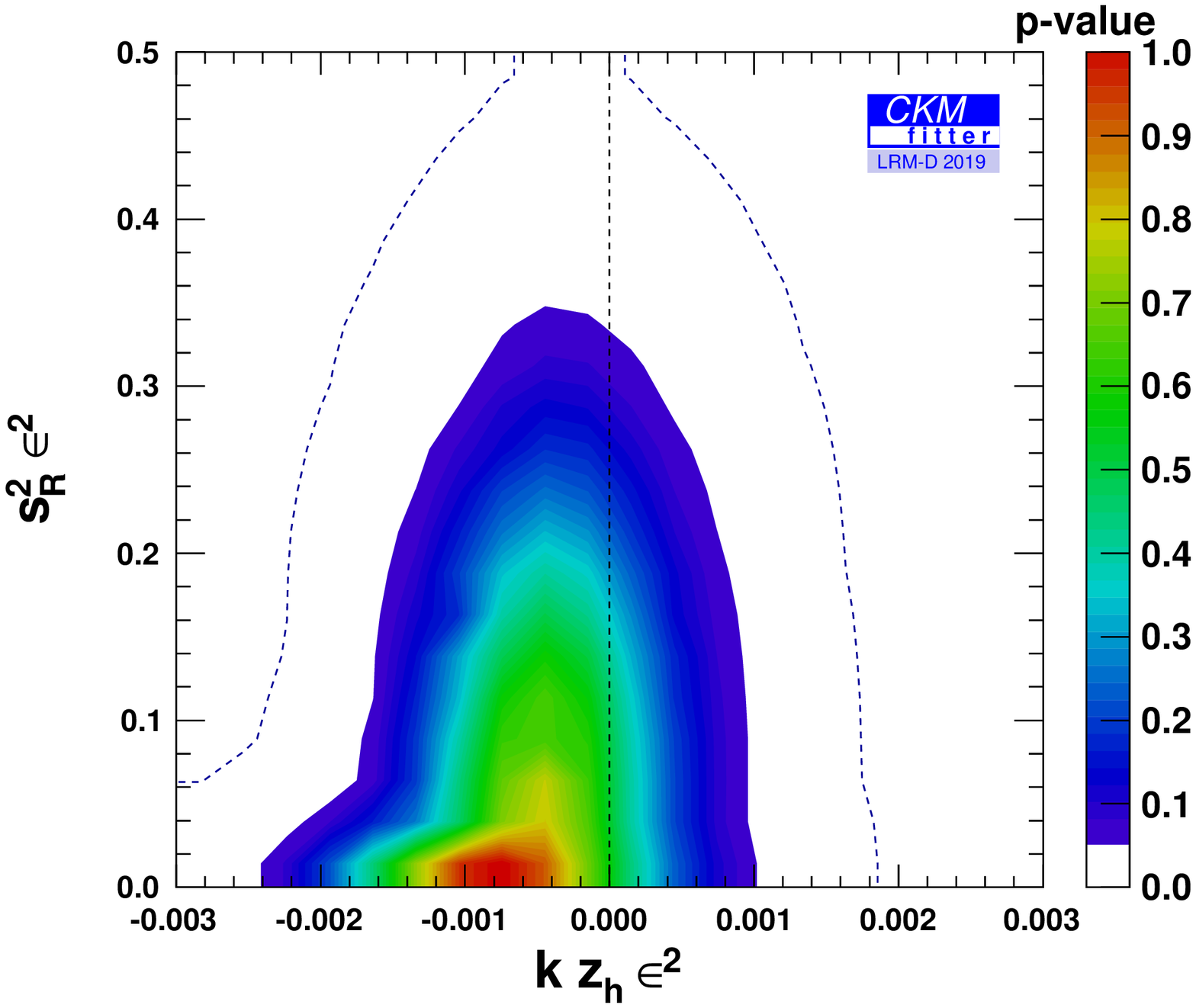, height=6cm}
\end{minipage}
\begin{minipage}[b]{0.5\linewidth}
\epsfig{figure=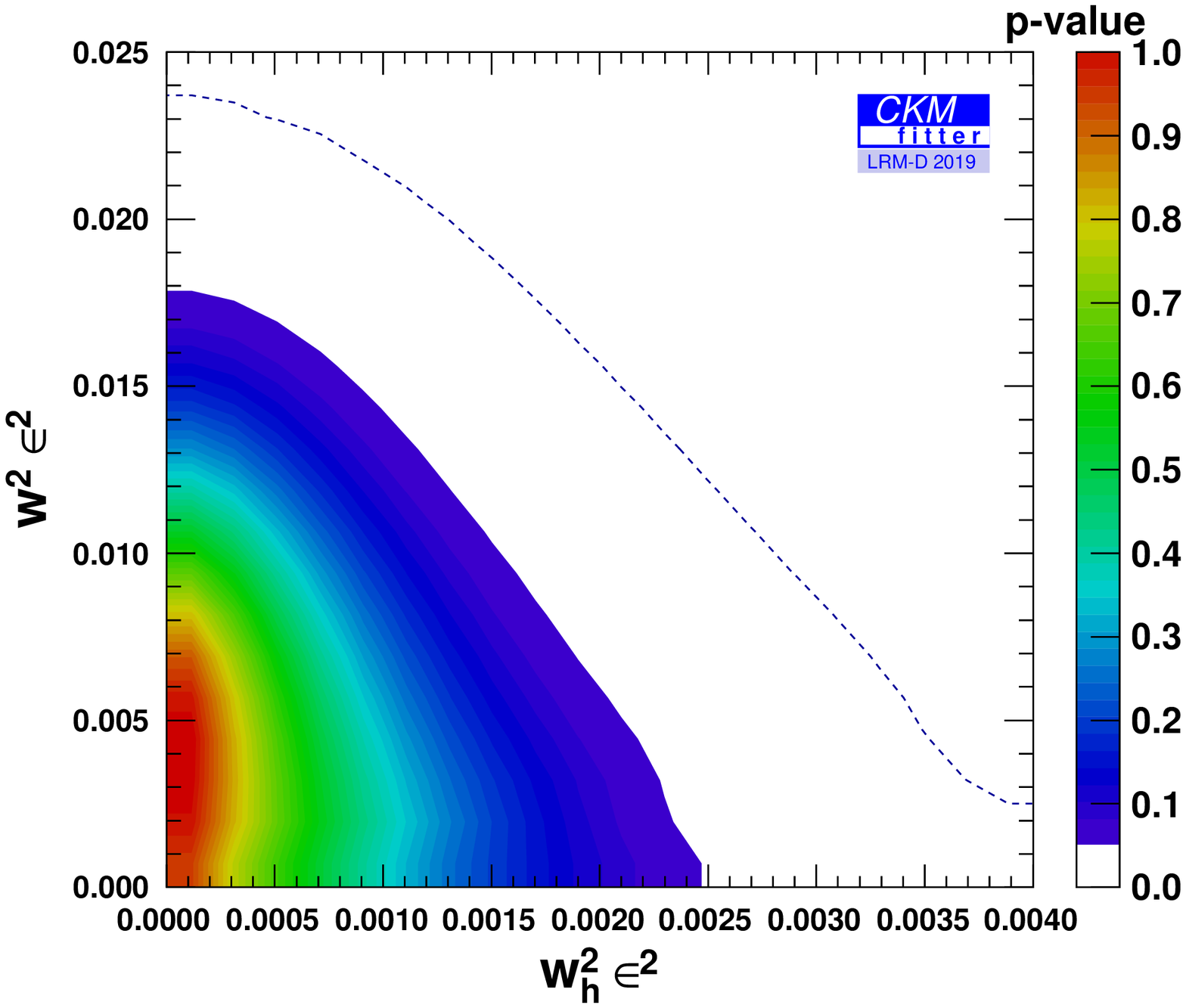,height=6cm}
\end{minipage}
\begin{minipage}[b]{0.5\linewidth}
\epsfig{figure=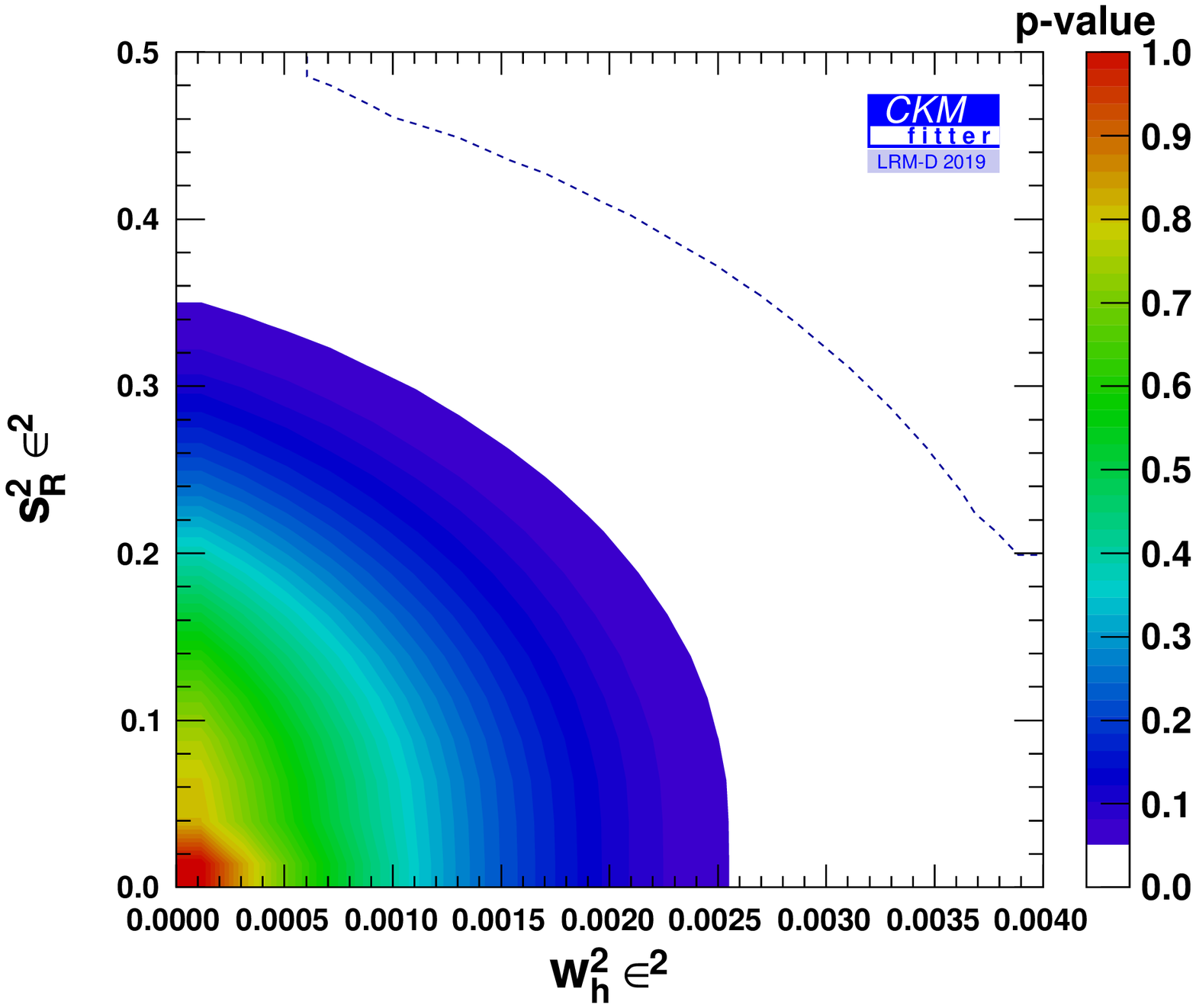, height=6cm}
\end{minipage}
\begin{minipage}[b]{0.5\linewidth}
\epsfig{figure=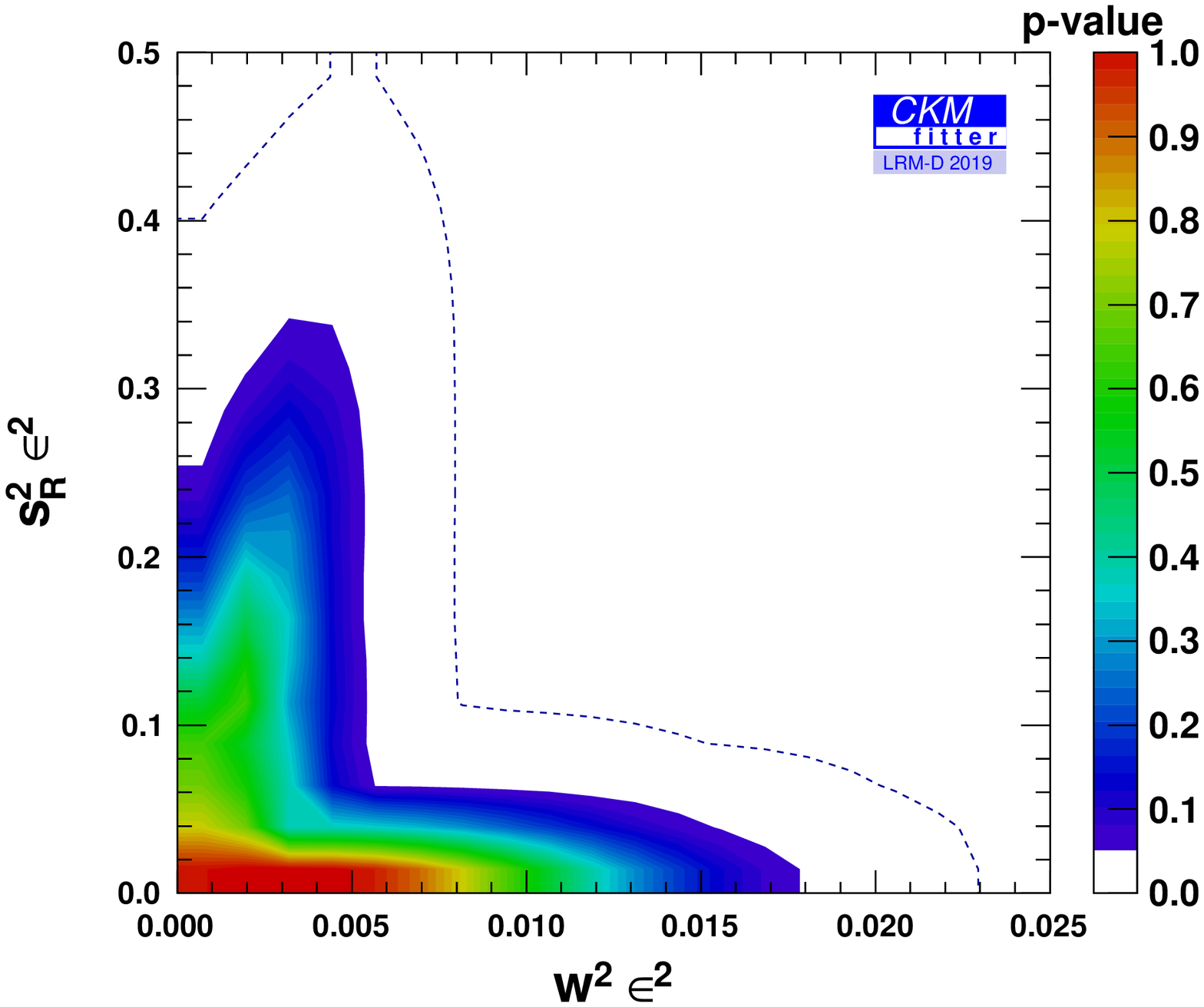, height=6cm}
\end{minipage}
}
\caption{Correlations between the combinations $ k z_h \epsilon^2 $, $ w_h^2 \epsilon^2 $, $ w^2 \epsilon^2 $ and $ s_R^2 \epsilon^2 $, where $ z_h $ and $ w_h $ are defined in Eq.~\eqref{eq:masswzLO}. The dotted lines set the $ 3\sigma $ region.}
\label{fig:param2DNew}
\end{figure}

The correlations between the four parameters $\epsilon, r, w$ and $c_R$ are shown in Fig.~\ref{fig:param2D}. As expected from the fact that $r, w$ and $c_R$
enter our expressions always multiplied by $\epsilon$ there are strong correlations between the latter, which sets the ratio of EW and LR symmetry breaking scales, and the former three. Much weaker correlations, or no correlations at all, are observed among $r, w$ and $c_R$.
In Figure~\ref{fig:paramNew}, we show the allowed values of the combinations $ k z_h \epsilon^2 $, $ w_h^2 \epsilon^2 $, $ w^2 \epsilon^2 $ and $ s_R^2 \epsilon^2 $ which are the natural quantities appearing in the EWPO, see Appendix~\ref{sec:expsEWPO} with $ z_h $ and $ w_h $ defined in Eq.~\eqref{eq:masswzLO}. Their correlations are given in Figure~\ref{fig:param2DNew}.

The set of observables in Eq.~\eqref{eq:data} alone is 
not sufficient
to set bounds on the Higgs sector of the theory specific to the DLRM. Indeed, the $p$-values obtained for the Higgs masses $ M_{H_{1,2,3}} $ and the parameters of the Higgs Lagrangian are very flat over a wide range of values, so that no stringent confidence interval can be deduced.
However, using the bound on $v_R$ given previously the unitarity relation from the analysis of $ Z' Z' \rightarrow Z' Z' $ scattering sets the bound $ M_{H_3} \lesssim 63 $~TeV at $ 1\sigma $.
This is analogous to the bound resulting from Eq.~\eqref{eq:a0ZZ} in the SM framework, where the knowledge of the EW scale $ v $ sets the bound $ M_h \lesssim 4 v $ on the mass of the light scalar.
More impact from the unitarity relations relies on constraining the parameters $ F_i, S_i \, w / k, c^{H_3} / k^2 $ ($ i=1,2 $) of Eq.~\eqref{eq:resumeUnitarity}. Note that $ F_i, S_i \, w / k $ ($ i=1,2 $) depend on the parameters $ r , w , M_{H_1} / M_{H_2} , \mu_1' / \mu_2' $, while $ c^{H_3} / k^2 $ depends also
on additional parameters of the scalar potential.  
The fit leads to:
\begin{equation}
-1.1 < F_1 < 0.2 \,, \quad -0.9 < F_2 < 0.5 \,, \quad -0.6 < S_1 \, w / k < 0.3 \,, \quad | S_2 \, w / k | < 0.7 \,, \quad | c^{H_3} / k^2 | < 250
\end{equation}
at $ 1\sigma $.
Apart from $c^{H_3} / k^2$ they are rather small and compatible with zero thus limiting the sensitivity
to the scalar spectrum.
If more information on the possible allowed values of the parameters of the scalar potential was 
 included, together with other indirect or direct bounds on part of the scalar spectrum, the remaining scalar masses 
would
 be probed more accurately 
by the unitarity 
constraints.

Let us come back to the $W'$ and $Z'$ masses and first discuss briefly what is 
known about them.  These heavy spin-1 charged and neutral particles (including sequential $W$ and $Z$ particles having the same couplings to fermions as the 
$W$ and $Z$, excited Kaluza-Klein modes, etc., which are also generically called $W'$ and $Z'$) can be looked for at hadron colliders as well as in 
measurements of processes at energies 
much below their masses. Active searches have been pursued by LEPII, ATLAS and 
CMS and some bounds
have been obtained, however we should stress that these bounds depend on the specific models considered, including different possible realizations of LR models.  
 Focusing on the latter a lower bound of $M_{W'}>715$ GeV  has been given at $90\%$
CL from an electroweak fit \cite{Czakon:1999ga}, while 
 in Ref.\cite{Aaboud:2018jux}  the constraint $M_{W'}>3.25$~TeV is found at $95\%$ CL from a search for vector-boson resonances decaying to a top quark and 
bottom quark  assuming a $ W' t b $ coupling similar to the SM $ W t b $ one: for a $ W' t b $ coupling $ \sim 10 $ times smaller, the analysis \cite{Aaboud:2018jux} is not sensitive to this decay channel of $ W' $.
On the theoretical side, Ref.~\cite{Bandyopadhyay:2019jzq} discusses 
contributions of the $ W' $ to the SM-like Higgs decaying into two photons within a LRM realized with doublets, reaching sensitivities to $ M_{W'} $ of $ \mathcal{O} (200) $~GeV for $g_L = g_R$.  Also, a thorough analysis   of flavour constraints has been performed in Ref.~\cite{Senjanovic:2015yea} in the case of TLRM with an additional discrete symmetry. Together with the
requirement of model perturbativity \cite{Maiezza:2016bzp} and constraints on CP violation \cite{Bertolini:2019out}, the $W'$ scale is pushed to at least 7 
TeV. This value is somewhat larger than the ones quoted before
and is similar to what is generally found in other models. For 
example, a lower 6 TeV limit has been obtained by ATLAS \cite{Aad:2019wvl} from the extraction of
the cross section for $pp \to W' \to l \nu$ 
 using a data sample of 139~fb$^{-1}$ of proton-proton collisions at 
 $\sqrt s = 13$ TeV  in the context of the Sequential Standard Model (SSM) where 
 the couplings of the $W'$ are assumed to be identical to the SM $W$ boson ones, whereas the couplings to the SM bosons are set to zero.
Concerning the $Z'$, assuming $g_L = g_R$
two lower bounds are quoted in \cite{Tanabashi:2018oca}: one from $ p \bar p $ direct search $M_{Z'}>630$ GeV \cite{Abe:1997fd}
and the other from an electroweak fit $M_{Z'}>1162$ GeV \cite{delAguila:2010mx}, both at $95\%$ CL.

\begin{figure}[t!]
\begin{minipage}[b]{0.5\linewidth}
\epsfig{figure=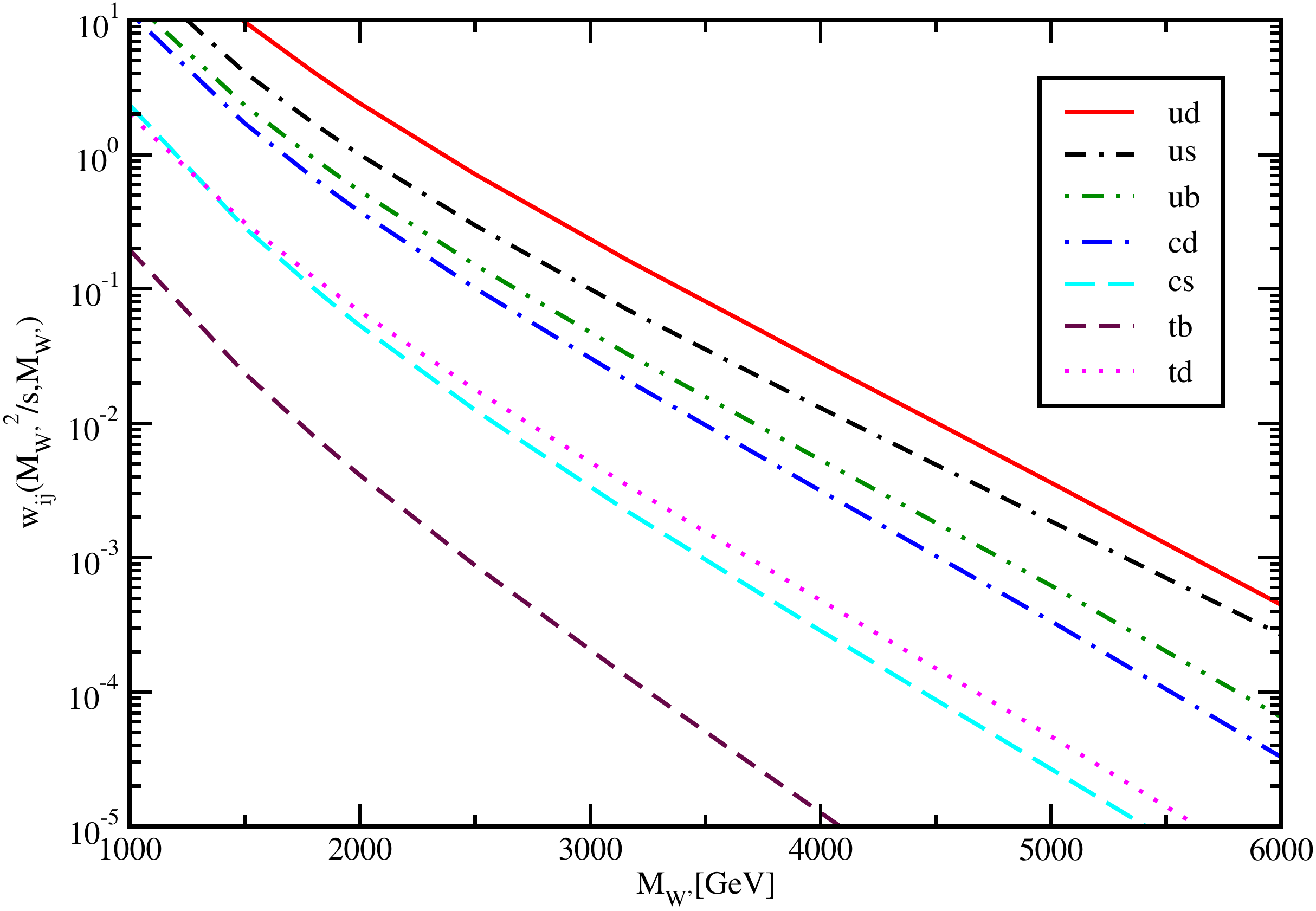, height=5.5cm}
\end{minipage}
\begin{minipage}[b]{0.5\linewidth}
\epsfig{figure=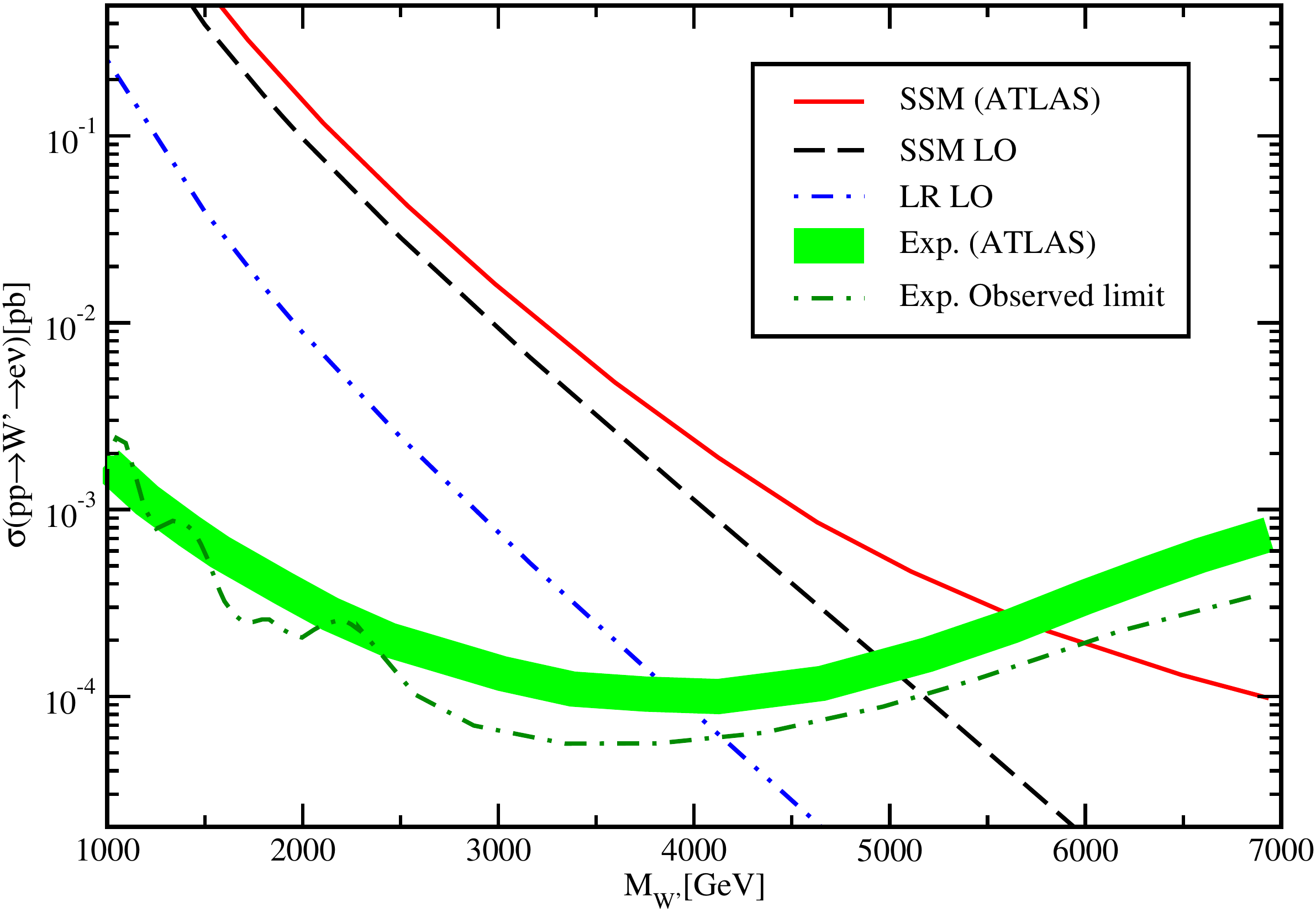, height=5.5cm}
\end{minipage}
\caption{(Left) The quantity $w_{ij}(M_{W'}^2/s,M_{W'})$, Eq.~\eqref{eq:pdf} for up and down type
quarks
of various generations as a function of the $W'$ mass. (Right) $pp \to W' \to e \nu$ cross section in the Sequential Standard Model (``SSM LO") and in the LR model (``LR LO") at LO. Also shown are the SSM result from ATLAS \cite{Aad:2019wvl} as well as their observed and expected upper limits at 95$\%$
CL.}
\label{fig:pdfatlas}
\end{figure}

In view of the above discussion we first leave $M_{W'}$  free in the fit. It 
favours masses  below $\sim 3$ TeV, see Table \ref{tab:fitlrm}, as well as
a value of $c_R$ at the best-fit point of $c_R \sim 0.5$, meaning that the 
$Z'$ mass is roughly twice as large as the $W'$ one (however, the uncertainty
on the latter result does not allow at that point to
differentiate between a  value of the $Z'$ mass 
comparable or much higher than the $W'$ one). While the value of $M_{W'}$ is in agreement 
with some of the bounds 
given above  it points towards a rather large suppression of the $W'$ production  in the direct
search from ATLAS discussed in the previous paragraph \cite{Aad:2019wvl}, and thus poses the question 
of the possible existence of  
couplings of the $W'$ to  the quarks which would allow it.
In order to answer it  we calculate  the production cross section at leading-order.  It is given by, see for example \cite{Campbell:2006wx,PDG2020}:
\begin{equation}
\sigma(pp \to W' X) = \frac{\pi}{6 s}  g_R^2 \sum_{i,j} | V^R_{ij} |^2 w_{ij}(M_{W'}^2/s,M_{W'})
\label{eq:sig}
\end{equation}
where $V_R$ is the analogue of the CKM matrix in the right-handed sector, and the functions $w_{ij}$ encode the information about the proton structure
\begin{equation}
w_{ij}(z,\mu)=\int_z^1 \frac{d x}{x} \biggl(u_i(x,\mu) \bar{d}_j \left( \frac{z}{x},\mu \right)
+ \bar{u}_i(x,\mu)  d_j \left( \frac{z}{x},\mu \right) \biggr)
\label{eq:pdf}
\end{equation} 
with $u_i(x,\mu)$ and  $d_i(x,\mu)$ the parton distribution functions (PDF) for the up
and down type quarks of the $i$th and $j$th generations in the proton, respectively,
at the factorization scale $\mu$ and parton momentum fraction $x$. In the following the results
have been obtained using the NNPDF23LO PDF set \cite{Ball:2012cx} 
and a constant 
branching fraction for $W'$ boson decays into leptons of one generation of 
8.2$\%$, see \cite{Aad:2019wvl}. The quantity $w_{ij}(M_{W'}^2/s,M_{W'})$ is shown 
on the left-panel of Fig.~\ref{fig:pdfatlas}  for various combinations of $ i $ and $ j $,
while the LO cross section, Eq.~\eqref{eq:sig} within the DLRM and SSM,\footnote{Note that at LO there is a rather large dependence of the result on the chosen factorization scale $\mu$.} and the results from ATLAS are displayed on the right-panel of Fig.~\ref{fig:pdfatlas}. The  curve from  ATLAS labelled ``SSM" on the figure  has been 
obtained using
the PYTHIA v8.183 event generator \cite{Sjostrand:2007gs} to produce the Monte-Carlo events in the
SSM at LO, as well as the A14 set of tuned parameters \cite{A14ATlas:2014}
for the parton showering and hadronization process. Clearly, in order to decrease the bound on the $W'$ mass one needs 
to suppress the largest contributions to the process under consideration, namely the ones from the PDF of the up and down quarks of the first generation. This can be achieved
partially by taking an anti-diagonal matrix $V_R$. The result is displayed on 
the figure for $g_R=g_L/\sqrt 3$, which corresponds approximately to the minimal possible value of $ g_R $ in the 
LR model, see Eq.~\eqref{eq:gRexpression}. A more refined calculation would be needed, which however goes  beyond the scope of the paper, but it seems difficult to lower the limit of 
the $W'$ mass much  more than 4 TeV.


\input{new_table_v3.tex}

Finally, the value for $\alpha_s(M_Z^2)$ used here corresponds to the world average value from PDG~\cite{Tanabashi:2018oca}. As illustrated by their review on QCD, obtaining a world average is not a trivial exercise.  There are various ways of determining $\alpha_s(M_Z^2)$ which can be grouped into certain sub-categories as $e^+ e^-$ into hadronic states, deep inelastic scattering (DIS), hadronic $\tau$ decay, lattice QCD, heavy quarkonia decays and hadron collider data. Actually the FLAG lattice average \cite{Aoki:2019cca} dominates 
the world average. Some of the non-lattice determinations are in good agreement with 
FLAG, but some are quite a bit lower. The EWPO are also  used to determine the strong coupling which 
results into central values slightly larger than the world average one, but compatible with it, the latest update of the global fit to electroweak precision data by Gfitter\cite{Baak:2014ora}   giving $\alpha_s(M_Z^2)=0.1196 \pm 0.0030$.  Considering a more conservative interval of $ 0.117 \pm 0.005 $ in our global fit in the DLRM clearly improves
the $\chi^2$ of the fit $ \chi^2_{min}|_{DLRM} = 18.9 $ (the SM $ \chi^2 $, however, remains $ \chi^2_{min} |_{SM} = 22 $).
This results mainly from  a better agreement of the DLRM predictions both of $ \sigma_{had} $ with its experimental value and of
$ \alpha_s (M_Z^2) $ with its more conservative input. The pulls for these two quantities  decrease respectively to $ 0.9 $ and $ 1.3 $, while
the ones of some other observables get only slightly smaller.

Note that if we disregard in the fit the information on $\alpha_s$ (column labelled ``prediction") the fit goes towards larger values of $\alpha_s$.

\section{Conclusions} \label{sec:concl}

In this article, we have considered a left-right symmetric extension of the Standard Model where the spontaneous breakdown of the left-right symmetry is triggered by doublets. The $\rho$ parameter is then protected at tree level from large corrections in this Doublet Left-Right Model (DLRM), contrary to the case where triplets are considered. This allows in principle for more diverse patterns of symmetry breaking.
There are, however, possibly large radiative corrections coming from the new 
scalar and vector particles to the $W$ and $Z$ self-energies that we investigated.
The new scalars can also be probed by unitarity constraints, exactly as 
considerations of unitarity in the scattering of longitudinally polarized gauge bosons helped setting theoretical bounds on the mass of the SM Higgs boson much before its direct discovery at LHC.


Combining unitarity, electroweak precision observables 
and the radiative corrections to the muon $\Delta r$ parameter within a frequentist (CKMfitter) approach, we see that the model is only mildly constrained:
the fit bounds DLRM corrections to remain small, pushing the LR scale to be of the order of a few TeV, thus limiting the sensitivity to the new fundamental parameters.
Nonetheless, a new qualitative feature, favoured by the data, emerges from our analysis of DLRM, which is the possibility of having spontaneous EW symmetry breaking triggered also by a doublet under $ SU(2)_L $, as opposed to the case mostly studied in which EWSB is triggered only by the bi-doublet under $ SU(2)_L \times SU(2)_R $. This possibility has not received much attention in the literature as it is not allowed in the triplet scenario, in which a triplet under $ SU(2)_L $ is considered.

The favoured masses of the new  $ W' $ and $ Z' $ gauge bosons are found in the range $ \mathcal{O} (1-3) $~TeV.
On the other hand, the large number of new possible scalar self-interactions limits the determination 
of the masses in the scalar sector, that could be much lighter or much heavier than the LR scale. The requirement of tree-level unitarity, establishing relations between the scalar masses and the LR scale, could be of much help once further constraints sensitive to the scalar sector are added to our analysis.
In that respect, our study 
can be improved by the consideration of the flavour sector of DLRM. Indeed, a large set of clean flavour observables is available and they are known to set important bounds on generic new interactions changing quark flavours, that in the DLRM are encoded in the CKM-like matrix and its right-handed sector analogue.
While these two matrices are closely related in LR models with an additional 
discrete symmetry (e.g., parity), it is not the case here, leading to a 
certain number of new parameters. Direct collider bounds on the $ W' $ 
gauge boson which exploit quark channel decays, or production mechanisms, 
depending on the same right-handed matrix can be used   
to determine these parameters as we have briefly discussed in the results 
section. Indeed we have seen that in order to be in agreement with the 
results from ATLAS \cite{Aad:2019wvl} on the production of  the $W'$, an anti-diagonal matrix would probably be necessary with a  bound on the
$W'$ mass  of  roughly 4 TeV, somewhat higher than what the 
EW fit gives at the  $1 \sigma$ level. However it would pass the EWP test at roughly the same level 
as the SM.
These additional flavour and collider constraints will clearly help in assessing the range of parameters allowed for the DLRM and the viability of the novel pattern of EW symmetry breaking that this model may embed, being the subject of future work
(a preliminary version of a full EW fit together with CP violation in kaon 
meson-mixing is given at \cite{ValeSilva:2016czu}).
\section*{Acknowledgments}

We are grateful to J\'{e}r\^{o}me Charles, Otto Eberhardt, Janusz Gluza,  Miha Nemev\v{s}ek, Tord Riemann and Jiang-Hao Yu for useful discussions and comments.
This project has received funding from the European Union's Horizon 2020 research and innovation programme under the Marie Sklodowska-Curie grant agreements No 690575 and 674896. LVS was supported in part by the Spanish Government and ERDF funds from the EU Commission [Grant FPA2017-84445-P] and the Generalitat Valenciana [Grant Prometeo/2017/053].

\appendix

\section{Corrections from DLRM to EW precision observables}\label{sec:expsEWPO}

Hereafter, we provide the expressions of the DLRM contributions to the EW precision observables that we have considered:

\begin{equation}
A_\ell = A_\ell^0 +k z_h \varepsilon ^2 \frac{8  s_W^2 \left(2 s_W^2-1\right) \left(1-2 c_R^2
   c_W^2\right)}{ \left(8 s_W^4-4 s_W^2+1\right)^2}
\end{equation}

\begin{equation}
A_c = A_c^0 + k z_h \varepsilon ^2 \frac{48 s_W^2 \left(4 s_W^2-3\right)  \left(1-4 c_R^2
   c_W^2\right)}{ \left(32 s_W^4-24 s_W^2+9\right)^2}
\end{equation}

\begin{equation}
A_b = A_b^0 + k z_h \varepsilon ^2 \frac{24  s_W^2 \left(2 s_W^2-3\right)  \left(-2 c_R^2
   c_W^2-1\right)}{ \left(8 s_W^4-12 s_W^2+9\right)^2}
\end{equation}

\begin{equation}
A_{FB} (\ell) = A_{FB}^0 (\ell) - k z_h \varepsilon ^2 \frac{12 s_W^2 \left(8 s_W^4-6 s_W^2+1\right)  \left(1-2 c_R^2
   c_W^2\right)}{ \left(8 s_W^4-4 s_W^2+1\right)^3}
\end{equation}

\begin{eqnarray}
&& A_{FB} (c) = A_{FB}^0 (c) + k z_h \varepsilon ^2 \frac{18 s_W^2}{ \left(8 s_W^4-4 s_W^2+1\right)^2 \left(32 s_W^4-24
   s_W^2+9\right)^2} \nonumber\\
&& \qquad \times \Big( -2 c_R^2 \left(s_W^2-1\right) \left(1024 s_W^8-1600 s_W^6+992 s_W^4-310 s_W^2+39\right)  \nonumber\\
&& \qquad\quad  +254 s_W^2-16 \left(48 s_W^4-76 s_W^2+49\right) s_W^4-33 \Big)
\end{eqnarray}

\begin{eqnarray}
&& A_{FB} (b) = A_{FB}^0 (b) + k z_h \varepsilon ^2 \frac{72 s_W^2}{
   \left(8 s_W^4-12 s_W^2+9\right)^2 \left(8 s_W^4-4 s_W^2+1\right)^2} \nonumber\\
&& \qquad \times \Big( c_R^2 \left(-91 s_W^2+4 \left(-16 s_W^6+56
   s_W^4-78 s_W^2+57\right) s_W^4+15\right) \nonumber\\
&& \qquad\quad  +25 s_W^2+8 \left(s_W^2-4\right) s_W^4-6 \Big)
\end{eqnarray}

\begin{eqnarray}
	&& \Gamma_Z = \Gamma_Z^0 -\frac{\alpha(M_Z^0) M_Z^0 }{144 k^2 c_W^2  s_W^2}  k z_h \varepsilon ^2 \Bigg( k z_h \left(8 (\xi+19) s_W^4-12 (\xi+9) s_W^2+9
   (\xi+6)\right) \nonumber\\
&& \qquad +4 k^2 \left(-2 s_W^2
   \left((\xi+49) s_R^2-3 (\xi+9)\right)-3 \xi s_R^2\right)\Bigg)
\end{eqnarray}

\begin{equation}
R_\ell = R_\ell^0 + k z_h \varepsilon ^2 \frac{8 \left(c_W^2 s_R^2 \left(8 (\xi+4) s_W^4-4 (4 \xi+13)
   s_W^2+3 (\xi+3)\right)-8 (\xi+4) s_W^6+3 (\xi+3) s_W^2\right)}{3 \left(8
   s_W^4-4 s_W^2+1\right)^2}
\end{equation}

\begin{equation}
R_u = R_u^0 + k z_h \varepsilon ^2 \frac{12 (\xi+2)  \left(s_R^2 \left(16 s_W^6-52 s_W^4+45 s_W^2-9\right)+16
   s_W^6-9 s_W^2\right)}{\left(8 (\xi+10) s_W^4-12 (\xi+6) s_W^2+9
   (\xi+4)\right)^2}
\end{equation}

\begin{equation}
R_d = R_d^0 + k z_h \varepsilon ^2 \frac{24 \xi   \left(s_R^2 \left(-16 s_W^6+52 s_W^4-45 s_W^2+9\right)-16
   s_W^6+9 s_W^2\right)}{\left(8 (\xi+10) s_W^4-12 (\xi+6) s_W^2+9
   (\xi+4)\right)^2}
\end{equation}

\begin{eqnarray}
&& \sigma_{had} = \sigma_{had}^0 + k z_h \varepsilon^2
\frac{36 \pi }{k^2 (M_Z^0)^2 \left(8 (\xi+19) s_W^4-12 (\xi+9) s_W^2+9
   (\xi+6)\right)^2} k^2  \nonumber\\
&& \quad \Bigg( \frac{8 k^2}{ \left(8 (\xi+19) s_W^4-12 (\xi+9) s_W^2+9
   (\xi+6)\right)} \left(-96 s_W^8 \left(\xi^2+(\xi (3 \xi+19)+49)
   s_R^2+7 \xi+12\right) \right. \nonumber\\
&& \quad\;\; +8 s_W^6 \left(6 \xi^2+(\xi (64 \xi+467)+1285) s_R^2+33
   \xi-27\right)+64 (\xi+1) (\xi+4) \left(s_R^2+1\right) s_W^{10} \nonumber\\
&& \quad\;\;\;\; -36 s_W^4 \left(\xi (13 \xi+101) s_R^2-\xi (\xi+9)+251 s_R^2-27\right) \nonumber\\
&& \quad\;\;\;\;\;\; \left. +9 s_W^2 \left((\xi (23 \xi+180)+410) s_R^2-3 (\xi (\xi+8)+18)\right)-27 (\xi
   (\xi+8)+18) s_R^2\right) \nonumber\\
&& \quad\;\;\;\;\;\;\;\; + k z_h  \left(8 s_W^4-4 s_W^2+1\right) \left(8 (\xi+10)
   s_W^4-12 (\xi+6) s_W^2+9 (\xi+4)\right) \Bigg)
\end{eqnarray}

\begin{equation}
\Gamma_W = \Gamma_W^0-  w_h^2\varepsilon ^2 \frac{3 \alpha(M_Z^0)  M_W^0 }{8  s_W^2}
\end{equation}

\begin{eqnarray}
&& Q_W (p) = Q_W^0 (p) +\varepsilon ^2 \frac{1}{k^2} \Big( \left(2 r^2 \left(2 s_W^2 \left(c_R^2 w^2+4\right)+\left(1-2 c_R^2\right) w^2-2\right) \right. \nonumber\\
&& \qquad \left. +w^2 \left(4 c_R^2 \left(s_W^2-1\right) \left(w^2+1\right)+\left(3-4 s_W^2\right) w^2+2\right)\right) \Big)
\end{eqnarray}

\begin{equation}
Q_W (n) = Q_W^0 (n)+ \varepsilon ^2 \frac{(2 +w^2)(2 r^2 +w^2)}{k^2}
\end{equation}

\noindent
where $ \xi = 0.989 $ is a kinematic correction for the channel $ Z \to b \bar{b} $, see e.g. \cite{Rosner:2001zy}.

\section{Scalar sector}\label{app:scala}

\subsection{Scalar mass eigenstates}\label{app:scalarLO}

The neutral scalar sector exhibits two Goldstone bosons, $G_Z^0$ and $G_{Z'}^0$,
\begin{eqnarray}
G_Z^0 &=& \frac{1}{k}(-\phi_{1,i}^0 + r \phi_{2,i}^0 + w \chi_{L,i}^0)
 +\frac{\epsilon}{k}(1+ r^2 - s_R^2 k^2) \chi_{R,i}^0 + \mathcal{O}(\epsilon^2)
\nonumber\\
G_{Z'}^0 & =& \chi_{R,i}^0+ \epsilon[c_R^2(\phi_{1,i}^0 - r \phi_{2,i}^0) +  w s_R^2\chi_{L,i}^0]+\mathcal{O}(\epsilon^2)
\end{eqnarray}
a light Higgs boson of mass $M_h \sim \mathcal{O}(\epsilon^2)$, 3 CP-even heavy scalar bosons of masses $M_i$ at LO 
defined in Eq.~\eqref{eq:dech}
and two CP-odd heavy scalar bosons 
\begin{equation}
A_i^0=\frac{1}{u_i}\biggl((  r + t_i  w) \phi_{1,i}^0+  \phi_{2,i}^0+t_i  \chi_{0,i}^L \biggr)
+{\cal O}(\epsilon^2)\qquad M_{A_i^0}^2=M_i^2 +{\cal O}(\epsilon^2)
\end{equation}
with at LO in $\epsilon$
\begin{eqnarray}
M_1^2+M_2^2 &=&
-\frac{v_R}{2 \left(r^2-1\right) w} \left(\left(r^2+1\right) v_R w \alpha_{34}+\sqrt{2} \left(\mu'_1 r^3+\mu'_2 r^2+\mu'_1 r \left(2 w^2-1\right)-\mu'_2 \left(2 w^2+1\right)\right)\right) \nonumber\\
(M_1^2-M_2^2)^2 &=& (M_1^2+M_2^2)^2
  -\frac{v_R^2 k^2}{\left(r^2-1\right) w} \left(\sqrt{2} (r \mu'_1 +\mu'_2)  v_R \alpha_{34}+2 w \left((\mu'_1)^2-(\mu'_2)^2\right)\right)
\nonumber\\
M_3^2&=& 2 \rho_1 v_R^2 \label{eq:M3rho1}
\end{eqnarray}
and $\alpha_{34}$ is defined in Eq.~\eqref{eq:param}. The following equalities hold:
\begin{eqnarray}
b\, q \,(w - r p ) &=&\frac{M_2^2}{M_1^2} d\, p\, (w -r q)\nonumber \\
 &=&2 M_2^2 w \left(\sqrt{2} (M_1^2 + M_2^2) r w+v_R \left(\mu'_2 r^3+\mu'_1 r^2+\mu'_2 r \left(w^2+2\right)+\mu'_1 w^2\right)\right)\nonumber \\
b &=&  v_R w \left(2 M_2^2 r w (\mu'_1-\mu'_2 r)+2 M_1^2 w (\mu'_2+\mu'_1 r)+\sqrt{2} v_R \left((r\mu'_1+\mu'_2)^2+w^2 \left((\mu'_1)^2+(\mu'_2)^2\right)\right)\right)\nonumber\\
d  &=& b(M_1^2\leftrightarrow M_2^2)
\end{eqnarray}
We have also the relation (which can be explicitly checked, but comes from the orthogonality of the various eigenvectors)
\begin{equation}
r^2+1+rw(p+q)+pq(w^2+1)=0
\end{equation}

In the charged sector, one has also 2 Goldstone bosons
\begin{eqnarray}
G_W^\pm &=& \frac{1}{k}\Bigg[
(-\phi_{1,i}^\pm + r \phi_{2,i}^\pm + w \chi_{L,i}^\pm)
+ 2r\epsilon\chi_{R,i}^0\nonumber\\
&&\qquad\qquad +\frac{2 r\epsilon^2}{k^2} (r^3 \phi_1^\pm - (1 + w^2) \phi_2^\pm + 
   r w (w \phi_1^\pm + \chi_L^\pm)
\Bigg]+\mathcal{O}(\epsilon^3)
\nonumber\\
G_{W'}^\pm &=& \chi_{R,i}^\pm +\epsilon (r \phi_{1,i}^\pm - \phi_{2,i}^\pm)+\mathcal{O}(\epsilon^2)
\end{eqnarray}
Note that there is a mass degeneracy among the charged and neutral scalars at LO in $\epsilon$.

The limit of small $w$ brings significant simplifications in the expressions, which we provide for illustration. We have\footnote{We have minimised the potential with respect to $\mu_1^2,\mu_2^2, \mu_3^2,\rho_3$. In the limit where $w$ is small, it proves however more natural to express the masses in terms of $\rho_3-2 \rho_1=-\sqrt{2}(r\mu'_1+\mu'_2)/(v_R w)$.}
\begin{eqnarray}
M_2^2&=&\frac{1}{2}\frac{1+r^2}{1-r^2} v_R^2 \alpha_{34} \qquad\qquad
M_1^2=\frac{1}{2} v_R^2 (\rho_3-2 \rho_1)
\nonumber \\
b&=&{\cal O}( w^2) \qquad
d=\sqrt{2} \mu'_2 \left(r^2+1\right) v_R^2 w (r\mu'_1+\mu'_2)
\end{eqnarray}

\subsection{Some useful relations}\label{app:scalarmasseigenstates}

The spin-0 states are linear combination of the various scalar field of the theory, given in Eq.~(\ref{eq:dech}).
Their coefficients obey useful relations:
\begin{eqnarray}
 d_{\chi_{R,r}}^{H_i}&=&\frac{k}{\sqrt 2} F_i 
\nonumber \\
-\frac{1}{u_i} d_{\chi_{R,r}}^{H_i}&=&(r + t_i w)d_{\phi_1}^{H_i}+ d_{\phi_2}^{H_i}+ t_i d_{\chi_{L,r}}^{H_i}
\end{eqnarray}
\begin{eqnarray}
c_{\chi_{R,r}}^{H_i}&=&-w S_i
+\frac{1}{2  (M_{H_i}^2-M_{H_3}^2)}\biggl(2 M_{H_2}^2 w S_i +(2 \frac{r}{u_i} +w S_i)v_R^2\alpha_{34}-(\alpha S_i +\beta \frac{w}{u_i})v_R \mu'_2\biggr)
\nonumber\\
- k c_{\chi_{R,r}}^{h^0}&=&c^{H_3}=\frac{1}{2  \rho_1}\left(  (1 + r^2)\alpha_1+ 2 r \alpha_2 + r^2 \alpha_3 +\alpha_4
+ 2 w^2 \rho_1\right)
\label{eq:coefb}
\end{eqnarray}
and
\begin{equation}
c^{{\cal H}_i} \equiv \pm c_{\phi_1}^{{\cal H}_i}+r c_{\phi_2}^{{\cal H}_i} + w c_{\chi_{L,r}}^{{\cal H}_i} \, , \quad i=1,..,3 
\end{equation}
with ${\cal H}$ stands for either a CP odd or a CP even neutral scalar boson and  the $+/-$ corresponds to the CP even/odd ones respectively.
One finds
\begin{equation}
c^{H_i}=-S_i w^3+\frac{v_R}{\sqrt 2 \delta^2 M_{H_1}^2}(\alpha S_i+ \beta \frac{w}{u_i}) w^2 \mu_2'
\end{equation}
and
\begin{equation}
\alpha = 1 +w^2 + r x \, , \quad \quad \beta= r-x
\end{equation}
A similar definition holds for $c^{h^0}$ and the charged scalar boson $d_{H_i^\pm}$ with again a nice 
relation between $c^{h^0} $ and $c^{H_3}$ necessary for unitarity to 
be fulfilled. 
\begin{equation}
(c^{H_3})^2= - 2 k c^{h^0}
\label{eq:cH3h0} 
\end{equation} 
Two additional relations between the $c_{\phi_1,\phi_2,\chi_{L,r}}^{H_i}$
which turn out to be useful for the calculation of the $W'$ self-energy are
\begin{eqnarray}
r  c_{\phi_1}^{H_i}- c_{\phi_2}^{H_i}&=&\frac{t_i^2 w (1+r^2+ t_i r w) (1+\delta^2)}{2 u_i^3(-1+\delta^2)}\biggl(-w \frac{3 k^2 -w^2}{1+r^2}\biggr.
\nonumber\\
&&\biggl. +\sqrt 2  \frac{\mu_2' v_R}{M_{H_1}^2(1+\delta^2)}\bigl( (r- x) (2- \frac{u_i}{1+r^2+ t_i r w})- w \frac{k^2}{1+r^2}(1 -\frac{2 k^2}{1-r^2})\bigr.\biggr.
\nonumber \\
&&\biggl. \bigl.
+(1+w^2+r x)(1+\frac{w^2}{1+r^2}(1-\frac{2 k^2}{1-r^2}))\bigr) \biggr)+\cdots
\nonumber\\
c^{H_i}|_{i=1,2}&=&-S_i w^3 +\frac{w^2}{\sqrt 2}\frac{\mu_2' v_R}{ M_{H_1}^2 \delta^2}\biggl((1+w^2+r x) S_i + (r-x) \frac{w}{u_i}\biggr)+\cdots 
\end{eqnarray}
where the dots refer to a lengthy expression which involves 
terms of a similar type

\section{Feynman rules}\label{app:feynmanrules}

\subsection{Couplings to fermions}\label{app:Hfermions}

The couplings are given in terms of the functions $F_i$  defined in Eqs.~(\ref{eq:F1}), as can be seen in Tabs.~\ref{tab:AZZpr-fermions}-\ref{tab:feynHHW}.
 It is easy to show that $F_1=0$ and $F_2=(-1+r^2)/(1+r^2)$
in the limiting case $w=0$, for $ \delta={\cal O}(1/\sqrt{w}) $, thus recovering the formulae given in \cite{Basecq:1985cr} in the case
$g_R=g_L$ and $r=0$.
These two quantities have the very important property that
\begin{equation}
\sum_{i=1,2} F_i(r,w,p) G_i(r,w,p)=1/k
\label{eq:propgauge}
\end{equation}
where the $G_i$ are similar functions appearing for similar couplings
\begin{equation}
G_i(r,w,p)=\frac{1}{(-1 + r^2)}\frac{1+r^2+ r w t_i}{\sqrt{1 + t_i^2 +(r + w t_i)^2}}  
\end{equation}
This property is required for example when showing the gauge invariance of the computation of meson mixing within the DLRM, however this goes beyond the scope of our article. In the limiting
case $w=0$, $G_1=\sqrt{1+r^2}/(-1+r^2)$ and $G_2=0$ so that $k F_1 G_1 =1$ independently
of the value  of $r$. Interestingly, the quantity
\begin{equation}
{\cal F}^r = -2 k^2 F_1 G_1 F_2 G_2
\end{equation}
is  proportional to $r w^3$ in the limit $ w \to 0$. 
One has in terms of the parameters of the model:
\begin{eqnarray}
k F_1 G_1  &=&\frac{1}{ 2  (1 - r^2)(1 +\beta(x) w^2) (1 - \delta^2)} 
 \biggl( \bigl(-k^2  + (k^2 -2(1 +  \nu(x) )) X \bigr) (1+ \delta^2) \biggr.
\nonumber \\
&&\biggl.+2 \bigl(1 + \nu(x) + \bigl(r^2 -\beta(x) w^2 (1 - r^2)  +  \nu(x)\bigr) \delta^2 \bigr) \biggr)
\end{eqnarray}
with
\begin{equation}
\nu(x) = w^2/(1+ r x )
\end{equation}
$F_2 G_2$ is obtained from this equation changing $\delta \to 1/\delta$.

Most of the couplings of the gauge bosons with the heavy scalars are proportional to $w$ so that they vanish in the limit  $w=0$ which is the case of the triplet left right models extensively used in the literature.

\subsection{Couplings to gauge bosons}\label{app:Hgauge}

The couplings to gauge bosons involve
\begin{equation}
u_i=\sqrt{1 + t_i^2 +(r + w t_i)^2}
\end{equation}
and $ w_h$ and $z_h$ are the quantities which appear in the masses of the light gauge bosons
at order $\epsilon^4$, see Eqs.~(\ref{eq:MgaugelightNLO}).
The various coefficients in the coupling of the
gauge bosons to two scalars, Table \ref{tab:feynHHW} and \ref{tab:feynHHW2} are defined as follows:
\begin{eqnarray}
{\tilde w}_h^2 &=&w_h^2 -\frac{c_{h^0}}{k} \quad \quad \quad {\tilde z}_h^2 =z_h^2 -\frac{c_{h^0}}{k} 
\nonumber\\
w_{{\cal H}_i}&=&\frac{8 r(r+t_i w)}{ u_i^2}\pm 2 b_{{\cal H}_i} \, \quad \quad  z_{{\cal H}_i}= 2 k z_h ( S_i^2-  c_R^2) \quad \quad \tilde{z}_{{\cal H}_i}= z_{{\cal H}_i}+ 2 b_{{\cal H}_i}
\nonumber\\
\quad c_{H_i^\pm}&=&\frac{1}{c_W^2-s_W^2} \bigl( z_{{\cal H}_i}+\frac{4 c_W^2 -3}{c_W^2-s_W^2} F_i^2 k^2 \bigr) 
\nonumber \\
b_{{\cal H}_i} &\equiv& \frac{1}{u_i}\bigl( \mp c_{\phi_1}^{{\cal H}_i}(r+ w t_i)
+ c_{\phi_2}^{{\cal H}_i}+c_{\chi_L}^{{\cal H}_i} t_i\bigr)
\nonumber \\
d_{H_3} &\equiv& (c_{\phi_1}^{H_3^0})^2
+( c_{\phi_2}^{H_3^0})^2+(c_{\chi_L}^{H_3^0})^2 
\label{eq:coeffGBH}
\end{eqnarray}
where ${\cal H}_i$ stands for any scalar boson $H_i$, $A_i$, $H_3^0$ and $h^0$ and the upper/lower signs correspond to the CP even/odd scalar.

Also the following relation holds
\begin{equation}
\sum_i\frac{(r+w t_i)}{1 +
t_i^2 +(r + w t_i)^2}=\frac{r^2}{k^2} 
\end{equation}

\section{More on renormalisation}\label{sec:append}

\vspace{0.25cm}

$\bullet$ {\bf Renormalisation of the charge}
 
\vspace{0.3truecm}
The bare charge reads 
\begin{equation}
\frac{1}{e^2}= \frac{1}{g_X^2}+\frac{1}{g_L^2}+\frac{1}{g_R^2}
\end{equation}
leading to 
\begin{equation}
\frac{\delta e}{e}=\delta Z_1^\gamma -\frac{3}{2} \delta Z_2^\gamma
\end{equation}

$\delta e/e$ can be re-expressed using three different relations. First the renormalisations of
 field and coupling constant for the $U(1)$ part are related
\begin{equation}
\delta Z_1^B =\delta Z_2^B
\end{equation}
due to the $U(1)_{B-L}$ Ward identity.\footnote{This relation is formally identical to
the QED one \cite{Hollik:1993cg}.} Second, the expressions for $\delta Z_i^{\gamma Z (Z')}$ are
given by the coefficient of the $\partial_\mu A^\nu \partial^\mu Z_\nu (Z'_\nu)$ term of the
renormalised Lagrangian 
\begin{eqnarray}
\delta Z_i^{\gamma Z}&=&\frac{c_W s_W}{c_W^2 -s_W^2}\bigl(\delta Z_i^Z -\delta Z_i^\gamma \bigr)+\epsilon^2  k z_h \frac{s_R^2 c_R^2 c_W}{s_W (c_W^2-s_W^2)^2(c_R^2-s_R^2)} \bigl(s_W^2\delta Z_i^Z -c_W^2 \delta Z_i^\gamma+ \delta Z_i^{Z'} \bigr) 
\nonumber\\
\delta Z_i^{\gamma Z'}&=&\frac{c_R s_R c_W}{(c_R^2-s_R^2)(c_W^2 -s_W^2)}
( (c_W^2 -s_W^2)\delta Z_i^{Z'}- c_W^2\delta Z_i^\gamma) 
-\frac{s_R s_W^2}{c_W^2 -s_W^2}
\delta Z_i^{Z}
\nonumber\\
&&+\epsilon^2  k z_h \frac{c_R c_W s_R}{(c_R^2-s_R^2)^2(c_W^2 -s_W^2)^2} \biggl(
(1+2c_W^2(-1+c_R^2(-1+3c_W^2)s_R^2))\delta Z_i^Z \biggr.
\nonumber\\
&&\biggl.+(-1+2c_W^2+2 c_R^2(2-6c_W^2+3 c_W^4)s_R^2) \delta Z_i^\gamma
\biggr.
\nonumber\\
&&\biggl.-2 s_R^2 c_R^2(-1+2 c_W^2)(-2+3 c_W^2)\delta Z_i^{Z'} \biggr) 
\end{eqnarray}
Finally, one can use the on-shell relations for $\hat \Sigma^{\gamma Z}(0)$
and  $\hat \Sigma^{\gamma Z'}(0)$ in Eq.~(\ref{eq:onsg2}). Combining these three relations yields:
\begin{equation}
\frac{\delta e}{e}=\frac{1}{2} \Pi^{\gamma}(0) + \frac{s_W}{c_W} \frac{\Sigma^{\gamma Z}(0)}{M_{Z}^2}\left(1+ \frac{s_R^2 k z_h}{ s_W^2}\epsilon^2\right) 
 + \frac{s_R}{c_R c_W} \frac{\Sigma^{\gamma Z'}(0)}{M_{Z'}^2}\left(1- \frac{s_W^2  z_h}{k s_R^2}(c_R^2 k z_h +s_R^2 w^2)\epsilon^2\right)
\end{equation}
with $ \Pi^{\gamma}$ defined in Eq.~\eqref{eq:ZfacW}.\footnote{This equation agrees with the one obtained in Ref.~\cite{Czakon:2002wm}, which was 
derived in the TLRM in the special case $w=0$ and $g_L=g_R$ when using the relation between $\Sigma^{\gamma Z^{(')}}$ and $\delta Z_{\gamma Z^{(')}}$. The agreement is obtained modulo a factor four which comes from the masses of the gauge bosons which differ by such a factor in the triplet and doublet models, i.e., $ M_{Z'}^2 = (g_R^2 + g_X^2) v_R^2 / 4 $ in the doublet case ($ M_{Z'}^2 = (g_R^2 + g_X^2) v_R^2 $ in the triplet case), where $ v_R / \sqrt{2} $ is the vacuum expectation value of the $ \chi_R $ doublet (respect., triplet).}

\vspace{0.7truecm}
$\bullet$ {\bf Renormalisation of $s_W$} 

\vspace{0.3truecm}
It is given by
\begin{equation}
\frac{\delta s_W^2}{s_W^2}=\delta Z_2^{W} + \Pi^\gamma(0)-2\frac{c_W}{s_W} \frac{\Sigma^{\gamma Z}(0)}{M_Z^2} +2 \frac{c_W z_h}{k c_R  s_R}(c_R^2 k z_h+ s_R^2 w^2) \frac{\Sigma^{\gamma Z'}(0)}{M_{Z'}^2} 
\end{equation}

\vspace{0.4truecm}
$\bullet$ {\bf Expressions at leading order in $\epsilon$  for the heavy particles}

\vspace{0.2truecm}
These are:
\begin{eqnarray}
\delta Z_2^{W'}&=&-\Pi^\gamma(0)+\delta Z_{Z' W' \gamma}-2 \frac{s_W^2}{c_W^2}\delta Z_{Z W \gamma}
  +{\cal O}(\epsilon^2) 
\nonumber\\
\delta Z_1^{W'}&=&-\Pi^\gamma(0) +  \frac{3 c_R^2+ s_R s_W}{ c_W c_R s_R} \frac{\Sigma^{\gamma Z'}(0)}{M_{Z'}^2} 
+ \frac{c_R^2}{s_R^2}\biggl(\frac{\delta M_{Z'}^2}{M_{Z'}^2}-\frac{\delta M_{W'}^2}{M_{W'}^2}\biggr)
\nonumber\\
&&-3 \frac{ s_W}{c_W} \frac{\Sigma^{\gamma Z}(0)}{M_Z^2}- \biggl(\frac{\delta M_Z^2}{M_Z^2}-\frac{\delta M_W^2}{M_W^2}\biggr) 
+{\cal O}(\epsilon^2)
\label{eq:renormh}
\end{eqnarray}
It is easy to see that at leading order in $\epsilon$ in the limit $s_W=0$ where Left and Right sectors decouple one recovers SM-like
expressions with $s_W \to s_R$ and $W \to W' \, , \, Z \to Z'$.
Indeed  due to the custodial symmetry the difference  $\delta M_Z^2/M_Z^2-\delta M_W^2/M_W^2$ is proportional to  $s_W$ and thus cancels in this limit.

\input{tables_couplings.tex}

\clearpage

\begin{sidewaystable}
\begin{center}
\begin{tabular}{|c|c|c|c|c|c|c|c|c|c|c|c|c|}
\hline
Obs. & $ c_{0} $ & $ c_{1} $ & $ c_{2} $ & $ c_{3} $ & $ c_{4} $ & $ c_{5} $ & $ c_{6} $ & $ c_{7} $ & $ c_{8}$& $ c_{9} $ & $ c_{10} $ & $ c_{11} $   \\
\hline
\hline
$ \Gamma_{Z} $ [MeV] & 2495.16 & -2.54 & 20.12 & 63.41 & -19.73 & -3.49 & -54.57 & 9217.30 & -0.33 & -14.14 & -3.53 & -93.70 \\
\hline
\hline
$ \sigma_{had} $ [pb] & 41478.8 & 1.81 & 49.24 & -630.62 & 27.07 & 2.02 & 86.56 & -85804.67 & -1.99 & 4.88 & 0.81 & 223.91 \\
\hline
\hline
$ R_{b} $ * & 215.833 & 0.03 & -3.07 & -0.04 & -0.67 & -0.07 & 0.76 & -20.44 & 0.02 & 0.06 & 0.20 & 0.47 \\
\hline
$ R_{c} $ * & 172.23 & -0.03 & 1.01 & 2.33 & 1.25 & 0.38 & -1.20 & 36.86 & -0.01 & -0.09 & -0.18 & -0.57 \\
\hline
$ R_{e} $ * & 20739.2 & -8.90 & -28.99 & 788.79 & -52.28 & -8.88 & -364.33 & 11635.59 & 0.80 & -25.15 & -5.80 & -413.52 \\
\hline
$ R_{\mu} $ * & 20739.4 & -8.90 & -28.99 & 788.79 & -52.27 & -8.88 & -364.34 & 11635.29 & 0.80 & -25.15 & -5.80 & -413.54 \\
\hline
$ R_{\tau} $ * & 20786.3 & -8.91 & -29.12 & 790.58 & -52.33 & -8.89 & -364.59 & 11549.52 & 0.80 & -25.16 & -5.80 & -414.50 \\
\hline
\hline
$ A_{FB} (b) $ * & 102.81 & -2.84 & 15.14 & -2.33 & -1.33 & -1.62 & -115.77 & 3571.84 & -0.22 & -1.12 & -1.14 & -22.73 \\
\hline
$ A_{FB} (c) $ * & 73.45 & -1.75 & 9.49 & -1.49 & -0.27 & -1.56 & -91.45 & 2800.34 & -1.29 & -0.13 & 1.23 & -10.99 \\
\hline
$ A_{FB} (\ell) $ * & 16.13 & -0.11 & 2.04 & -0.31 & -1.46 & -0.12 & -34.68 & 980.77 & 0.36 & -1.16 & -0.21 & 40.07 \\
\hline
\hline
$ \mathcal{A}_{b} $ * & 934.637 & -0.32 & 0.55 & -0.18 & -1.21 & -0.09 & -13.39 & 423.59 & 1.2 $ \times 10^{-3} $ & -0.95 & -0.35 & -11.17 \\
\hline
$ \mathcal{A}_{c} $ * & 667.717 & -1.75 & 9.42 & -1.45 & -6.96 & -1.02 & -71.76 & 2264.66 & -0.15 & -5.53 & -1.84 & -48.56 \\
\hline
$ \mathcal{A}_{\ell} $ * & 146.673 & -4.00 & 21.44 & -3.30 & -17.73 & -2.41 & -162.98 & 5149.15 & -0.36 & -13.90 & -4.36 & -141.83 \\
\hline
\hline
$ M_{W} $ [GeV] & 80.3644 & -0.06 & 0.53 & -0.08 & -0.10 & -0.05 & -1.09 & 114.74 & -0.01 & -0.09 & -0.07 & -0.79 \\
\hline
$ \Gamma_{W} $ [MeV] & 2090.86 & -4.66 & 41.95 & 47.91 & -19.03 & -4.11 & -84.25 & 8942.90 & -0.51 & -14.92 & -8.32 & -106.34 \\
\hline
\hline
$ Q_{W} (Cs) $ & -72.9586 & -0.10 & 0.10 & -0.23 & -0.92 & -0.15 & -5.22 & 166.44 & -0.01 & -0.70 & -0.25 & -4.88 \\
\hline
$ Q_{W} (Tl) $ & -115.469 & -0.15 & 0.10 & -0.35 & -1.38 & -0.24 & -7.74 & 246.99 & -0.02 & -1.04 & -0.38 & -7.28 \\
\hline
$ Q_{W} (p) $ * & 73.0053 & -2.10 & 11.34 & -1.63 & -12.37 & -1.18 & -85.55 & 2700.33 & -0.17 & -9.68 & -2.87 & -73.06 \\
\hline
\end{tabular}
\end{center}
\caption{Values of the first 12 coefficients  in the parametrization of various EWPO, see Eq.~\eqref{eq:paramZfitter}. Above, $ * = 10^{3} $.}
\label{tab:coef1}
\end{sidewaystable}

\clearpage

\clearpage

\begin{sidewaystable}
\begin{center}
\begin{tabular}{|c|c|c|c|c|c|c|c|c|c|c|c|c|}
\hline
Obs. & $ c_{12} $ & $ c_{13} $ & $ c_{14} $ & $ c_{15} $ & $ c_{16} $ & $ c_{17} $ & $ c_{18} $ & $ c_{19} $ & $ c_{20}$& $ c_{21} $ & $ c_{22} $ & $ c_{23} $   \\
\hline
\hline
$ \Gamma_{Z} $ [MeV] & 11510.0 & 133.8 & 16.6 & 3377.0 & 362.5 & 11700.0 & -2377.0 & -1102.0 & 3304.0 & -168.8 & -3672.0 & 166900.0 \\
\hline
\hline
$ \sigma_{had} $ [pb] & 341600.0 & 548.4 & 1642.0 & -1344.0 & 53970.0 & 214100.0 & -1.8 $\times 10^6$ & -184600.0 & -2.1 $\times 10^6$ & -13570.0 & -982300.0 & 445000.0 \\
\hline
\hline
$ R_{b} $ * & -73.3 & -1.4 & -0.2 & -14.1 & -14.5 & -81.5 & 712.3 & 99.5 & 987.5 & 5.6 & 516.9 & -41.7 \\
\hline
$ R_{c} $ * & 101.4 & -3.2 & 0.9 & 19.6 & 20.5 & 120.7 & -1054.0 & -147.6 & -1478.0 & -8.0 & -776.1 & 980.6 \\
\hline
$ R_{e} $ * & -151300.0 & 587.3 & -1251.0 & 5756.0 & -39890.0 & -146400.0 & 720200.0 & 112200.0 & 931400.0 & 9315.0 & 506900.0 & -79820.0 \\
\hline
$ R_{\mu} $ * & -151300.0 & 587.3 & -1251.0 & 5756.0 & -39890.0 & -146400.0 & 720200.0 & 112200.0 & 931400.0 & 9315.0 & 506900.0 & -79930.0 \\
\hline
$ R_{\tau} $ * & -151800.0 & 584.7 & -1256.0 & 5760.0 & -40050.0 & -147000.0 & 724700.0 & 112800.0 & 937100.0 & 9355.0 & 509600.0 & -80730.0 \\
\hline
\hline
$ A_{FB} (b) $ * & 3596.6 & 3.3 & -25.6 & 1800.0 & 371.3 & 5303.9 & 12443.0 & 1596.6 & 5560.0 & 435.5 & 2023.4 & -354.0 \\
\hline
$ A_{FB} (c) $ * & 6586.8 & -18.2 & 127.5 & 1407.8 & 238.4 & 5979.6 & -48140.7 & -4440.0 & -50758.5 & -1595.7 & -25437.1 & 113211.1 \\
\hline
$ A_{FB} (\ell) $ * & 14385.0 & -15.1 & 98.8 & 495.0 & 2504.5 & 12994.3 & -16064.1 & -33899.1 & -140962.4 & -1469.0 & -135554.3 & -59309.6 \\
\hline
\hline
$ \mathcal{A}_{b} $ * & -1993.0 & 1.5 & -7.3 & 212.1 & -557.3 & -1781.0 & 8844.0 & 1418.0 & 11250.0 & 114.5 & 6110.0 & 1780.0 \\
\hline
$ \mathcal{A}_{c} $ * & -10690.0 & 5.9 & -42.5 & 1141.0 & -3018.0 & -9620.0 & 40190.0 & 6491.0 & 50090.0 & 589.0 & 27720.0 & 106100.0 \\
\hline
$ \mathcal{A}_{\ell} $ * & -25700.0 & 4.6 & -36.7 & 2594.0 & -7211.0 & -23250.0 & -16150.0 & 5536.0 & -16030.0 & 1150.0 & 6526.0 & 1.6 $\times 10^6$ \\
\hline
\hline
$ M_{W} $ [GeV] & -66.6 & -0.049 & 0.2 & 17.7 & -30.0 & -65.9 & 239.9 & 58.6 & 328.3 & 4.2 & 200.0 & 322.5 \\
\hline
$ \Gamma_{W} $ [MeV] & 7131.0 & 99.0 & 25.5 & 3465.0 & -761.3 & 7405.0 & 9060.0 & 1443.0 & 18570.0 & 89.7 & 6349.0 & 185300.0 \\
\hline
\hline
$ Q_{W} (Cs) $ & -392.3 & -0.4 & -1.3 & 83.4 & -121.3 & -306.9 & 4069.0 & 619.9 & 5892.0 & 23.5 & 3185.0 & -64540.0 \\
\hline
$ Q_{W} (Tl) $ & -579.7 & -0.7 & -1.8 & 123.7 & -179.3 & -452.8 & 6120.0 & 931.5 & 8875.0 & 34.9 & 4797.0 & -97900.0 \\
\hline
$ Q_{W} (p) $ * & -6804.0 & 1.2 & -27.7 & 1363.0 & -2072.0 & -5419.0 & 51620.0 & 8026.0 & 72530.0 & 386.0 & 39410.0 & -676800.0 \\
\hline
\end{tabular}
\end{center}
\caption{Same as in Table \ref{tab:coef1} but for the next 12 coefficients.}
\label{tab:coef2}
\end{sidewaystable}

\clearpage
\clearpage

\begin{sidewaystable}
\begin{center}
\begin{tabular}{|c|c|c|c|c|c|c|c|c|c|c|c|c|}
\hline
Obs. & $ c_{24} $ & $ c_{25} $ & $ c_{26} $ & $ c_{27} $ & $ c_{28} $ & $ c_{29} $ & $ c_{30} $ & $ c_{31} $ & $ c_{32}$& $ c_{33} $ & $ c_{34} $ & max. \\
&&&&&&&&&&& & dev. $[\%]$ \\
\hline
\hline
$ \Gamma_{Z} $ [MeV] & 661800.0 & 1.9 $\times 10^6$ & -57.1 & 132.4 & 65.2 & 2.0 $\times 10^6$ & -42.4 & -186.2 & 592900.0 & 2.8 $\times 10^6$ & -1.0 $\times 10^6$ & 0.005 \\
\hline
\hline
$ \sigma_{had} $ [pb] & 3.0 $\times 10^6$ & 8.4 $\times 10^6$ & -6709.0 & 44420.0 & 25210.0 & 8.1 $\times 10^6$ & -2578.0 & 118100.0 & -32260.0 & 1.3 $\times 10^7$ & -8.6 $\times 10^6$ & 0.007 \\
\hline
\hline
$ R_{b} $ * & -2019.0 & -5477.0 & 2.7 & -44.6 & -16.5 & -4874.0 & 2.0 & -62.8 & -1549.0 & -6782.0 & 8356.0 & 0.002 \\
\hline
$ R_{c} $ * & 3014.0 & 8255.0 & -4.0 & 66.2 & 25.0 & 6374.0 & -2.9 & 93.5 & 1612.0 & 10250.0 & -10310.0 & 0.002 \\
\hline
$ R_{e} $ * & -1.1 $\times 10^6$ & -2.4 $\times 10^6$ & 4678.0 & -40130.0 & -15960.0 & -2.1 $\times 10^6$ & 1964.0 & -58390.0 & 687300.0 & -3.4 $\times 10^6$ & 1.6 $\times 10^6$ & 0.01 \\
\hline
$ R_{\mu} $ * & -1.1 $\times 10^6$ & -2.4 $\times 10^6$ & 4679.0 & -40130.0 & -15970.0 & -2.1 $\times 10^6$ & 1965.0 & -58390.0 & 687300.0 & -3.4 $\times 10^6$ & 1.6 $\times 10^6$ & 0.01 \\
\hline
$ R_{\tau} $ * & -1.1 $\times 10^6$ & -2.4 $\times 10^6$ & 4698.0 & -40360.0 & -16040.0 & -2.1 $\times 10^6$ & 1975.0 & -58750.0 & 688400.0 & -3.4 $\times 10^6$ & 1.7 $\times 10^6$ & 0.01 \\
\hline
\hline
$ A_{FB} (b) $ * & 3568.8 & 2181.3 & 219.6 & -3.9 & -164.8 & 28008.9 & -79.8 & 12.0 & 7046.5 & 6590.4 & -14429.2 & 0.066 \\
\hline
$ A_{FB} (c) $ * & 320931.0 & 422936.3 & -752.8 & -1358.9 & 250.6 & 544022.1 & -5193.3 & 3974.9 & 88399.4 & 705771.5 & 191032.3 & 11.0  \\
\hline
$ A_{FB} (\ell) $ * & -208918.8 & 371316.2 & -645.1 & 1609.5 & 573.0 & 2.1 $\times 10^6$ & 2840.4 & 1203.5 & 20210.1 & 2.2 $\times 10^6$ & -103658.3 & 11.5 \\
\hline
\hline
$ \mathcal{A}_{b} $ * & -19660.0 & -44070.0 & 57.5 & -491.8 & -196.5 & -33490.0 & 25.0 & -713.0 & 24000.0 & -65120.0 & 63380.0 & 0.002 \\
\hline
$ \mathcal{A}_{c} $ * & 474200.0 & 1.1 $\times 10^6$ & 295.5 & 17.2 & -40.6 & 593100.0 & -99.3 & -785.6 & 150800.0 & 1.3 $\times 10^6$ & -576900.0 & 0.007 \\
\hline
$ \mathcal{A}_{\ell} $ * & 1.4 $\times 10^6$ & -6.4 $\times 10^6$ & 579.6 & 3559.0 & 336.0 & 2.4 $\times 10^7$ & 6449.0 & -3344.0 & 1.1 $\times 10^6$ & 5.8 $\times 10^6$ & -1.3 $\times 10^6$ & 0.02 \\
\hline
\hline
$ M_{W} $ [GeV] & -553.8 & -900.2 & 2.7 & -15.1 & -6.4 & -465.3 & -0.7 & -18.5 & 1841.0 & -1650.0 & -926.5 & 0.0008 \\
\hline
$ \Gamma_{W} $ [MeV] & 666200.0 & 1.6 $\times 10^6$ & 87.2 & 205.6 & -78.3 & 1.7 $\times 10^6$ & 54.6 & -83.2 & 692100.0 & 2.2 $\times 10^6$ & -1.3 $\times 10^6$ & 0.007 \\
\hline
\hline
$ Q_{W} (Cs) $ & -558000.0 & -1.7 $\times 10^6$ & 11.9 & 101.5 & 17.4 & -1.0 $\times 10^6$ & 25.7 & 38.7 & 31300.0 & -2.1 $\times 10^6$ & -103300.0 & 0.005 \\
\hline
$ Q_{W} (Tl) $ & -845800.0 & -2.5 $\times 10^6$ & 17.7 & 156.2 & 27.2 & -1.5 $\times 10^6$ & 39.0 & 61.9 & 47170.0 & -3.2 $\times 10^6$ & -156200.0 & 0.005 \\
\hline
$ Q_{W} (p) $ * & -5.9 $\times 10^6$ & -1.8 $\times 10^7$ & 194.2 & 681.9 & 53.9 & -1.1 $\times 10^7$ & 253.7 & -145.2 & 382200.0 & -2.3 $\times 10^7$ & -1.2 $\times 10^6$ & 0.04 \\
\hline
\end{tabular}
\end{center}
\caption{Same as in Table \ref{tab:coef1} but for the last 11 coefficients. The maximum deviation is given in the last column.}
\label{tab:coef3}
\end{sidewaystable}

\clearpage

\input{biblio_6.tex}

\end{document}

%% file: new_table_v3.tex
We have thus studied  how the quality of the fit would deteriorate if we would increase the lower bound on  $M_{W'}$.  The result is  shown on
Table \ref{tab:MW'} where the minimum of the $\chi^2$  gets increasingly close
to the SM limit (equivalent to $ \epsilon = 0 $). 
As an example, we show in Table~\ref{tab:fit_w_MWR_bound} the parameters of the DLRM when the mass of the $ W' $ is constrained to be larger than  
$ 10 $~TeV.
As expected from Table~\ref{tab:MW'}, this reduces the quality of the fit, increasing the $ \chi^2_{min} $ to $ 22.0 $. Moreover, since $ M_{W'} $ helps probing the scale of spontaneous breaking of the LR symmetry, Eq.~\eqref{eq:masswpzpLO}, the ratio $ \epsilon $ is much smaller than in the case of Table~\ref{tab:fitlrm}, its constraint at $ 1\sigma $ being now $ \epsilon < 0.02 $. Since the NP contributions scale with the size of $ \epsilon $, the extracted central values and ranges of the various quantities get in general closer to the results extracted from the SM fit of Table~\ref{tab:fitsm}, and in particular the values of the observables at LO ($ M_Z^{(0)} $, etc.) are  closer to their experimental values  (respectively, $ M_Z $, etc.). The constraints on the spontaneous LR and EW breaking parameters $ c_R, r, w $ are still similar to those obtained with
$ M_{W'} $ left free, which is consistent with Fig.~\ref{fig:param2D} for small values of $ \epsilon $, while many additional parameters of the DLRM remain unconstrained. Also note that, following Eq.~\eqref{eq:masswpzpLO}, the mass of the $ Z' $ gauge boson is necessarily  larger or equal to  $ 10 $~TeV and, following Eq.~\eqref{eq:resumeUnitarity}, the mass of the scalar $ H_3 $, the heavier analogue of the SM-like Higgs boson, is less constrained since the DLRM scale $ v_R $ is pushed to higher values.







\begin{table}
\renewcommand{\arraystretch}{1.1}
\begin{center}
\begin{tabular}{|c|c|c|c||c|}
\hline
Observable  & input & full fit ($ 1~\sigma $) & prediction ($ 1~\sigma $) & pull \\
\hline
\hline
  $ 10^5 \times G_F^{(0)} $ [GeV$ ^{-2} $]  & -- & $ 1.167 \pm 0.001 $ & -- & - \\
  $ M_Z^{(0)} $ [GeV] & -- & $ 91.188^{+0.014}_{-0.003} $ & -- & - \\
  $ M_h^{(0)} $ [GeV] &  $125.1 \pm 12 $ & $ 125 \pm 12 $ & $ 90^{+160}_{-50} $ & - \\
  $ M_W^{(0)} $ [GeV] & -- & $ 80.37^{+0.02}_{-0.01} $ & -- & - \\
  $ m_{top}^{pole,(0)} $ [GeV]  & $ 172.47 \pm 17 $ & $ 174^{+5}_{-4} $ & $ 174 ^{+5}_{-4} $ & - \\
  $ \Delta \alpha (M_Z^2) $ & $ 0.05898 \pm 0.0032 $ & $ 0.059 \pm 0.001 $ & $ 0.059 \pm 0.001 $ & - \\
  $ \alpha_s(M_Z^2) $ &  $ 0.1184 \pm 0 \pm 0.0012 $ & $ 0.1187^{+0.0009}_{-0.0015} $ & $ 0.119 \pm 0.003 $ & 0.0 \\
  $ c_R $ & $ [0.1, 0.99] $ & no bound & -- & - \\
  $ \epsilon$  & $ \geq 0 $ & $ 0.001^{+0.018}_{-0.001} $ & -- & - \\
  $ r $ & $ [0, 0.99] $ & no bound & -- & - \\
  $ w $ & $ \geq 0 $ & $ 5.7^{+1.5}_{-5.7} $ & -- & - \\
  $ \alpha_{124}, \alpha_2, \lambda_{1}, \lambda^\pm_{23}, \lambda_{4} $ & $ [-20, 20] $ & no bound & -- & - \\
  $ M_{H_1}, M_{H_2} $ & $ \geq 0 $ & no bound & -- & - \\
  $ M_{H_3} $ [TeV] & $ \geq 0 $ & $ 50^{+70}_{-50} $ & -- & - \\
  $ x_{} = \mu'_1 / \mu'_2 $ & -- & no bound & -- & - \\
\hline
\hline
  $ M_Z $ [GeV]  & $91.1876 \pm 0.0021$ & $ 91.187 \pm 0.002 $ & $ 91.16 \pm 0.03 $ & 1.0 \\
\hline
\hline
  $ M_{Z'}^2 $ [TeV$^2$] & -- & $ > 10^2 $ & -- & - \\
\hline
\end{tabular}
\end{center}
\vspace{-6mm}
\caption{Same scenario of Table~\ref{tab:fitlrm}, but with the constraint $ M_{W'} > 10 $~TeV imposed.}\label{tab:fit_w_MWR_bound}
\end{table}

%% file: tables_couplings.tex
\begin{table}
\begin{center}
\def\arraystretch{2.5}
\begin{tabular}{|c|c|c|c|}
\hline
 & $A^0_\mu$ & $Z^0_\mu$ & $Z'^0_\mu$\\
 \hline
$\bar{u}^i_L u^i_L$ 
& ${i} \frac{2}{3}e\gamma^\mu$ 
& ${i} \frac{e}{c_Ws_W}\gamma^\mu\left(\frac{1}{2}-\frac{2}{3}s_W^2-\frac{1}{6}\epsilon^2[(1+r^2)s_R^2-k^2s_R^4]\right)$ 
& ${i} \frac{es_R}{c_Rc_W}\left(-\frac{1}{6}\right)\gamma^\mu$\\
$\bar{u}^i_R u^i_R$ 
& ${i} \frac{2}{3}e\gamma^\mu$ 
& ${i} \frac{e}{c_Ws_W}\gamma^\mu\left(-\frac{2}{3}s_W^2+\frac{(4c_R^2-1)}{6}\epsilon^2[-w^2+c_R^2k^2]\right)$ 
& ${i} \frac{e}{s_Rc_Rc_W}\left(\frac{1}{2}-\frac{2}{3}s_R^2\right)\gamma^\mu$\\
$\bar{d}^i_L d^i_L$ 
& $-{i} \frac{1}{3}e\gamma^\mu$ 
& ${i} \frac{e}{c_Ws_W}\gamma^\mu\left(-\frac{1}{2}+\frac{1}{3}s_W^2-\frac{1}{6}\epsilon^2[(1+r^2)s_R^2-k^2s_R^4]\right)$ 
& ${i} \frac{es_R}{c_Rc_W}\left(-\frac{1}{6}\right)\gamma^\mu$\\
$\bar{d}^i_R d^i_R$ 
& $-{i} \frac{1}{3}e\gamma^\mu$ 
& ${i} \frac{e}{c_Ws_W}\gamma^\mu\left(\frac{1}{3}s_W^2-\frac{(2c_R^2+1)}{6}\epsilon^2[-w^2+c_R^2k^2]\right)$ 
& ${i} \frac{e}{s_Rc_Rc_W}\left(-\frac{1}{2}+\frac{1}{3}s_R^2\right)\gamma^\mu$\\
$\bar{\nu}^i_L \nu^i_L$ 
& $0$ 
& ${i} \frac{e}{c_Ws_W}\gamma^\mu\left(\frac{1}{2}+\frac{1}{2}\epsilon^2[(1+r^2)s_R^2-k^2s_R^4]\right)$ 
& ${i} \frac{es_R}{c_Rc_W}\left(\frac{1}{2}\right)\gamma^\mu$\\
$\bar{\nu}^i_R \nu^i_R$ 
&$0$ 
& ${i} \frac{e}{c_Ws_W}\gamma^\mu\left(\frac{1}{2}\epsilon^2[(1+r^2)-k^2s_R^2]\right)$ 
& ${i} \frac{e}{s_R c_R c_W}\left(\frac{1}{2}\right)\gamma^\mu$\\
$\bar{\ell}^i_L \ell^i_L$ 
& $-{i} e\gamma^\mu$ 
& ${i} \frac{e}{c_Ws_W}\gamma^\mu\left(-\frac{1}{2}+s_W^2+ \frac{1}{2} \epsilon^2[(1+r^2)s_R^2-k^2s_R^4]\right)$ 
& ${i} \frac{es_R}{c_Rc_W}\left(\frac{1}{2}\right)\gamma^\mu$\\
$\bar{\ell}^i_R \ell^i_R$ 
& $-{i} e\gamma^\mu$ 
& ${i} \frac{e}{c_Ws_W}\gamma^\mu\left(s_W^2-\frac{(2c_R^2-1)}{2}\epsilon^2[-w^2+c_R^2k^2]\right)$ 
& ${i} \frac{e}{s_Rc_Rc_W}\left(-\frac{1}{2}+s_R^2\right)\gamma^\mu$
\\
\hline
\end{tabular}
\end{center}
\caption{Couplings of neutral gauge bosons to quarks and leptons.}\label{tab:AZZpr-fermions}
\end{table}

\begin{table}
\begin{center}
\def\arraystretch{2.5}
\begin{tabular}{|c|c|c|}
\hline
 & $W^+_\mu$ & $W'^+_\mu$\\
 \hline
$\bar{u}^i_L d^j_L$ &  ${i} \frac{e}{\sqrt{2}s_W} V_L^{ij} \gamma_\mu $ & $ \mathcal{O} (\epsilon^2) $ \\
$\bar{u}^i_R d^j_R$ & ${i} \frac{e\sqrt{2}}{s_W}  r \epsilon^2 V_R^{ij} \gamma_\mu $ & ${i} \frac{e}{\sqrt{2}s_Rc_W}V_R^{ij} \gamma_\mu$
\\
\hline
\end{tabular}
\end{center}
\caption{Couplings of charged gauge bosons to quarks (the adaptation to leptons is straightforward).}\label{tab:WWpr-fermions}
\end{table}


\begin{table}
\begin{center}\small
\def\arraystretch{2.5}
\begin{tabular}{|c|c|c|}
\hline
 & $H_i^0$ & $A_i^0$  \\
 \hline
$\bar{u}^i_R u^j_L$ & $-\frac{i}{\epsilon v_R}\biggl(\frac{1}{1- r^2}\biggl(\frac{2r}{u_i} + w S_i \biggr)m_u^i \delta^{ij}+G_i V_R^{ia}m^a_d V_L^{ja*}\biggr)$
& $\frac{1}{\epsilon v_R}\biggl(\frac{1}{1- r^2}\biggl(\frac{2r}{u_i} + w S_i \biggr)m_u^i \delta^{ij}+G_i V_R^{ia}m^a_d V_L^{ja*}\biggr) $\\
$\bar{u}^i_L u^j_R$ & $-\frac{i}{\epsilon v_R}\biggl(\frac{1}{1- r^2}\biggl(\frac{2r}{u_i} + w S_i \biggr) m_u^i \delta^{ij}+G_i V_L^{ia}m^a_d V_R^{ja*}\biggr)$
& $-\frac{1}{\epsilon v_R}\biggl(\frac{1}{1- r^2}\biggl(\frac{2r}{u_i} + w S_i \biggr) m_u^i \delta^{ij}+G_i V_L^{ia}m^a_d V_R^{ja*}\biggr)$\\
$\bar{d}^i_R d^j_L$ & $\frac{i}{\epsilon v_R}\biggl( G_i V_R^{*ai}m^a_u V_L^{aj}- \frac{1}{1- r^2}\biggl(\frac{2r}{u_i} + w S_i \biggr) m^i_d \delta^{ij}\biggr)$
&  $\frac{1}{\epsilon v_R}\biggl( G_i V_R^{*ai}m^a_u V_L^{aj}- \frac{1}{1- r^2}\biggl(\frac{2r}{u_i} + w S_i \biggr) m^i_d \delta^{ij}\biggr)$ \\
$\bar{d}^i_L d^j_R$ &  $\frac{i}{\epsilon v_R}\biggl( G_i V_L^{*ai}m^a_u V_R^{aj}- \frac{1}{1- r^2}\biggl(\frac{2r}{u_i} + w S_i \biggr) m^i_d \delta^{ij}\biggr)$
& $-\frac{1}{\epsilon v_R}\biggl( G_i V_L^{*ai}m^a_u V_R^{aj}- \frac{1}{1- r^2}\biggl(\frac{2r}{u_i} + w S_i \biggr) m^i_d \delta^{ij}\biggr)$
\\
\hline
\end{tabular}
\caption{Couplings of neutral scalar bosons to quarks (the adaptation to leptons is straightforward). $H^0_3$ does not couple to fermions at this order.}\label{tab:HA-fermions}
\end{center}
\end{table}

\begin{table}
\begin{center}
{\small \def\arraystretch{2.5}
\begin{tabular}{|c|c|}
\hline
 & $H_i^+$   \\
 \hline
$\bar{u}^i_R d^j_L$ & $\sqrt{2} \times \frac{i}{\epsilon v_R}\biggl(\frac{1}{1- r^2}\biggl(\frac{2r}{u_i} + w S_i \biggr)m_u^i V_L^{ij}+G_i V_R^{ij}m_d^j\biggr)$  \\
$\bar{u}^i_L d^j_R$ &  $-\sqrt{2}\times \frac{i}{\epsilon v_R}\biggl(G_i m_u^i V_R^{ij} - \frac{1}{1- r^2}\biggl(\frac{2r}{u_i} + w S_i \biggr) V_L^{ij}m_d^j\biggr)$
\\
\hline
\end{tabular}}
\end{center}
\caption{Couplings of charged scalar bosons to quarks (the adaptation to leptons is straightforward).}\label{tab:H-fermions}
\end{table}

\begin{table}
\begin{center}
\def\arraystretch{2.5}
\begin{tabular}{|c|c|c|}
\hline
 & $G^+_W$ & $G^+_{W'}$ \\
 \hline
$\bar{u}^i_R d^j_L$ & $\frac{-i \sqrt{2}}{\epsilon v_Rk}
\left[m^{i}_u V_L^{ij}\left(1+\frac{2r^2}{k^2}\epsilon^2\right)-2r\epsilon^2 V_R^{ij}m^j_d\right]$ & $\frac{i \sqrt{2}}{v_Rk} V_R^{ij} m^{j}_d$ \\
$\bar{u}^i_L d^j_R$ &  $\frac{i \sqrt{2}}{\epsilon v_Rk}
\left[V_L^{ij}m^j_d\left(1+\frac{2r^2}{k^2}\epsilon^2\right)-2r\epsilon^2 m^{i}_u V_R^{ij}\right]
$ & $ \frac{-i \sqrt{2}}{v_Rk} m^i_u V_R^{ij} $ \\
\hline
\end{tabular}
\caption{Couplings of Golstdone bosons associated with charged gauge bosons to quarks (the adaptation to leptons is straightforward). }\label{GWGWpr-fermions}
\end{center}
\end{table}

\begin{table}
\begin{center}
\def\arraystretch{2.5}
\begin{tabular}{|c|c|c|c|}
\hline
 & $G^0_Z$ & $G^0_{Z'}$ & $h^0$\\
 \hline
$\bar{u}^i_L u^i_R$ & $\frac{m^i_u}{\epsilon v_Rk}
\left[1+\frac{w^4-c_R^4k^2}{2k^4}\epsilon\right]
$ & $ -c_R^2 \frac{m^i_u}{v_Rk} $ & $i \frac{m^i_u}{\epsilon v_Rk} +\mathcal{O}(\epsilon^2)$ \\
$\bar{d}^i_L d^i_R$ &  $-\frac{m^i_d}{\epsilon v_Rk}\left[1+\frac{w^4-c_R^4k^2}{2k^4}\epsilon\right]$ & $ c_R^2 \frac{m^i_d}{v_Rk} $ & $i \frac{m^i_d}{\epsilon v_Rk}+\mathcal{O}(\epsilon^2)$
\\
\hline
\end{tabular}
\end{center}
\caption{Couplings of Golstdone bosons associated with neutral gauge bosons to quarks (the adaptation to leptons is straightforward). The $\mathcal{O}(\epsilon^2)$ corrections to $h^0$ have a rather complicated expression, involving various contributions from the scalar potential.
}\label{tab:GZGZprh-fermions}
\end{table}

\begin{table}[h!]
\begin{center}
\begin{tabular}{|c| c| c |}
\hline
 & $H_i^0$& $ A_i^0$ \\
\hline
&&\\
$ W^\pm {W'}^\mp $ & $i g_R M_W F_i(r,w,t_i)$ &$ \mp g_R M_W F_i(r,w,t_i)$ \\
&&\\
$W^\pm G_{W'}^\mp$ & $i \frac{g_R}{2} \frac{M_W}{M_W'} F_i(r,w,t_i) (p'-p)_\mu$  & $ \frac{g_R}{2} \frac{M_W}{M_W'} F_i(r,w,t_i) (p'-p)_\mu$    \\
&&\\
$G^\pm_{W} W'^\mp$ & $i \frac{g_R}{2}  F_i(r,w,t_i) (p'-p)_\mu$  & $ \frac{g_R}{2}  F_i(r,w,t_i) (p'-p)_\mu $  \\
&&\\
$G_W^\pm G_{W'}^\mp$ & $- i \frac{g_R}{2} \frac{{M^2_{H{_i^0}}}}{M_W'} F_i(r,w,t_i)$  &$ \frac{g_R}{2} \frac{{M^2_{A{_i^0}}}}{M_W'} F_i(r,w,t_i)$    \\
&&\\
$Gh \; Gh'$ &$ -i \xi  \frac{g_R}{\sqrt 2} M_W F_i(r,w,t_i)$ &$ \xi \frac{g_R}{\sqrt 2} M_W F_i(r,w,t_i)$ \\
&&\\
\hline
\end{tabular}
\end{center}
\caption{Couplings of neutral (pseudo)scalar bosons to gauge bosons, Goldstone bosons and ghosts. }\label{tab:HA-GaugeBosons}
\label{tab:feyn}
\end{table}

\begin{table}
\vspace{1.truecm}
\begin{center}
\begin{tabular}{|c| c| c |c |c|}
\hline
 &$h^0$& $H_i^0 (i=1,2)$ & $A_i^0 (i=1,2)$ &$H_3^0$\\
\hline
&&&&\\
 $ W_\alpha W_\beta$& $i \frac{2} {v_R k \epsilon}(1- 2(w_h^2 -\frac{c^{h^0}}{k}) \epsilon^2)  $& $i \frac{2}{v_R k} \epsilon (4 r F_i(r,w,t_i)+ \frac{c^{H_i}}{k})$ &$0$ &$\frac{2 i}{k^2 v_R} c^{H_3}$ \\
&&&&\\
$W^\mp_\alpha {W_\beta '}^\pm$ & $i\frac{2 w_h}{v_R k } $ & $i \frac { 2 F_i(r,w,t_i)}{ v_R }$ &  $ \pm \frac {2 F_i(r,w,t_i)}{ v_R }$&${\cal O}(\epsilon)$ \\
&&&&\\
$W_\alpha ' W_\beta '$ & $i\frac{2( 1 + r^2+k c_{\chi_{R,r}}^{h^0})\epsilon}{v_R k} $ & $i\frac{2 \epsilon} {v_R } (-w S_i+c^{H_i}_{\chi_{R,r}})   $ &$ 0$& $i \frac{2} {v_R}$  \\
&&&&\\
$Z_\alpha Z_\beta$ & $i\frac{ 2} {v_R k \epsilon}(1- 2( z_h^2-\frac{c^{h^0}}{k})  \epsilon^2)  $ & $i\frac{2} {v_R k}(2 w S_i z_h+\frac{c^{H_i}}{k})\epsilon$ & $0$ &$\frac{2 i}{k^2 v_R} c^{H_3}$\\
&&&&\\
$Z_\alpha Z_\beta '$ &$-i \frac{2 z_h} {v_R k }   $ & $i\frac{2} {v_R k} w S_i  $ & $0$ &${\cal O}(\epsilon)$\\
&&&&\\
$Z_\alpha ' Z_\beta '$ &$i\frac{2 \epsilon} {v_R k} (c_R^2 k z_h + s_R^2 w^2 +k c_{\chi_{R,r}}^{h^0})  $ & $i\frac{2 \epsilon} {v_R  } (-(c_R^2 -s_R^2) w S_i+c^{H_i}_{\chi_{R,r}})  $ &$0$& $i \frac{2} {v_R}$\\
&&&&\\
\hline
\end{tabular}
\end{center}
\caption{Feynman rules relevant for setting unitarity bounds in the left-right model. All the couplings have to be multiplied by $g_{\alpha \beta}$. Only the LO terms up
to ${\cal O}(\epsilon)$ are shown except for the light Higgs boson where the
NLO are also given. The couplings to $H_3^0$ which are not needed are not given explicitly. The results are given in units  of the masses of the two 
interacting gauge boson given in the first column. 
 }\label{tab:hHAH-GaugeBosons}
\label{tab:feynGBH}
\end{table}

\begin{table}
\begin{center}
\hspace{-0.2cm}
\begin{tabular}{|c c c c c c|}
\hline
&$h^0$& $H_i^0$ &$ H_3^0 $& $A_i^0$ &$ H_i^{\pm}$\\
\hline
$ W$ &$i\frac{g_L^2}{2 }(1- 2 \tilde {w}_h^2 \epsilon^2)$& $i \frac{g_L^2}{2 }(1+w_{H_i} \epsilon^2)$& $i \frac{g_L^2}{2} d_{H_3} \epsilon^2$& 0 &$i\frac{g_L^2}{2}(1 - k^2 F_i^2 \epsilon^2) $ \\
&&&&&\\
$Z$ &$i \frac{g_L^2}{2 c_W^2} (1 -2  \tilde{z}_{h}^2  \epsilon^2  )$& $i \frac{g_L^2}{2 c_W^2}(1+ \tilde{z}_{H_i} \epsilon^2 )  $&$i \frac{g_L^2}{2 c_W^2} d_{H_3} \epsilon^2$& 0 & $i \frac{g_L^2 (c_W^2-s_W^2)^2}{2 c_W^2}( 1- c_{H_i^\pm}\epsilon^2)$ \\
\hline
\end{tabular}
\end{center}
\caption{Couplings of $W_\alpha W_\beta$ and $Z_\alpha Z_\beta$ with various scalars. The expressions 
of the various  coefficients of the $\epsilon^2$ terms are given in Eq.~(\ref{eq:coeffGBH}). All these couplings are multiplied by $g_{\alpha, \beta}$. 
}
\label{tab:feynHHW2}
\end{table}

\clearpage

\begin{table}
\vspace{1.truecm}
\begin{center}
\begin{tabular}{|c| c| c|}
\hline
 $W_\alpha Z_\beta '$& $ W_\alpha ' Z_\beta $&$ W_\alpha ' Z_\beta ' $ \\
\hline
&&\\
 $-i\frac{ 2} {v_R k}   w S_i   $& $-i\frac{ 2 } {v_R } F_i $&$-i\frac{ 2 s_R^2 } {v_R } k  F_i $\\
&&\\
\hline
\end{tabular}
\end{center}
\caption{Same as in Table~\ref{tab:feynGBH} but for the two charged scalars $H_i^{\pm}$. }
\label{tab:feynGBWZ}
\end{table}


\begin{table}
\begin{center}
\begin{tabular}{|c| c| c |c |}
\hline
 &$Z_\mu$& $Z'_\mu$ &$\gamma $\\
\hline
&&&\\
 $ W_\alpha^- W_\beta^+$& $ i e\frac{c_W} {s_W}  $& $- i e \frac{c_R s_R c_W} {  s_W^2} k z_h \epsilon^2 $
& $ie$ \\
&&&\\
$W_\alpha^- W_\beta^{'+}$ & $-i e 2 r \frac{  s_R} { s_W^2} \epsilon^2$ & $i e  2 r  \frac{  c_R} { s_W}\epsilon^2$& $0$ \\
&&&\\
$W_\alpha^{'-} W_\beta^{'+}$ & $-i e ( \frac{s_W}{c_W} - \frac{c_R^2}{  c_W s_W} k z_h \epsilon ^2) $ & $ i e
 \frac{c_R}{c_W}( \frac{1}{s_R} + s_R k z_h\epsilon ^2)  $ & $i e$ \\
&&&\\
\hline
\end{tabular}
\end{center}
\caption{Triple gauge couplings involving either one $Z$ or one $Z'$ boson. These couplings have to be multiplied by $g_{\alpha \beta}(p_+-p_-)_\mu+g_{\mu \beta}(q-p_+)_\alpha+g_{\mu \alpha}(p_--q)_\beta$ where ($p_+$, $p_-$, $q$) are the incoming momenta of the positively, negatively charged and neutral particles respectively. }
\label{tab:feyn3GB}
\end{table}

\begin{table}
\begin{tabular}{|c| c |c|}
\hline
&$Z_\alpha $&$A_\alpha$ \\
\hline
&&\\
 $ G^\pm W_\beta '$& $ -i  g_L \frac{w_h} { c_W} M_{W'} \epsilon  $
& $0$\\
&&\\
$ G'^\pm W_\beta$ & $0 $ &$ 0$\\
&&\\
$ G'^\pm W_\beta '$ & $- i g_L\frac{s_W^2}{c_W}(1-\frac{1}{s_W^2}(1+r^2-s_R^2 k z_h)\epsilon^2) M_{W'}$ &$ i  g_L s_W M_{W'}$\\
&&\\
$ G^\pm G'^\mp$& $i  g_L \frac{1}{2 c_W} w_h \epsilon $ &$ 0$\\
& &\\
$ G'^\pm G'^\mp$ & $ i  g_L \frac{s_W^2}{c_W}\biggl(1- \frac{1}{2 s_W^2} \bigr(
1+r^2+( c_R^2-s_R^2) k z_h \bigr) \epsilon^2 \biggr) $&$- ig_L s_W$ \\
&&\\
$ g_{W'} g_{W'}$ &$g_L\frac{s_W^2}{c_W}(1- \frac{c_R^2}{s_W^2}k z_h \epsilon^2)$
& $-g_L {s_W}$\\
&&\\
$ g_{W'} g_{W}$ & $0$& $-2 g_L r s_R c_W  \epsilon^2$\\
&&\\
\hline
\hline
&&\\
$Z_\beta  W_\gamma '^\pm W_\delta '^\mp$ &$i g_L^2\frac{ s_W^4}{c_W^2}\biggl(1-2 \frac{c_R^2}{s_W^2} k z_h \epsilon^2 \biggr)$& - \\ 
&&\\
$Z_\beta  G'^\pm G'^\mp$ & $i 2 g_L^2 \frac{s_W^4}{c_W^2}\biggl(1+\frac{1}{4 s_W^4}(c_g - 4 s_W^2 d_g )\epsilon^2\biggr)$ & $-2 i g_L^2 \frac{s_W^3}{c_W}(1-\frac{1}{2 s_W^2}(1+r^2+ d_g)\epsilon^2)$\\
&& \\
$  Z_\beta  G'_0 G'_0$ & $i\frac{g_L^2} { 2 c_W^2}(c_R^2(1+r^2)-s_R^2 k z_h)\epsilon^2$
&$0$\\
&&\\
$ A_\beta   G'^\pm G'^\mp$ & $- 2 i g_L^2 \frac{s_W^3}{c_W}(1-\frac{1}{2 s_W^2}(1+r^2 + (c_R^2-s_R^2) k z_h)\epsilon^2) $& $2 i g_L^2 s_W^2$ \\
&&\\
$A_\beta   W_\gamma '^\pm W_\delta '^\mp$ & $ -i g_L^2 \frac{s_W^3}{c_W}(1-\frac{c_R^2}{ s_W^2} k z_h \epsilon^2)  $& $ i g_L^2 s_W^2$ \\
&&\\
\hline
\end{tabular}
\caption{Triple and quadruple gauge couplings involving  a $Z$  boson and at least
one unphysical Goldstone boson. $c_g=(1- 4 s_W^2)(1+r^2)$, $d_g=(c_R^2 - s_R^2) k z_h$ and $M_{W'}$ is the mass of the $W'$ up to order $\epsilon^2$. All the 
couplings are multiplied by $g_{\alpha \beta}$ except the quadruple couplings 
involving only physical gauge bosons, the triple gauge couplings with two 
unphysical gauge bosons or two ghosts. The former should be multiplied by 
$g_{\gamma \alpha} g_{\delta \beta} + g_{\gamma \beta} g_{\delta \alpha} 
-2 g_{\gamma \delta} g_{\alpha \beta}$ while the latter with unphysical 
gauge bosons should be multiplied $(p_--p_+)_\alpha$ with $p_\pm$ the incoming 
momenta of $G^\pm$. The triple couplings with ghosts are multiplied by $p_\alpha$, the outgoing momenta of the ghost. }
\label{tab:feynZGG}
\end{table}
\begin{table}
\begin{center}
\begin{tabular}{|c| c |}
\hline
&$Z_\alpha '$ \\
\hline
&\\
 $ G^\pm W_\beta '$& $ -i  g_L w_h \frac{s_R s_W} {c_R c_W} M_{W'} \epsilon  $\\
&\\
$ G'^\pm W_\beta $ & $ -i \frac{ 2  r}{c_R} \epsilon^2$ \\
&\\
$ G'^\pm W_\beta '$ & $- i g_L\frac{s_R s_W}{c_R c_W}(1-\frac{1}{2}(1+r^2+c_R^2 k z_h)\epsilon^2) M_{W'}$ \\
&\\
$ G^\pm G'^\mp$& $i  g_L \frac{s_R s_W}{c_R c_W} \frac{w_h}{2} \epsilon $\\
& \\
$ G'^\pm G'^\mp$ & $- i  g_L \frac{s_W(c_R^2-s_R^2)}{2 c_R c_W s_R}\biggl(1- \frac{s_R^2}{c_R^2-s_R^2} \bigr(
1+r^2+ 2 c_R^2 k z_h \bigr) \epsilon^2 \biggr) $ \\
&\\
$ g_{W'} g_{W'}$ &$-g_L\frac{c_R s_W}{c_W s_R}(1+ s_R^2 k z_h \epsilon^2)$
\\
&\\
\hline
\hline
&\\
$Z_\beta  W_\gamma '^\pm W_\delta '^\mp$ &  $- i g_L^2 \frac{s_W^3 c_R}{c_W^2 s_R }(1-\frac{1}{s_W^2} (c_R^2- s_R^2 s_W^2) k z_h \epsilon^2)$\\
& \\
$A_\beta  W_\gamma '^\pm W_\delta '^\mp$ &  $ i g_L^2 \frac{s_W^2 c_R}{c_W s_R }(1+s_R^2  k z_h \epsilon^2)$\\
& \\
$A_\beta  G'^\pm G'^\mp$ &  $ i g_L^2 (c_R^2-s_R^2) \frac{s_W^2}{c_R c_W s_R}(1+\frac{s_R^2}{c_R^2-s_R^2}\bigl(1+r^2+ 2 c_R^2 k z_h\bigr)\epsilon^2)$\\
& \\
$Z_\beta  G'^0 G'^0$ &  $ i g_L^2 \frac{c_R s_R s_W}{2 c_W^2}(1+r^2 -w^2+ k z_h)\epsilon^2 
$\\
& \\
&  $- i g_L^2 (c_R^2-s_R^2) \frac{s_W^3}{c_R c_W^2 s_R}(1+\frac{1}{c_R^2-s_R^2}\bigl(1+r^2\bigr. $\\
$Z_\beta  G'^\pm G'^\mp$ & \\
&
$\bigl.+ \frac{s_R^2 w^2}{2 s_W^2}(-3+2 c_W^2)-\frac{-2 +c_W^2}
{s_W^2}(-1+ 2 c_R^2 s_R^2) k z_h\bigr)\epsilon^2)$\\
& \\
\hline
\end{tabular}
\caption{Triple and quadruple gauge couplings involving  a $Z'$  boson and at least
one unphysical Goldstone boson. For the factor multiplying these couplings see Table~\ref{tab:feynZGG}. }
\label{tab:feynZpGG}
\end{center}
\end{table}

\begin{table}
\begin{center}
\begin{tabular}{|c| c |}
\hline
&\\
 $W_\alpha G_0 W_\beta'$& $ i g_L w_h  \epsilon M_{W'} $
 \\
&\\
$W_\alpha G^\pm Z_\beta '$ & $ i \frac{g_L}{c_R} z_h \epsilon M_{W'}$  \\
&\\
$W_\alpha G^\pm G'_0$& $- g_L \frac{z_h}{2} \epsilon $ \\
& \\
$W_\alpha G'^\pm G_0$& $- g_L \frac{w_h}{2} \epsilon  $ \\
& \\
\hline
\hline
&\\
$ W_\alpha W_\beta G'^\pm G'^\mp$ & $i \frac{g_L^2}{2} (1 + r^2) \epsilon^2$ \\
&\\
$ W_\alpha W_\beta G'_0 G'_0$ & $i  \frac{g_L^2}{2}  (c_R^2(1+r^2)-s_R^2 k z_h)\epsilon^2$
\\
&\\
\hline
\end{tabular}
\end{center}
\caption{Triple and quadruple  gauge couplings involving  a $W$  boson and at least
one unphysical Goldstone boson. Only the couplings contributing at $\epsilon^2$
to $\Delta r$ are shown. All these couplings are multiplied by $g_{\alpha \beta}$ except the triple couplings to two unphysical gauge bosons which are 
multiplied by $(p_\pm-p_0)_\alpha$ with $p_{\pm,0}$ the incoming momenta of 
$G_{\pm,0}$. } 
\label{tab:feynWGG}
\end{table}

\begin{table}
\begin{center}
\begin{tabular}{|c c|| c c|}
\hline
$W_\alpha H_i^{\pm}$&&$Z_\alpha$&\\
\hline
$ h^0$ &${\cal O}(\epsilon^2)$& $ A_i^{0} h^0$& ${\cal O}( \epsilon^2)$\\
&&&\\
$ H_i^{0}$ &$i\frac{g_L}{2}\bigl(1+ \frac{1}{2}(w_{H_i}- F_i^2 k^2)\epsilon^2 \bigr)  $& $ A_i^{0} H_i^0$&$ \frac{g_L}{2 c_W} \bigl(1+\frac{1}{2}(z_H+ 2 (b_{A_i}+b_{H_i}))\epsilon^2\bigr)$\\
&&&\\
$ A_i^{0}$ &$ \frac{g_L}{2}\bigl(1- \frac{1}{2}(w_{A_i}+ F_i^2 k^2)\epsilon^2 \bigr) $&$ H_i^{\pm} H_i^{\mp}$ &$i \frac{g_L (c_W^2-s_W^2)}{2 c_W}\biggl(1-\frac{1}{c_W^2-s_W^2}(\frac{z_H}{2}+  F_i^2 k^2)  \epsilon^2 \biggr)$ \\
&&&\\
$H_3^0$ &$ i\frac{g_L}{2} b_{H_3}^i \epsilon$&$A_i^0 H_3^0$ &$ \frac{g_L}{2 c_W} b_{H_3}^i  \epsilon$ \\
\hline
\end{tabular}
\end{center}
\caption{Couplings of $W_\alpha$ and $Z_\alpha$ with two scalars ${\cal H}_1 {\cal H}_2$. The couplings with $i \neq j$ are ${\cal O} (\epsilon^2)$ and thus irrelevant here. These couplings are multiplied by the difference of the incoming momenta of the two scalars $(p_{{\cal H}_2}-p_{{\cal H}_1})_\alpha$ where ${\cal H}_1=H^0$ and ${\cal H}_2=H^+$ in the former case. }
\label{tab:feynHHW} 
\end{table}

%% file: biblio_6.tex
\newpage{\pagestyle{empty}\cleardoublepage}